\documentclass[reprint,superscriptaddress,amsmath,amssymb,aps,pre]{revtex4-1}

\usepackage{amsmath}
\usepackage{graphicx}
\usepackage{dcolumn}
\usepackage{bm}
\usepackage{color}
\usepackage{bbold}
\usepackage{soul}
\usepackage{subfigure}
\newcommand{\newc}{\newcommand}
\newc{\beq}{\begin{equation}}
\newc{\eeq}{\end{equation}}
\newc{\kt}{\rangle}
\newc{\br}{\langle}
\newc{\beqa}{\begin{eqnarray}}
\newc{\eeqa}{\end{eqnarray}}
\newc{\longra}{\longrightarrow}

\makeatletter
\let\Hy@backout\@gobble
\makeatother

\begin{document}

\title{Homoclinic orbit expansion of arbitrary trajectories in chaotic systems: classical action function and its memory}

\author{Jizhou Li}
\affiliation{Department of Physics and Astronomy, Washington State University, Pullman, Washington 99164-2814, USA}
\affiliation{Department of Physics, Tokyo Metropolitan University, Minami-Osawa, Hachioji, Tokyo 192-0397, Japan}
\author{Steven Tomsovic}
\affiliation{Department of Physics and Astronomy, Washington State University, Pullman, Washington 99164-2814, USA}

\date{\today}

\begin{abstract}
Special subsets of orbits in chaotic systems, e.g. periodic orbits, heteroclinic orbits, closed orbits, can be considered as skeletons or scaffolds upon which the full dynamics of the system is built. In particular, as demonstrated in previous publications [Phys.~Rev.~E {\bf 95}, 062224 (2017), Phys.~Rev.~E {\bf 97}, 022216 (2018)], the determination of homoclinic orbits is sufficient for the exact calculation of classical action functions of unstable periodic orbits, which have potential applications in semiclassical trace formulas. Here this previous work is generalized to the calculation of classical action functions of arbitrary trajectory segments  in multidimensional chaotic Hamiltonian systems. The unstable trajectory segments' actions are expanded into linear combinations of homoclinic orbit actions that shadow them in a piece-wise fashion.  The results lend themselves to an approximation with controllable exponentially small errors, and which demonstrates an exponentially rapid loss of memory of a segment's classical action to its past and future.  Furthermore, it does not require an actual construction of the trajectory segment, only its Markov partition sequence. An alternative point of view is also proposed which partitions the trajectories into short segments of transient visits to the neighborhoods of successive periodic orbits, giving rise to a periodic orbit expansion scheme which is equivalent to the homoclinic orbit expansion. This clearly demonstrates that homoclinic and periodic orbits are equally valid skeletal structures for the tessellation of phase-space dynamics.
\end{abstract}


\maketitle

\section{Introduction}
\label{Introduction}

Following some insights and work of Poincar\'e~\cite{Poincare99} and cycle expansions~\cite{Artuso90a,Artuso90b}, Cvitanovi{\'c} discussed periodic orbits as a skeleton of classical and quantum chaos~\cite{Cvitanovic91}.  Crudely speaking, the idea is that calculating convergent expressions of dynamical averages in chaotic systems can be expressed in terms of unstable periodic orbits and be dominated by the shorter ones.  In effect, the infinity of orbits in the hyperbolic flow resummed in averages can be reduced to this skeleton.  It turns out that other special sets of orbits, such as closed orbits~\cite{Du88a,Du88b,Friedrich89} or homoclinic (heteroclinic) orbits~\cite{Tomsovic91b,Tomsovic93}, can also be thought of as providing a skeleton, depending on the circumstances.  This is not surprising in the sense that exact relations can be established between such special sets.  For example, the relations between periodic orbits and homoclinic orbits for classical action functions were given in Refs.~\cite{Li17a,Li18} and stability exponents in Ref.~\cite{Li19a}.  In a sense, the work in this article is the opposite of that in the calculation of dynamical averages, i.e.~such skeletons can also be used to predict every microscopic feature of the dynamics.  In particular, it is shown here that the classical action function for any arbitrary unstable trajectory segment can be exactly related either to properties of particular homoclinic  or periodic orbits; this gives a generalization of~Ref.~\cite{Li18}.

An exact scheme is introduced to expand unstable trajectories into sequences of simple homoclinic orbits that shadow them in a piece-wise fashion, and express their classical action functions in terms of sums of the homoclinic orbit actions plus certain phase-space areas as connectors between successive homoclinic orbits. An added benefit is that the action functions of extremely long unstable trajectory segments can be accurately calculated, even for lengths beyond which the trajectories can be directly followed in detail due to exponential propagation of errors.  The method relies on a collection of simple homoclinic orbits with relatively short transit times~\cite{Rom-Kedar90}, which can be computed effortlessly with existing stable algorithms~\cite{Li17}.  Most importantly, the results developed here may provide a general scheme for the computation of unstable trajectories in chaotic systems from the sole knowledge of homoclinic orbits, which may provide an alternative route to the investigation of chaotic phenomena in Hamiltonian systems.  An application to the calculation of periodic orbit actions is also introduced, which may enable a resummation of the Gutzwiller trace formula~\cite{Gutzwiller71} (or the semiclassical $\zeta$ function) in terms of homoclinic orbits, although more work in this direction is currently under investigation. 

An alternative route is also introduced which is the same in spirit as the cycle expansion~\cite{Chaosbook1}.  It  treats the trajectories as shadowed by short periodic orbits in a piecewise fashion instead of the homoclinic orbits. The resulting formulas provide an exact expression for the action correction to the original cycle expansion, which is expressed in terms of the same collection of simple homoclinic orbits. Therefore, periodic and homoclinic orbits are equivalent fundamental structures upon which the full dynamics is built.

This article is organized as follows. Sec.~\ref{Basic Concepts and Definitions} introduces the basic concepts and definitions. The main contents of the current work are introduced in Sec.~\ref{Homoclinic expansion}.  Exact relations are derived along with an asymptotic form that does not require construction of the orbit. This is applied to the computation of long periodic orbit actions and a study of its accuracy. Sec.~\ref{Conclusions} summarizes the article and points to the direction of future research. 

\section{Basic Concepts}
\label{Basic Concepts and Definitions}

\subsection{Covariant Lyapunov vectors}
\label{Covariant Lyapunov vectors}

Let us consider an $(f+1)$-degree-of-freedom Hamiltonian system for which the dynamics is highly chaotic. With energy conservation and applying the Poincar\'{e} surface of section technique \cite{Poincare99}, the Hamiltonian flow is reduced to a discrete symplectic map $M$ on the $2f$-dimensional phase space $z=(\mathbf{q},\mathbf{p})=(q_1, \cdots, q_{f}, p_1, \cdots, p_{f})$. The \textit{trajectory} (or \textit{orbit}) of a phase-space point $z_0$, denoted by $\lbrace z_0 \rbrace$, is the bi-infinite collection of all $M^{n}(z_0)$: 
\begin{equation}\label{eq:Orbit}
\lbrace z_0 \rbrace =\lbrace \cdots,z_{-1},z_0,z_1,\cdots \rbrace  \nonumber
\end{equation}
where $z_n = M^n(z_0)$ for all integers $n$. Assuming the dynamics is almost everywhere hyperbolic, then the corresponding Poincar\'{e} map $M$ is also hyperbolic. Intuitively speaking, the map stretches any given phase-space region along its unstable directions and compresses along the stable directions, then folds and remixes it with other parts of the phase space.  The stability matrix at $z_0$, denoted by $DM(z_0)$, is defined by
\begin{equation}\label{eq:Stability matrix at y}
DM(z_0) = \frac{\partial M}{\partial z}(z_0)
\end{equation}
where $DM(z_0)$ characterizes the tangent dynamics at $z_0$ under one iteration of $M$. The stability matrix of the $n$-th compound mapping is 
\begin{equation}\label{eq:Stability matrix at y compound}
DM^n(z_0) = \frac{\partial M^n}{\partial z}(z_0)
\end{equation}
that characterizes the tangent dynamics at $z_0$ under $n$ iterations of $M$. Under the hyperbolicity assumption, there exists an invariant Oseledec splitting~\cite{Oseledec68} of the tangent space of $z_0$ into exponentially expanding and contracting directions under the asymptotic map $\lim_{n\to \infty} DM^{n}(z_0)$. These directions are given by the \textit{covariant} \textit{Lyapunov} \textit{vectors} (CLV) \cite{Ginelli07,Ginelli13,Kuptsov12} at $z_0$. Denote the CLV at $z_n$ by $\mathbf{v_{i}}(z_n)$, which are covariant in the sense that
\begin{equation}\label{eq:CLV}
DM(z_n) \mathbf{v_i}(z_n) = \lambda_i(z_n) \mathbf{v_i}(z_{n+1})
\end{equation}
where $\mathbf{v_{i}}(z_n)$ and $\mathbf{v_i}(z_{n+1})$ are vectors with unit norms, and $\lambda_i(z_n)$ is the local expanding (for $1 \leq i \leq f$) or contracting (for $ f+1 \leq i \leq 2f$) factor of $z_n$. $\mathbf{v_i}(z_n)$ ($1\leq i \leq f$) yields the $i$-th most rapidly expanding direction along the unstable manifold of $z_n$, and $\mathbf{v_{f+i}}(z_n)$ ($1\leq i \leq f$) yields the $(f-i+1)$-th most rapidly contracting direction along the stable manifold of $z_n$. Consequently, the unstable manifold of $z_n$, denoted by $U(z_n)$, is the $f$-dimensional hyper-surface spanned by the streamlines of $[\mathbf{v_1}(z_n), \cdots, \mathbf{v_{f}}(z_n)]$, and the stable manifold of $z_n$, denoted by $S(z_n)$, is the $f$-dimensional hyper-surface spanned by the streamlines of $[\mathbf{v_{f+1}}(z_n), \cdots, \mathbf{v_{2f}}(z_n)]$.    

The Lyapunov spectrum of $z_0 $, namely $\mu^{(i)}(z_0)$ ($1 \leq i \leq 2f$), is 
\begin{equation}
\label{eq:Lyapunov exponent}
\mu^{(i)}(z_0) = \lim_{N\to \infty} \frac{1}{N} \sum_{n=0}^{N-1} \ln |\lambda_i(z_n)| .
\end{equation}
For Hamiltonian systems, the Lyapunov exponents come in pairs: $\mu^{(i)}(z_0) = -\mu^{(2f+1-i)}(z_0)$ for $i=1, \dots, f$.  Assume the positive Lyapunov spectrum of all relevant points $z_0$ are bounded away from zero:
\begin{equation}
\label{eq:Lyapunov spectrum}
\mu^{(1)}(z_0) \geq  \cdots \geq \mu^{(f)}(z_0) >0\ . 
\end{equation} 
Typically, $\mu^{(i)}(z_0)$ is almost-everywhere independent of $z_0$ for ergodic trajectories, and therefore the $z_0$-dependence can be removed. 

For periodic orbits: $y_0 = M^{T}(y_0)$, the Lyapunov exponents reduce to 
\begin{equation}
\label{eq:Lyapunov exponent periodic}
\mu^{(i)}_{\lbrace y_0\rbrace} \equiv \mu^{(i)}(y_0) =  \frac{1}{T} \sum_{n=0}^{T-1} \ln |\lambda_i(y_n)| ,
\end{equation}
which has a dependence on initial conditions since they are not ergodic.  Other special sets may also have this property.  To emphasize the difference between periodic orbits and ergodic trajectories, the $\mu^{(i)}_{\lbrace y_0\rbrace}$ are referred to as \textit{stability} \textit{exponents} of $\lbrace y_0 \rbrace$.  

\subsection{Symbolic dynamics}
\label{Symbolic dynamics}

Let $x$ be a hyperbolic fixed point, i.e., $M(x)=x$. Denote the unstable and stable manifolds of $x$ by $U(x)$ and $S(x)$, respectively.  Typically, $U(x)$ and $S(x)$ are $f$-dimensional invariant hypersurfaces that intersect infinitely many times and form a complicated pattern called a homoclinic tangle~\cite{Poincare99,Easton86,Rom-Kedar90}. It is well-known that generating Markov partitions to the phase space \cite{Bowen75,Gaspard98} exist, under which the mapping $M$ is conjugate to a subshift of finite type on bi-infinite symbolic strings \cite{Hadamard1898,Birkhoff27a,Birkhoff35,Morse38} representing phase-space itineraries of orbits. The cells of the partition $\mathcal{V} = [V_0, V_1, \cdots, V_K]$ are $2f$-dimensional curvilinear parallelograms, which are dubbed  ``vertical slabs" in \cite{Wiggins88}. Under the mapping $M$, the $\mathcal{V}$ cells are stretched along the unstable directions, contracted along the stable directions, and mapped into a set of cells $\mathcal{H} = [H_0, H_1, \cdots, H_K]$ (where $H_{i}=M(V_{i})$) that intersect with $\mathcal{V}$ to create mixing (see upper panel of Fig.~\ref{fig:Markov}). Letting $s_i \in \lbrace 0,\cdots, K \rbrace$ ($\forall i \in \mathbb{Z}$) be the integer digits that labels the cells, then under successive inverse mappings, the intersections 
\begin{equation}\label{eq:Markov partition intersections vertical}
V_{s_0s_1\cdots s_{n-1}}\equiv \bigcap_{i=0}^{n-1}M^{-i}(V_{s_i})
\end{equation}
become a family of cells whose widths along $U(x)$ decrease exponentially with $n$. Similarly, under successive forward mappings, the intersections 
\begin{equation}\label{eq:Markov partition intersections horizontal}
H_{s_{-n}s_{-n+1}\cdots s_{-1}} \equiv \bigcap_{i=1}^{n}M^{i-1}(H_{s_{-i}})
\end{equation}
become a family of cells whose widths along $S(x)$ decrease exponentially with $n$. From their definitions it is easy to see that $H_{\gamma}=M^n(V_{\gamma})$, where $\gamma = s_0 \cdots s_{n-1}$ denotes an arbitrary finite string of length $n$. 

For systems with $f=1$, both $U(x)$ and $S(x)$ are $1$-dimensional curves. Together they form the boundaries of $V_i$ and $H_i$; see Figs.~\ref{fig:horseshoe} and~\ref{fig:Markov} for illustrations. The intersection $H_{s_{-n}\cdots s_{-1}} \cap V_{s_0\cdots s_{n-1}}$ therefore localizes an exponentially small region in phase space, as demonstrated by Fig.~\ref{fig:Shrinking_Cells}. In the limiting case of $n \to \infty$, such intersections create a Cantor set of points on which the dynamics is topologically conjugate to a subshift of finite type on bi-infinite symbolic strings. Each point $z_0$ from the Cantor set is assigned a symbolic string
\begin{equation}\label{eq:symbolic code}
z_0 = \lim_{n \to \infty} H_{s_{-n}\cdots s_{-1}} \cap V_{s_0\cdots s_{n-1}}  \Rightarrow \cdots s_{-2}s_{-1} \cdot s_{0}s_{1}s_{2}\cdots  
\end{equation}
where each character $s_n$ in the sequence denotes the cell to which $M^{n}(z_0)$ belongs: $M^{n}(z_0) = z_n \in V_{s_n}$, $s_n \in \lbrace 0,\cdots, K \rbrace$. The separation dot in the middle indicates the current iteration: $z_0 \in V_{s_0}$.  The symbolic code gives an ``itinerary" of $z_0$ under successive forward and backward iterations, in terms of the Markov cells in which each iteration lies. The mapping $M$ under the symbolic dynamics is then reduced to a simple shift of the dot in the code:
\begin{equation}\label{eq:Shift map}
M^n(z_0)=z_n \Rightarrow \cdots s_{n-1} \cdot s_{n}s_{n+1} \cdots  \nonumber.
\end{equation}
Points along the same trajectory have the same symbolic strings but shifting separation dots. Therefore, a trajectory can be represented by the symbolic string without the dot:
\begin{equation}\label{eq:symbolic code general orbits}
\lbrace z_0 \rbrace \Rightarrow \cdots s_{-2}s_{-1} s_{0}s_{1}s_{2}\cdots
\end{equation}  

\begin{figure}[ht]
\centering
{\includegraphics[width=9cm]{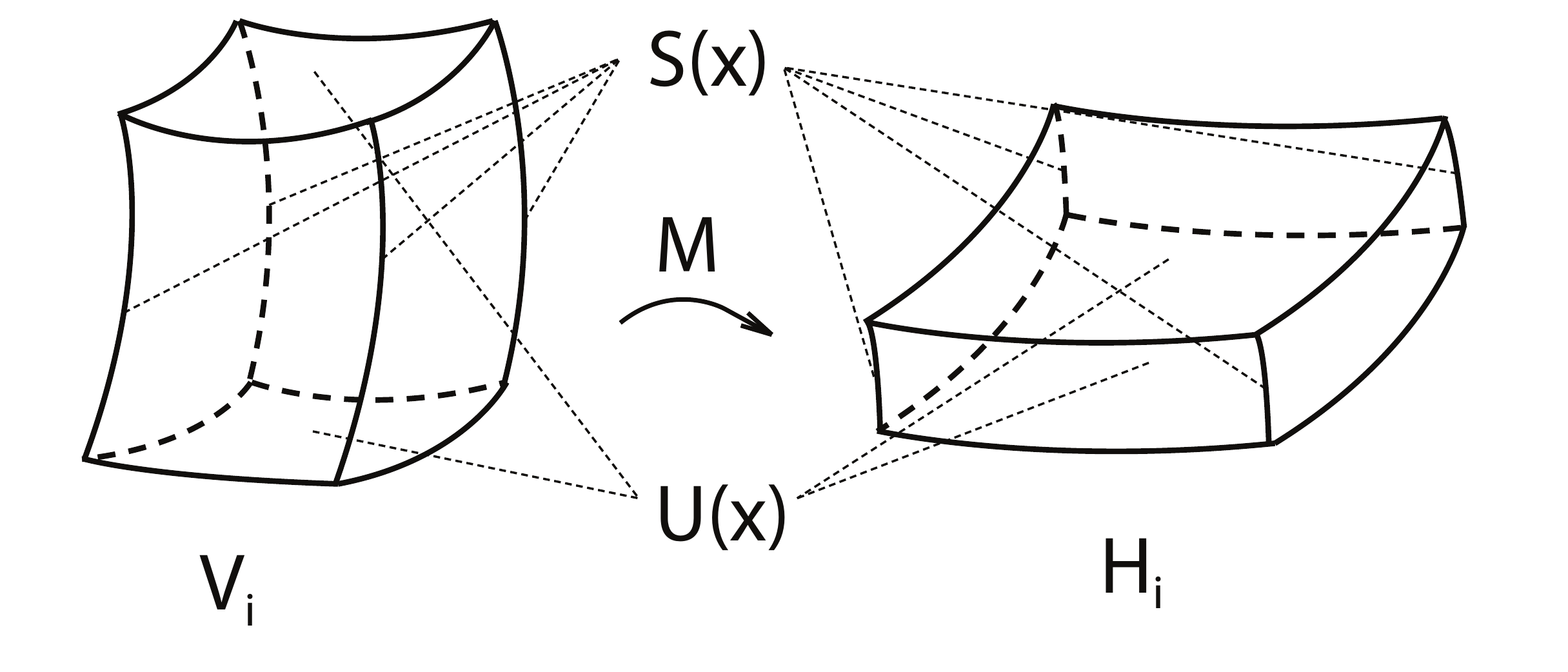}}
 \caption{Vertical slabs $V_i$ and horizontal slabs $H_i=M(V_i)$ as generating Markov partition for the symbolic dynamics of multidimensional chaotic systems. Notice that $V_i$ and $H_i$ are actually high-dimensional curvy ``parallelograms'', and only a $3$-dimensional illustration is possible. The horizontal direction is aligned with $U(x)$, and the vertical aligned with $S(x)$. Both $U(x)$ and $S(x)$ are $f$-dimensional surfaces.  The expectation is that the bounding side surfaces of $V_i$ and $H_i$ are spanned by two portions of each of the surfaces formed by the set of CLVs missing the $j^{th}$ member, i.e.~$[\mathbf{v_1}(x), \cdots, \mathbf{v_{j-1}}(x),\mathbf{v_{j+1}}(x), \cdots, \mathbf{v_{2f}}(x)]$, for $j=1,\cdots,2f$, successively.  }
\label{fig:Markov_Partition_3D}
\end{figure}

For systems with $f \geq 2$, the partition cells $\mathcal{V} = [V_0, V_1, \cdots, V_K]$ become ``vertical slabs" which are mapped into ``horizontal slabs" $\mathcal{H} = [H_0, H_1, \cdots, H_K]$ that intersect $\mathcal{V}$~\cite{Wiggins88}.  As illustrated schematically by Fig.~\ref{fig:Markov_Partition_3D},
some edges of $V_i$ are located on $S(x)$, and others located on $U(x)$. Under one iteration of $M$, $V_i$ is contracted along the stable edges and expanded along the unstable edges, and deformed into $H_i=M(V_i)$. To the authors' knowledge, there has not been an explicit study of the bounding surfaces of $V_i$ and $H_i$ in multidimensional systems.  A reasonable conjecture is that the side surfaces of $V_i$ and $H_i$ are spanned by two portions of each of the invariant manifolds beginning from the CLVs $[\mathbf{v_1}(x), \cdots, \mathbf{v_{j-1}}(x),\mathbf{v_{j+1}}(x), \cdots, \mathbf{v_{2f}}(x)]$, i.e.~excluding $\mathbf{v_j}(x)$, and by letting $j=1, \cdots, 2f$. In such a way $2f$ codimension-1 surfaces can be created, which are the natural extension of the side surfaces of $V_i$ and $H_i$ in the $f=1$ case. The contracting and expanding directions of $V_i$ under the mapping are therefore governed by the intricate way that the CLV surfaces form its side surfaces, which is left for future studies.  
\begin{figure}[ht]
\centering
{\includegraphics[width=8cm]{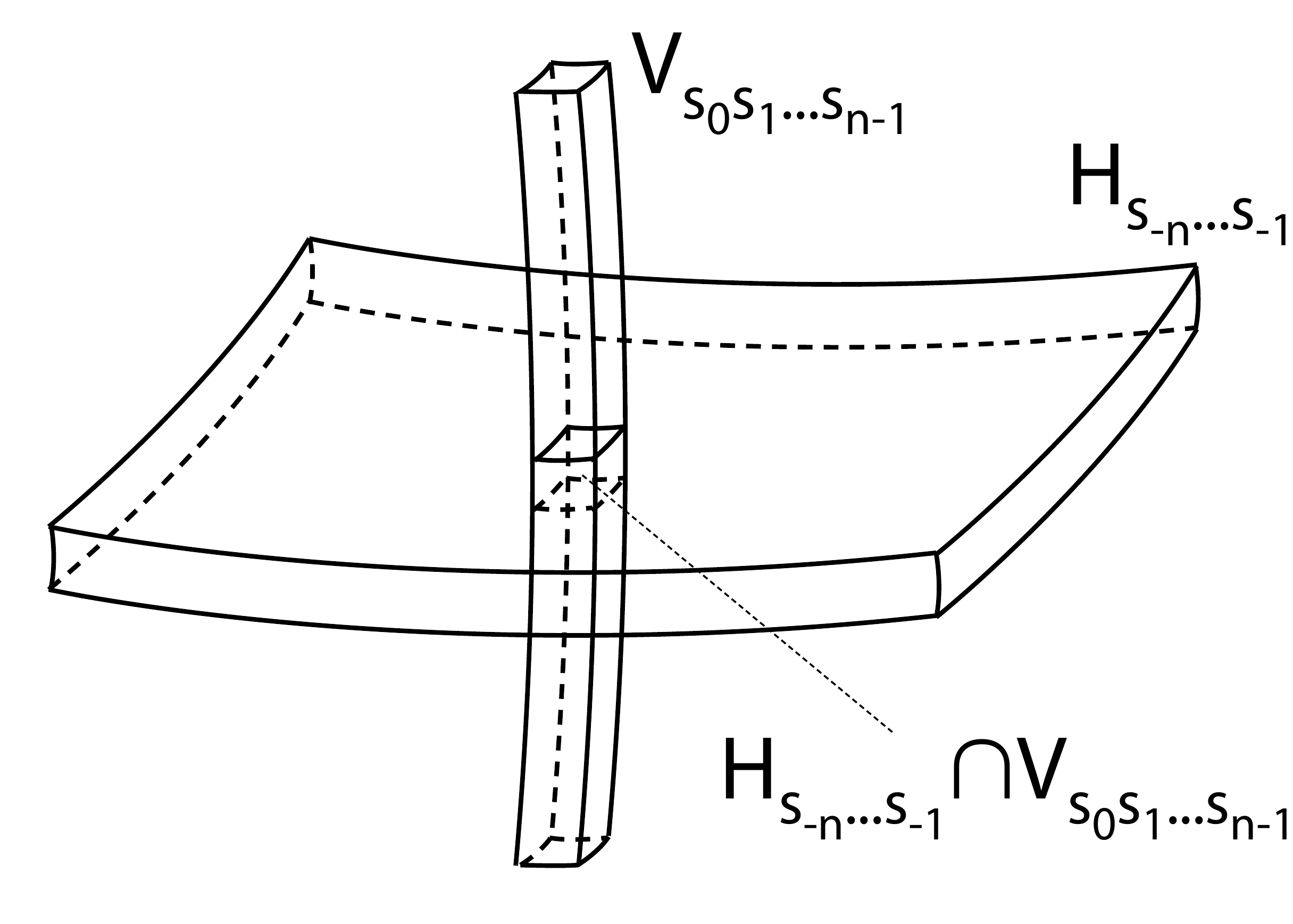}}
 \caption{$V_{s_0s_1\cdots s_{n-1}}$ is exponentially thin in the ``horizontal" (unstable) directions, and $H_{s_{-n}\cdots s_{-1}}$ is exponentially thin in the ``vertical" (stable) directions. The intersection $H_{s_{-n}\cdots s_{-1}} \cap V_{s_0\cdots s_{n-1}}$ is an exponentially small region, which under the $n \to \infty$ limit gives rise to a unique phase-space point.  }
\label{fig:Shrinking_Cell_3D}
\end{figure}     
As schematically illustrated by Fig.~\ref{fig:Shrinking_Cell_3D}, the intersections $H_{s_{-n}\cdots s_{-1}} \cap V_{s_0\cdots s_{n-1}}$ again localize an exponentially small volume in phase space, which under the $n \to \infty$ limit gives rise to a Cantor set of points on which the dynamics is topologically conjugate to a subshift of finite type on a bi-infinite symbolic string.  More details on the multidimensional formulation can be found in Chap.~2.3 of~\cite{Wiggins88}.  Here it is assumed that the symbolic dynamics permits all possible transitions $s_i s_{i+1}$ for $s_i, s_{i+1} \in \lbrace 0, 1, \cdots, K \rbrace$.  However, the results derived ahead carry over into more complicated systems possessing a pruning front~\cite{Cvitanovic88a,Cvitanovic91}.  

Under the symbolic dynamics, a period-$T$ point $y_0$, where $M^T(y_0)=y_0$, can always be associated with a symbolic string with infinite repetitions of a substring with length $T$: 
\begin{equation}\label{eq:Periodic point}
 y_0  \Rightarrow \cdots s_0 s_1 \cdots s_{T-1} \cdot s_0 s_1 \cdots s_{T-1} \cdots =\overline{\gamma} \cdot \overline{\gamma}
\end{equation}
where $\gamma =s_0 \cdots s_{T-1}$ is the finite substring and $\overline{\gamma} \cdot \overline{\gamma}$ denotes its infinite repetition (on both sides of the dot). Notice that the cyclic permutations of $s_0 \cdots s_{T-1}$ can be associated with the successive mappings of $y_0$, generating a one-to-one mapping to the set of points on the orbit. Since an orbit can be represented by any point on it, the position of the dot does not matter, therefore we denote the periodic orbit $\lbrace y_0 \rbrace$ as
\begin{equation}\label{eq:Periodic orbit}
\lbrace  y_0 \rbrace \Rightarrow \overline{\gamma} 
\end{equation}  
with the dot removed. Correspondingly, the stability exponents $\mu^{(i)}_{\lbrace y_0 \rbrace}$ can be written alternatively as $\mu^{(i)}_{\gamma}$.

Ahead an approximation scheme is developed to simplify the exact action formulas.  The scaling relation of an upper bound for the remainder (error term) is associated with the slowest shrinking dimension of $H_{s_{-n}\cdots s_{-1}} \cap V_{s_0\cdots s_{n-1}}$.  For $V_{s_0\cdots s_{n-1}}$ the slowest shrinking dimension of the  ``horizontal" ($U(x)$) direction is governed by the smallest positive Lyapunov exponent.  For $H_{s_{-n}\cdots s_{-1}}$ the scaling for the ``vertical" ($S(x)$) direction is governed by the least negative Lyapunov exponent.  A reasonable estimate for this process is provided by the stability exponents of the periodic orbits $\overline{\gamma_1}$ and $\overline{\gamma_2}$, where $\gamma_1 = s_{-n}\cdots s_{-1}$ and $\gamma_2 = s_0\cdots s_{n-1}$:
\begin{equation}\label{eq:Width estimate refined}
\begin{split}
&\text{scaling of } V_{\gamma_2} \sim O(e^{-n \mu^{(f)}_{\gamma_2}}) \\
&\text{scaling of } H_{\gamma_1} \sim O(e^{-n \mu^{(f)}_{\gamma_1}}).
\end{split}
\end{equation}
The scaling of $H_{\gamma_1} \cap V_{\gamma_2} $, as measured by the maximum component-wise phase-space separations between two points $z=(\mathbf{q},\mathbf{p})$ and $z^{\prime}=(\mathbf{q^{\prime}},\mathbf{p^{\prime}})$ located inside $H_{\gamma_1} \cap V_{\gamma_2}$, i.e.,
\begin{equation}
\label{eq:Size definition}
\text{scaling of } H_{\gamma_1} \cap V_{\gamma_2} \equiv \max_{\substack{z,z^{\prime} \in H_{\gamma_1} \cap V_{\gamma_2}; \\ 1 \leq i \leq f}} \big\lbrace 	|p_i - p^{\prime}_i|, |q_i - q^{\prime}_i| \big\rbrace\ ,
\end{equation}
is then estimated by the greater of two widths in Eq.~\eqref{eq:Width estimate refined}:
\begin{equation}\label{eq:Dimension estimate refined}
\text{scaling of } H_{\gamma_1} \cap V_{\gamma_2} \sim O \big( \max \lbrace e^{-n \mu^{(f)}_{\gamma_1}}, e^{-n \mu^{(f)}_{\gamma_2}} \rbrace \big) \ .
\end{equation}

Here, the hyperbolic fixed point with symbolic code $x \Rightarrow \overline{0} \cdot \overline{0}$\  and its orbit $\lbrace x \rbrace \Rightarrow \overline{0}$ are chosen as the reference. The action functions of all arbitrary hyperbolic trajectories will be calculated based on the knowledge of $x$, $S(x)$, and $U(x)$ only.  The intersections between the $f$-dimensional $S(x)$ and $f$-dimensional $U(x)$ in $2f$-dimensional phase space give rise to homoclinic points, which are asymptotic to $x$ under both $M^{\pm\infty}$.  A homoclinic point $h_0$ of $x$ has symbolic code of the form~\cite{Hagiwara04}: 
\begin{equation}\label{eq:Homoclinic point}
h_0 \Rightarrow \overline{0}  s_{-m}\cdots s_{-1} \cdot s_0 s_1 \cdots s_n  \overline{0}\ .
\end{equation}
Similar to the periodic orbit case, the homoclinic orbit can be represented as
\begin{equation}\label{eq:Homoclinic orbit}
\lbrace h_0 \rbrace \Rightarrow  \overline{0}  s_{-m}\cdots s_{-1}s_0 s_1 \cdots s_n  \overline{0}
\end{equation}
with the dot removed, as compared to Eq.~\eqref{eq:Homoclinic point}. 

\subsection{Generating function and classical action}
\label{Generating function and classical action}

For any phase space point $z_n=(\mathbf {q_n}, \mathbf{p_n})$ and its image $M(z_n)=z_{n+1}=(\mathbf{q_{n+1}},\mathbf{p_{n+1}})$, the mapping $M$ can be viewed as a canonical transformation that maps $z_n$ to $z_{n+1}$ while preserving the symplectic area, therefore a $\mathit{generating}$ ($\mathit{action}$) $\mathit{function}$ $F(\mathbf{q_{n}},\mathbf{q_{n+1}})$ can be associated with this process such that~\cite{MacKay84a,Meiss92}:
\begin{equation}\label{eq:Definition generating function}
\begin{split}
&\mathbf{p_{n}}=-\partial F/\partial \mathbf{q_{n}}\\ 
&\mathbf{p_{n+1}}=\partial F/\partial \mathbf{q_{n+1}}.
\end{split}
\end{equation}
Despite the fact that $F$ is a function of $\mathbf{q_n}$ and $\mathbf{q_{n+1}}$, it is convenient to denote it as $F(z_n,z_{n+1})$. This should cause no confusion as long as it is kept in mind that it is the $\mathbf{q}$ variables of $z_n$ and $z_{n+1}$ that go into the expression of $F$.  The compound mapping $M^{k}$, which maps $z_n$ to $z_{n+k}$, then has the generating function:
\begin{equation}\label{eq:Definition generating function compound map}
F(z_n,z_{n+k}) \equiv \sum_{i=n}^{n+k-1}F(z_{i},z_{i+1})
\end{equation}  
which, strictly speaking, is a function of the $\mathbf{q}$ variables. 

For periodic orbits $\lbrace y_0 \rbrace \Rightarrow \overline{\gamma}$ with primitive period $T$, the primitive period $\mathit{classical}$ $\mathit{action}$ ${\cal F}_{\gamma}$ of the orbit is:
\begin{equation}
\label{eq:Definition generating function primitive periodic orbits}
{\cal F}_{\gamma}  \equiv \sum_{i=0}^{T-1}F(y_{i},y_{i+1})\ .
\end{equation}  
For the special case of the fixed point $x$, Eq.~\eqref{eq:Definition generating function primitive periodic orbits} reduces to:
\begin{equation}\label{eq:Definition generating function fixed points}
{\cal F}_{0}  = F(x,x) 
\end{equation} 
where $F(x,x)$ is the generating function that maps $x$ into itself in one iteration.

\begin{figure}[ht]
\centering
{\includegraphics[width=7cm]{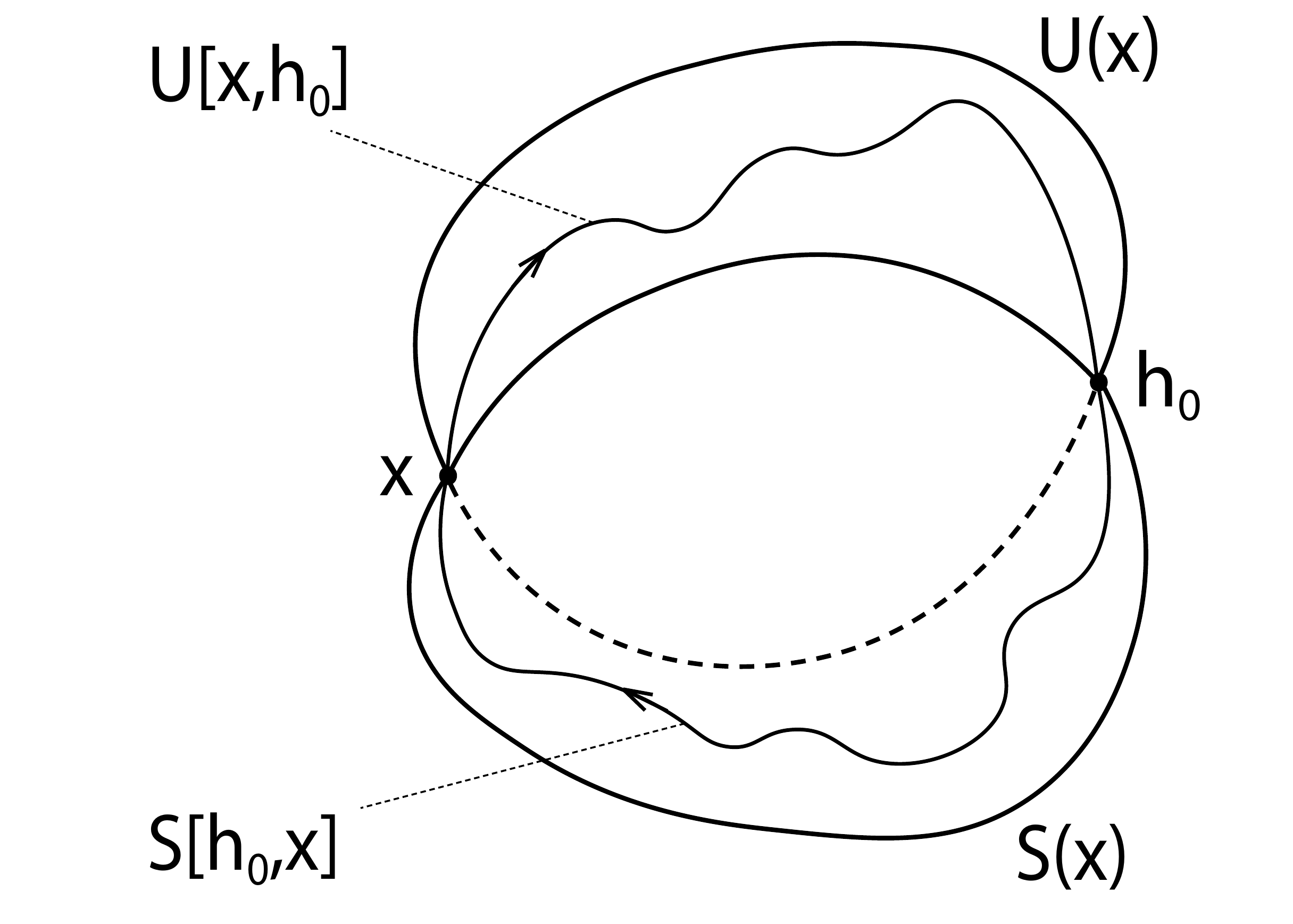}}
 \caption{Integration paths in Eq.~\eqref{eq:relative action homoclinic}. Both $U(x)$ and $S(x)$ are $f$-dimensional manifolds in the $2f$-dimensional phase space, and their intersections give rise to homoclinic points such as $h_0$. $U[x,h_0]$ is an arbitrary path directed from $x$ to $h_0$, and similarly for $S[h_0,x]$. The $ \mathbf{p}\cdot \mathrm{\mathbf{d}}\mathbf{q}$ integrals along them are independent of the paths. }
\label{fig:Homoclinic_Loop_3D}
\end{figure}     

For non-periodic orbits $\lbrace y_0 \rbrace$, the classical action is the sum of the generating functions over infinite successive mappings:
\begin{equation}
\label{eq:full orbit action in general}
{\cal F}_{\lbrace y_0 \rbrace} \equiv \lim_{N \to \infty} \sum_{i=-N}^{N-1} F(y_{i},y_{i+1})=\lim_{N \to \infty} F(y_{-N},y_{N})
\end{equation}
and is divergent in general. However, the MacKay-Meiss-Percival action principle~\cite{MacKay84a,Meiss92} can be applied to obtain well-defined action differences for particular pairs of orbits. Refer to Appendix~\ref{MacKay-Meiss-Percival} for a brief introduction of the action principle. An important and simple case is the $\mathit{relative}$ $\mathit{action}$ $\Delta {\cal F}_{\lbrace h_0 \rbrace  \lbrace x \rbrace}$ between a fixed point $x$ and its homoclinic orbit $\lbrace h_{0} \rbrace$, where $h_{\pm \infty}\to x$:
\begin{eqnarray}
\label{eq:relative action homoclinic}
\Delta {\cal F}_{\lbrace h_0 \rbrace  \lbrace x \rbrace} &\equiv & \lim_{N \to \infty} \sum_{i=-N}^{N-1}\left[F(h_i,h_{i+1})-F(x,x)\right] \nonumber \\
&=& \int\limits_{U[x,h_{0}]} \sum_{j=1}^{f} p_j \mathrm{d}q_j+\int\limits_{S[h_{0},x]} \sum_{j=1}^{f} p_j\mathrm{d}q_j \nonumber \\
&=& \int\limits_{U[x,h_{0}]} \mathbf{p}\cdot\mathrm{\mathbf{d}}\mathbf{q}+\int\limits_{S[h_{0},x]}\mathbf{p}\cdot\mathrm{\mathbf{d}}\mathbf{q}\nonumber \\
 &=& \oint_{US[x,h_{0}]}  \mathbf{p}\cdot\mathrm{\mathbf{d}}\mathbf{q} \nonumber \\
&=& {\cal A}^\circ_{US[x,h_{0}]}
\end{eqnarray}
where the notation $U[a,b]$ is introduced to denote the finite segment of an arbitrary $1$-dimensional curve on $U(x)$ extending from $a$ to $b$, both of which are points on $U(x)$, and similarly for $S(x)$. For systems with $f=1$, $U[a,b]$ and $S[a,b]$ are unique segments since both $U(x)$ and $S(x)$ are $1$-dimensional curves themselves. For systems with $f\geq 2$ the choices of $U[a,b]$ and $S[a,b]$ are not unique since there are infinitely many $1$-dimensional curves connecting $a$ and $b$ on multidimensional hyper-surfaces, as demonstrated by Fig.~\ref{fig:Homoclinic_Loop_3D}. However, due to the fact that both $U(x)$ and $S(x)$ are Langrangian manifolds, the phase-space integrals $ \int\limits_{ U[a,b] } \mathbf{p}\cdot \mathrm{\mathbf{d}}\mathbf{q}$ and $ \int\limits_{ S[a,b] } \mathbf{p}\cdot \mathrm{\mathbf{d}}\mathbf{q}$ are independent of the paths and are uniquely determined by the endpoints $a$ and $b$. The $\circ$ superscript on the last line indicates that the symplectic area is evaluated for a path that forms a closed loop, and the subscript indicates the path: $US[x,h_{0}]=U[x,h_{0}]+S[h_{0},x]$. 

 $\Delta {\cal F}_{\lbrace h_0 \rbrace \lbrace x \rbrace} $ gives the action difference between the homoclinic orbit segment $[ h_{-N},\cdots,h_{N} ]$ and the length-$(2N+1)$ fixed point orbit segment $[ x, \cdots, x ]$ in the limit $N \to \infty$. In later sections, upon specifying the symbolic code of the homoclinic orbit $\lbrace h_0 \rbrace \Rightarrow \overline{0} \gamma \overline{0}$, we also denote $\Delta {\cal F}_{\lbrace h_0 \rbrace  \lbrace x \rbrace}$ alternatively as
\begin{equation}\label{eq:relative action homoclinic symbolic notation}
\Delta {\cal F}_{\lbrace h_0 \rbrace  \lbrace x \rbrace} = \Delta {\cal F}_{\overline{0} \gamma \overline{0},  \overline{0}}
\end{equation}
by replacing the orbits in the subscript with their symbolic codes. 

Another case of great interest here is the $\mathit{relative}$ $\mathit{action}$ $\Delta {\cal F}_{\lbrace h^{\prime}_0 \rbrace \lbrace h_0 \rbrace}$ between a pair of homoclinic orbits $\lbrace h^{\prime}_0 \rbrace$ and $\lbrace h_0 \rbrace$, such that $h^{\prime}_{\pm \infty} = h_{\pm \infty} = x$: 
\begin{equation}\label{eq:Area-action homoclinic pair}
\begin{split}
& \Delta {\cal F}_{\lbrace h^{\prime}_0 \rbrace \lbrace h_0 \rbrace} \equiv \lim_{N \to \infty} \sum_{i=-N}^{N-1} \left[ F(h^{\prime}_i, h^{\prime}_{i+1}) - F(h_{i}, h_{i+1}) \right] \\
&= \int\limits_{ U[h_0, h^{\prime}_0] } \mathbf{p}\cdot\mathrm{\mathbf{d}}\mathbf{q} +\int\limits_{ S[h^{\prime}_0, h_0] } \mathbf{p}\cdot\mathrm{\mathbf{d}}\mathbf{q} = {\cal A}^{\circ}_{US[h_0,h^{\prime}_0]}.
\end{split}
\end{equation}
Similar to the notation adopted in Eq.~\eqref{eq:relative action homoclinic symbolic notation}, upon the specification of their symbolic codes $\lbrace h_0 \rbrace \Rightarrow \overline{0} \gamma \overline{0}$ and $\lbrace h^{\prime}_0 \rbrace \Rightarrow \overline{0} \gamma^{\prime} \overline{0}$, $\Delta {\cal F}_{\lbrace h^{\prime}_0 \rbrace \lbrace h_0 \rbrace}$ can also be denoted alternatively as
\begin{equation}\label{eq:relative action homoclinic pair symbolic notation}
\Delta {\cal F}_{\lbrace h^{\prime}_0 \rbrace \lbrace h_0 \rbrace} = \Delta {\cal F}_{\overline{0} \gamma^{\prime} \overline{0} ,\overline{0} \gamma \overline{0}}
\end{equation}
by replacing the orbits in the subscript with their symbolic codes. From the definitions of the relative actions it is easy to check that
\begin{equation}\label{eq:Two-point relative action difference symbolic notation}
\Delta {\cal F}_{\overline{0} \gamma^{\prime} \overline{0} ,\overline{0} \gamma \overline{0}} = \Delta {\cal F}_{\overline{0} \gamma^{\prime} \overline{0},  \overline{0}} - \Delta {\cal F}_{\overline{0} \gamma \overline{0},  \overline{0}} .
\end{equation}

A further useful generalization of Eq.~\eqref{eq:Area-action homoclinic pair} applies to four arbitrary homoclinic orbits of $x$, namely $\lbrace a_0 \rbrace$, $\lbrace b_0 \rbrace$, $\lbrace c_0 \rbrace$, and $\lbrace d_0 \rbrace$ \cite{Meiss92}:
\begin{equation}\label{eq:Area-action two homoclinic pairs}
\begin{split}
&( \Delta {\cal F}_{\lbrace a_0 \rbrace \lbrace x \rbrace} - \Delta {\cal F}_{\lbrace b_0 \rbrace \lbrace x \rbrace} ) - ( \Delta {\cal F}_{\lbrace c_0 \rbrace \lbrace x \rbrace} - \Delta {\cal F}_{\lbrace d_0 \rbrace \lbrace x \rbrace} ) \\
& = {\cal A}^{\circ}_{SUSU[ a_0, c_0, d_0, b_0 ]}
\end{split}
\end{equation}
where 
\begin{equation}\label{eq:Parallelogram area definition}
\begin{split}
& {\cal A}^{\circ}_{SUSU[ a_0, c_0, d_0, b_0 ]} \equiv \int\limits_{ S[a_0, c_0] } \mathbf{p}\cdot\mathrm{\mathbf{d}}\mathbf{q} + \int\limits_{ U[c_0, d_0] } \mathbf{p}\cdot\mathrm{\mathbf{d}}\mathbf{q} \\
& \quad + \int\limits_{ S[d_0, b_0] } \mathbf{p}\cdot\mathrm{\mathbf{d}}\mathbf{q} + \int\limits_{ U[b_0, a_0] } \mathbf{p}\cdot\mathrm{\mathbf{d}}\mathbf{q}
\end{split}
\end{equation}
is the symplectic area of a loop formed by alternating curve segments from $S(x)$ and $U(x)$ connecting the four homoclinic points. Same with the previous cases, the choice of the paths does not matter here since the integrals are uniquely fixed by the four homoclinic points. 

\section{Homoclinic expansion}
\label{Homoclinic expansion}

\subsection{Exact expansion}
\label{Exact general expansion}

This subsection is dedicated to the derivation of a general formula for the action(generating) function of unstable trajectories. The approach is to cut a long trajectory into several segments, and replace each segment with a homoclinic orbit that has a similar phase-space excursion (therefore a similar symbolic code). Consider an arbitrary unstable trajectory $\lbrace y_0 \rbrace$ with the symbolic code $y_0  \Rightarrow \alpha \cdot \beta \delta$, where the Greek letters $\alpha$ and $\delta$ here denote the left- and right-infinite strings of digits, respectively:
\begin{equation}
\begin{split}
\alpha & = \cdots s^{\prime}_{-1}s^{\prime}_{0} \\
\delta & = s^{\prime\prime}_{0}s^{\prime\prime}_{1} \cdots
\end{split}
\end{equation} 
and $\beta$ denotes a finite string of digits with length $N$:
\begin{equation}\label{eq:Beta}
\beta = s_0 s_1 \cdots s_{N-1}.
\end{equation}
The main interest here is the actions of long orbit segments, thus the integer $N$ is assumed to be very large.  Conceptually, cut the orbit segment of $\lbrace y_0 \rbrace$ corresponding to $\beta$  into $L$ pieces:
\begin{equation}\label{eq:Beta truncation}
\beta = \beta_1 \beta_2 \cdots \beta_L
\end{equation}     
where each piece $\beta_i$ has length $n_i $ and $\sum_{i=1}^{L}n_i = N$. For the sake of simplicity, define a  cumulative index
\begin{equation}\label{eq:Index m_k}
m_k \equiv \sum_{i=1}^k n_i,
\end{equation}
note that $m_1 = n_1$ and $m_L = N$. 

Setting up the symbolic code of $y_0$ this way, the codes of forward images of $y_0$, namely $y_{m_k}=M^{m_k}(y_0)$ ($1 \leq k \leq L-1$), are determined by
\begin{equation}\label{eq:Symbolic codes y}
y_{m_k}  \Rightarrow \alpha \beta_1 \beta_2 \cdots \beta_k \cdot \beta_{k+1} \cdots \beta_{L-1} \beta_{L} \delta.
\end{equation}
In particular, two special cases of Eq.~\eqref{eq:Symbolic codes y} under $k=1$ and $k=L-1$ yield
\begin{equation}
\label{eq:Symbolic codes y end points}
\begin{split}
&y_{m_1}  \Rightarrow \alpha \beta_1 \cdot \beta_2 \cdots \beta_{L} \delta \\
&y_{m_{(L-1)}}  \Rightarrow \alpha \beta_1 \beta_2 \cdots \beta_{L-1} \cdot \beta_{L} \delta .
\end{split}
\end{equation}
The generating function of interest, $F(y_{m_1}, y_{m_{L-1}})$, corresponds to the orbit segment $\lbrace y_{m_1}, \cdots, y_{m_{L-1}}  \rbrace$ and the action function $F(y_{m_1}, y_{m_{L-1}})$ is cut in the same way:
\begin{equation}
\label{eq:Action partition general}
F(y_{m_1}, y_{m_{L-1}}) = \sum_{k=1}^{L-2} F(y_{m_k},y_{m_{k+1}})
\end{equation}
where $F(y_{m_k},y_{m_{k+1}})$ is the generating function that maps from the beginning to the ending point of the $k$-th piece. 

The key point of the scheme is to replace the trajectory pieces by suitable homoclinic orbits that mimic their phase-space excursions, thereby avoiding the numerical construction of the trajectories themselves. To be more specific, the phase-space behavior of each piece in $\lbrace y_{m_k}, \cdots, y_{m_{k+1}}  \rbrace$ is characterized by the substring $\beta_{k+1}$, which has a similar excursion with a finite segment of an auxiliary homoclinic orbit, namely $\lbrace h^{(k+1)}_0 \rbrace$, identified by the code $\lbrace h^{(k+1)}_0 \rbrace \Rightarrow \overline{0} \beta_{k+1} \overline{0}$ and for which
\begin{equation}\label{eq:Auxiliary homoclinic points}
\begin{split}
& h^{(k+1)}_0 \Rightarrow \overline{0} \cdot \beta_{k+1} \overline{0} \\
& h^{(k+1)}_{n_{(k+1)}} \Rightarrow \overline{0}  \beta_{k+1} \cdot \overline{0}\ .
\end{split}
\end{equation}
This shadowing of the pieces of the homoclinic orbit and the trajectory piece gives rise to an exact relation between $F(y_{m_k},y_{m_{k+1}})$ and the homoclinic orbit relative action $\Delta {\cal F}_{\lbrace h^{(k+1)}_0 \rbrace  \lbrace x \rbrace}$, which is the building block of the scheme. 

To expose this relation, consider the homoclinic orbit relative action $\Delta {\cal F}_{\lbrace h^{(k+1)}_0 \rbrace  \lbrace x \rbrace}$ (defined in Eq.~\eqref{eq:relative action homoclinic}) split into three parts
\begin{equation}
\label{eq:Infinite homoclinic action partition}
\begin{split}
\Delta {\cal F}_{\lbrace h^{(k+1)}_0 \rbrace  \lbrace x \rbrace} = &  \lim_{N \to \infty} \sum_{i=-N}^{N-1}\left[F \left(h^{(k+1)}_i,h^{(k+1)}_{i+1}\right)-{\cal F}_0\right]  \\
= & \lim_{N \to \infty} \sum_{i=-N}^{-1}\left[F \left(h^{(k+1)}_i,h^{(k+1)}_{i+1}\right)-{\cal F}_0\right] \\
& + F \left(h^{(k+1)}_0,h^{(k+1)}_{n_{k+1}}\right) - n_{k+1} {\cal F}_0 \\
& + \lim_{N \to \infty} \sum_{i= n_{k+1}}^{N-1}\left[F \left(h^{(k+1)}_i,h^{(k+1)}_{i+1}\right)-{\cal F}_0\right].
\end{split}
\end{equation}
The difference between $F(y_{m_k},y_{m_{k+1}})$ and $\Delta {\cal F}_{\lbrace h^{(k+1)}_0 \rbrace  \lbrace x \rbrace}$ can thus be expressed as
\begin{equation}
\label{eq:Difference between trajectory and homoclinic}
\begin{split}
& F(y_{m_k},y_{m_{k+1}}) - \Delta {\cal F}_{\lbrace h^{(k+1)}_0 \rbrace  \lbrace x \rbrace}\\
& = -  \lim_{N \to \infty} \sum_{i=-N}^{-1}\left[F \left(h^{(k+1)}_i,h^{(k+1)}_{i+1}\right)-{\cal F}_0\right]\\
& \quad + \left[ F(y_{m_k},y_{m_{k+1}}) - F \left(h^{(k+1)}_0,h^{(k+1)}_{n_{k+1}}\right) + n_{k+1} {\cal F}_0 \right] \\
& \quad -  \lim_{N \to \infty} \sum_{i= n_{k+1}}^{N-1}\left[F \left(h^{(k+1)}_i,h^{(k+1)}_{i+1}\right)-{\cal F}_0\right].
\end{split}
\end{equation}
Invoking the MacKay-Meiss-Percival action principal \cite{MacKay84a,Meiss92}, the three terms on the right-hand side of Eq.~\eqref{eq:Difference between trajectory and homoclinic} can be converted into phase-space integrals along certain manifold segments.  For the first term, Eq.~\eqref{eq:Homoclinic action difference infinite past appendix} with $b_i = h^{(k+1)}_i$ and $a_i = x$, gives
\begin{equation}
\label{eq:Difference between trajectory and homoclinic left connector}
\begin{split}
&-  \lim_{N \to \infty} \sum_{i=-N}^{-1}\left[F\left(h^{(k+1)}_i,h^{(k+1)}_{i+1}\right)-{\cal F}_0\right]\\
& =  \int\limits_{U[h^{(k+1)}_0,x]} \mathbf{p}\cdot\mathrm{\mathbf{d}}\mathbf{q}. 
\end{split}
\end{equation}
Similarly, for the third term Eq.~\eqref{eq:Homoclinic action difference infinite future appendix} with $b_i = h^{(k+1)}_i$ and $a_i = x$ gives
\begin{equation}\label{eq:Difference between trajectory and homoclinic right connector}
\begin{split}
& -  \lim_{N \to \infty} \sum_{i= n_{k+1}}^{N-1}\left[F\left(h^{(k+1)}_i,h^{(k+1)}_{i+1}\right)-{\cal F}_0\right]\\
 &= \int\limits_{S\left[x, h^{(k+1)}_{n_{k+1}}\right]}\mathbf{p}\cdot\mathrm{\mathbf{d}}\mathbf{q}.
\end{split}
\end{equation}

The procedure for the second term is less straightforward because there are no stable or unstable manifolds directly connecting points on the $\lbrace y_0 \rbrace$ orbit with points on the $\lbrace h^{(k+1)}_0 \rbrace$ orbit.  It is necessary to look for connecting curves exclusive of $S(x)$ and $U(x)$. 
\begin{figure}[ht]
\centering
{\includegraphics[width=7cm]{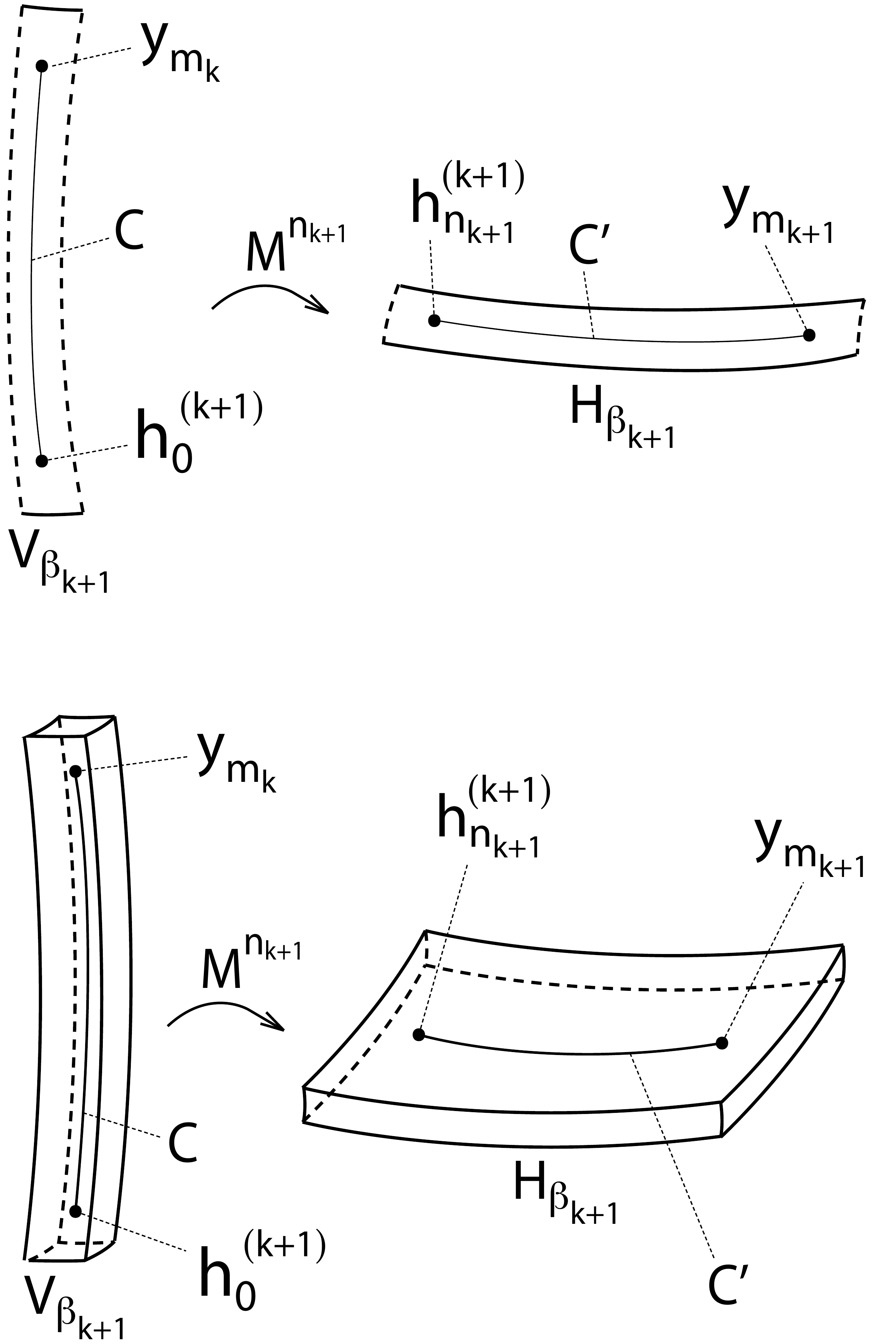}}
 \caption{Dynamics under $n_{k+1}$ iterations; $y_{m_k}$,$h^{(k+1)}_0 \in V_{\beta_{k+1}}$, and $y_{m_{k+1}}$,$h^{(k+1)}_{n_{k+1}} \in H_{\beta_{k+1}}$, where $H_{\beta_{k+1}} = M^{n_{k+1}} (V_{\beta_{k+1}})$.The ``horizontal" width of $V_{\beta_{k+1}}$ and the ``vertical" width of $H_{\beta_{k+1}}$ decrease exponentially with the length of the $\beta_{k+1}$ string. Under $n_{k+1}$ iterations of the map, $V_{\beta_{k+1}}$ is compressed along its stable directions and stretched along its unstable directions into $H_{\beta_{k+1}}$. Points inside $V_{\beta_{k+1}}$ follow a similar dynamics uniformly. $C$ is a curve inside $V_{\beta_{k+1}}$ connecting $y_{m_k}$ and $h^{(k+1)}_0$ that is approximately parallel to $S(x)$. $C^{\prime} = M^{n_{k+1}}(C)$, similarly, $C^{\prime}$ is a curve inside $H_{\beta_{k+1}}$ connecting $y_{m_{k+1}}$ and $h^{(k+1)}_{n_{k+1}}$ that is roughly parallel to $U(x)$. Upper panel: the $f=1$ case. $V_{\beta_{k+1}}$ and $H_{\beta_{k+1}}$ are both phase-space cells bounded by $S(x)$ (thick dashed curve) and $U(x)$ (thick solid curve). Lower panel: schematic of the $f \geq 2$ case. }
\label{fig:General_Trajectory_partitions}
\end{figure}     
By construction the symbolic codes of $y_{m_k}$ and $h^{(k+1)}_0$ share a common substring $\beta_{k+1}$ on the right-hand sides of the separation dots.  This indicates that they are located within the same ``vertical" slab $V_{\beta_{k+1}}$, as illustrated by Fig.~\ref{fig:General_Trajectory_partitions}.  As indicated in  Eq.~\eqref{eq:Width estimate refined}, the slowest shrinking scale of $V_{\beta_{k+1}}$ contracts as $\sim O(e^{-n_{k+1}\mu^{(f)}_{\beta_{k+1}}})$, where $\mu^{(f)}_{\beta_{k+1}}$ is the smallest positive stability exponent of the periodic orbit $\overline{\beta}_{k+1}$.  

Therefore, $y_{m_k}$ and $h^{(k+1)}_0$ are confined within the exponentially thin phase-space cell $V_{\beta_{k+1}}$, which is illustrated schematically in Fig.~\ref{fig:General_Trajectory_partitions}. Under $n_{k+1}$ iterations of the map, $V_{\beta_{k+1}}$ is compressed along its stable directions and stretched along its unstable directions, and eventually mapped into $H_{\beta_{k+1}}$. Points inside $V_{\beta_{k+1}}$ follow a similar dynamics uniformly. Therefore, successive forward images of $y_{m_k}$ and $h^{(k+1)}_0$ first approach, then separate from each other, making a near fly-by somewhere in the middle. Just like $V_{\beta_{k+1}}$, the 	``vertical" width of $H_{\beta_{k+1}}$ is also estimated to be $\sim O(e^{-n_{k+1}\mu^{(f)}_{\beta_{k+1}}})$. The initial separation between $y_{m_k}$ and $h^{(k+1)}_0$ is almost entirely along the stable manifold direction, while the final separation between $y_{m_{k+1}}$ and $h^{(k+1)}_{n_{k+1}}$ is almost entirely along the unstable manifold direction. 

As shown in Fig.~\ref{fig:General_Trajectory_partitions}, in spite of the fact that $y_{m_k}$ and $h^{(k+1)}_0$ are not directly connected by $U(x)$ nor $S(x)$, a curve $C$ can be constructed to connect them, where $C$ is chosen to be approximately parallel to $S(x)$. Then under $n_{k+1}$ iterations another curve $C^{\prime} = M^{n_{k+1}}(C)$ is created that is approximately parallel to $U(x)$. With the help of $C$ and $C^{\prime}$, the MacKay-Meiss-Percival action principle can again be applied to calculate the second term on the right-hand side of Eq.~\eqref{eq:Difference between trajectory and homoclinic}. This is done with Eq.~\eqref{eq:Meiss92 multidimensional}, setting $b = y_{m_k}$, $b^{\prime} = y_{m_{k+1}}$, $a = h^{(k+1)}_0$, $a^{\prime} = h^{(k+1)}_{n_{k+1}}$, and $c$ and $c^{\prime}$ to $C$ and $C^{\prime}$, respectively: 
\begin{equation}
\label{eq:Difference between trajectory and homoclinic middle segment}
\begin{split}
& F(y_{m_k},y_{m_{k+1}}) - F(h^{(k+1)}_0,h^{(k+1)}_{n_{k+1}}) \\
& = \int\limits_{C^{\prime}\left[h^{(k+1)}_{n_{k+1}},y_{m_{k+1}}\right]}\mathbf{p}\cdot\mathrm{\mathbf{d}}\mathbf{q} - \int\limits_{C\left[h^{(k+1)}_{0},y_{m_k}\right]}\mathbf{p}\cdot\mathrm{\mathbf{d}}\mathbf{q} \\
& = \int\limits_{C^{\prime}\left[h^{(k+1)}_{n_{k+1}},y_{m_{k+1}}\right]}\mathbf{p}\cdot\mathrm{\mathbf{d}}\mathbf{q} + \int\limits_{C\left[y_{m_k},h^{(k+1)}_{0}\right]}\mathbf{p}\cdot\mathrm{\mathbf{d}}\mathbf{q}
\end{split}
\end{equation}
where the $C$ and $C^{\prime}$ segments are illustrated by Fig.~\ref{fig:General_Trajectory_partitions}. Substituting Eqs.~\eqref{eq:Difference between trajectory and homoclinic left connector}, \eqref{eq:Difference between trajectory and homoclinic right connector}, and \eqref{eq:Difference between trajectory and homoclinic middle segment} into Eq.~\eqref{eq:Difference between trajectory and homoclinic} yields 
\begin{equation}\label{eq:Difference between trajectory and homoclinic each piece}
\begin{split}
& F(y_{m_k},y_{m_{k+1}}) = \int\limits_{C\left[y_{m_k},h^{(k+1)}_{0}\right]}\mathbf{p}\cdot\mathrm{\mathbf{d}}\mathbf{q} + \int\limits_{U[h^{(k+1)}_0,x]}\mathbf{p}\cdot\mathrm{\mathbf{d}}\mathbf{q} \\ 
& \quad + n_{k+1} {\cal F}_0 + \Delta {\cal F}_{\lbrace h^{(k+1)}_0 \rbrace  \lbrace x \rbrace} \quad + \int\limits_{S\left[x, h^{(k+1)}_{n_{k+1}}\right]}\mathbf{p}\cdot\mathrm{\mathbf{d}}\mathbf{q} \\
& + \int\limits_{C^{\prime}\left[h^{(k+1)}_{n_{k+1}},y_{m_{k+1}}\right]}\mathbf{p}\cdot\mathrm{\mathbf{d}}\mathbf{q} .
\end{split}
\end{equation}

\begin{figure}[ht]
\centering
{\includegraphics[width=7cm]{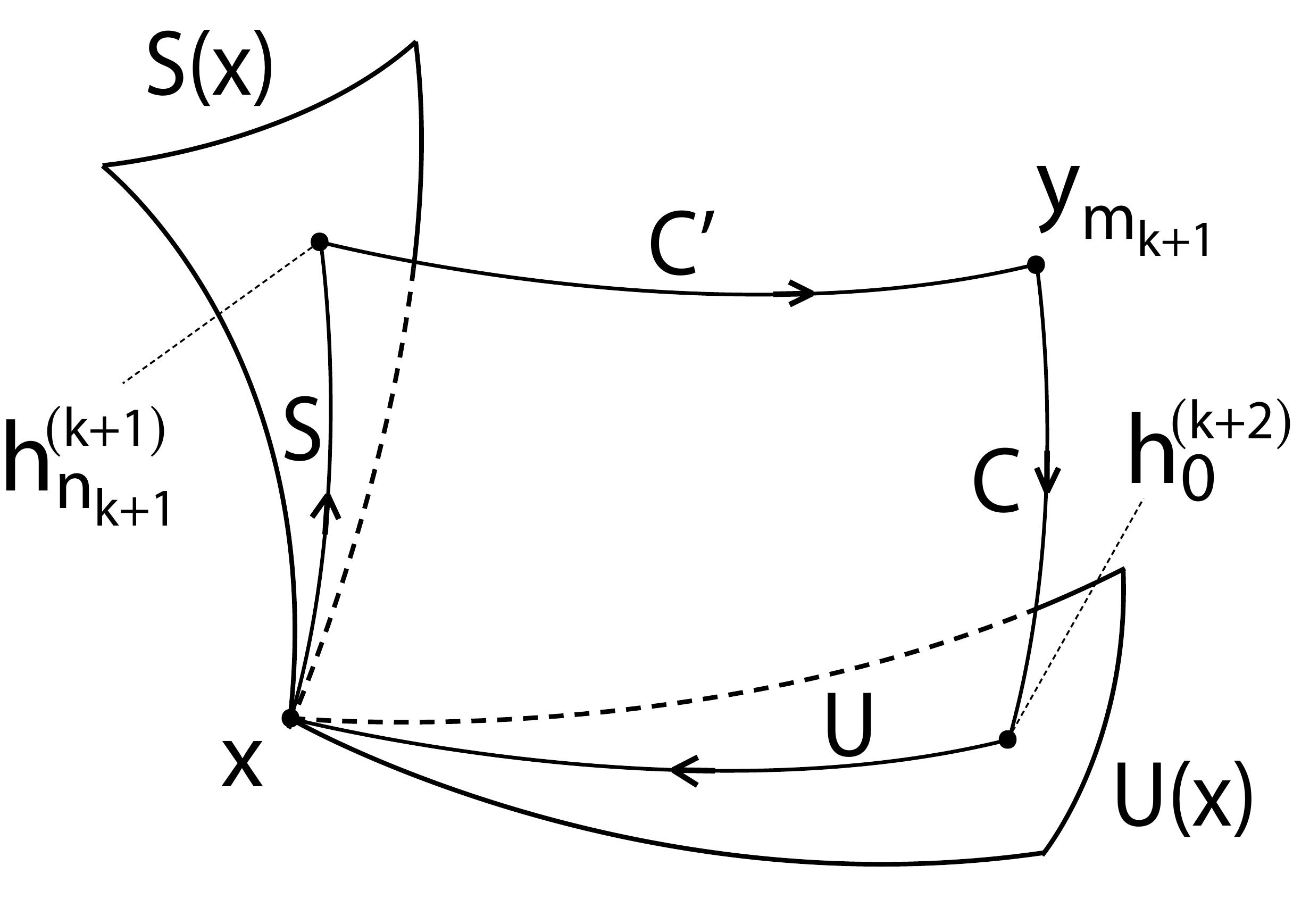}}
 \caption{The ${\cal A}^\circ_{CUSC^{\prime}\left[ y_{m_{k+1}},h^{(k+2)}_0,x,h^{(k+1)}_{n_{k+1}}\right]}$ term in Eq.~\eqref{eq:General trajectory action exact} yields the symplectic area of the loop shown in the figure. The curves labeled by $U$ and $S$ are arbitrary paths on $U(x)$ and $S(x)$, respectively. The curves $C$ and $C^{\prime}$ are defined in the same way as those in Fig.~\ref{fig:General_Trajectory_partitions}. Therefore, $C$ is nearly parallel to the nearby $S(x)$ that goes through $h^{(k+2)}_0$ (not shown here), and $C^{\prime}$ is nearly parallel to the nearby $U(x)$ that goes through $h^{(k+1)}_{n_{k+1}}$ (not shown here). This symplectic area is the action connector between $\lbrace h^{(k+1)}_0 \rbrace$ and $\lbrace h^{(k+2)}_0 \rbrace$. } \label{fig:Action_Connector_Exact}
\end{figure}

Having obtained the expression for the generating function of each piece, the total action function of the trajectory segment $\lbrace y_{m_1}, \cdots, y_{m_{L-1}} \rbrace$ is then just the sum of all the pieces:
\begin{equation}
\label{eq:General trajectory action exact}
\begin{split}
& F(y_{m_1}, y_{m_{L-1}}) = \sum_{k=1}^{L-2} F(y_{m_k},y_{m_{k+1}}) \\
& = \int\limits_{C\left[y_{m_1},h^{(2)}_{0}\right]}\mathbf{p}\cdot\mathrm{\mathbf{d}}\mathbf{q} +  \int\limits_{U\left[h^{(2)}_0,x\right]}\mathbf{p}\cdot\mathrm{\mathbf{d}}\mathbf{q}\\
& \quad  + \sum_{k=1}^{L-2} \left[ n_{k+1} {\cal F}_0 +  \Delta {\cal F}_{\lbrace h^{(k+1)}_0 \rbrace  \lbrace x \rbrace} \right] \\
& \quad + \sum_{k=1}^{L-3} {\cal A}^\circ_{CUSC^{\prime}\left[y_{m_{k+1}},h^{(k+2)}_0,x,h^{(k+1)}_{n_{k+1}}\right]}\\
& \quad + \int\limits_{S\left[x, h^{(L-1)}_{n_{L-1}}\right]}\mathbf{p}\cdot\mathrm{\mathbf{d}}\mathbf{q} +  \int\limits_{C^{\prime}\left[h^{(L-1)}_{n_{L-1}},y_{m_{L-1}}\right]}\mathbf{p}\cdot\mathrm{\mathbf{d}}\mathbf{q}
\end{split}
\end{equation}
where
\begin{equation}
\begin{split}
&{\cal A}^\circ_{CUSC^{\prime}\left[y_{m_{k+1}},h^{(k+2)}_0,x,h^{(k+1)}_{n_{k+1}}\right]}\\
&=\int\limits_{C\left[y_{m_{k+1}},h^{(k+2)}_0\right]}\mathbf{p}\cdot\mathrm{\mathbf{d}}\mathbf{q}+\int\limits_{U\left[h^{(k+2)}_0,x\right]}\mathbf{p}\cdot\mathrm{\mathbf{d}}\mathbf{q}\\
& \quad +  \int\limits_{S\left[x,h^{(k+1)}_{n_{k+1}}\right]}\mathbf{p}\cdot\mathrm{\mathbf{d}}\mathbf{q}+\int\limits_{C^{\prime}\left[h^{(k+1)}_{n_{k+1}},y_{m_{k+1}}\right]}\mathbf{p}\cdot\mathrm{\mathbf{d}}\mathbf{q}
\end{split}
\end{equation}
yields the symplectic area of the loop schematically depicted in Fig.~\ref{fig:Action_Connector_Exact}. 

\begin{figure}[ht]
\centering
{\includegraphics[width=8cm]{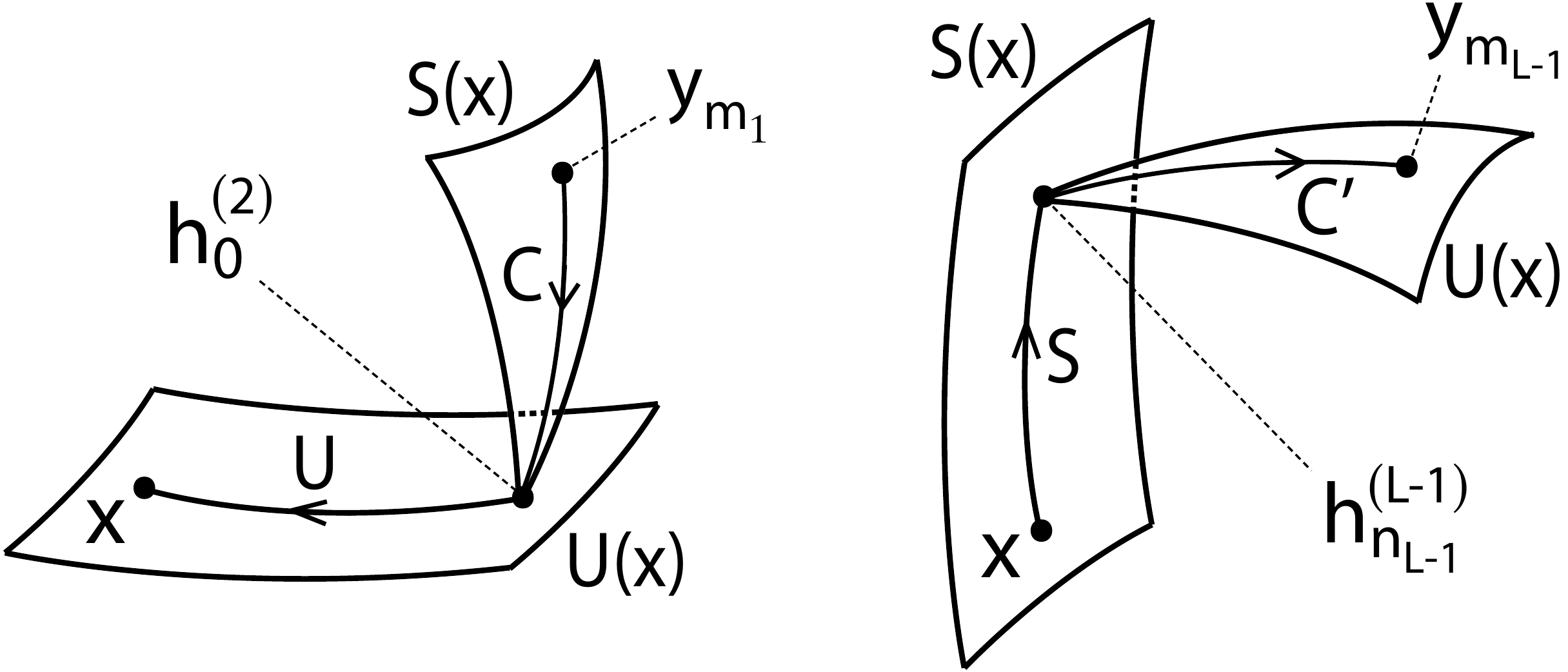}}
 \caption{Phase-space integrals as orbit action connectors. Left panel: paths for Eq.~\eqref{eq:Exact formula left connector}, which is the connector between $y_{m_1}$ and $h^{(2)}_0$, or equivalently, the connector between the $\beta_1$ and $\beta_2$ segments in Eq.~\eqref{eq:Symbolic codes y}.  Note that $y_{m_1}$ is located exponentially close to, but not a member of $S(x)$.  The resulting path $C$ is approximately parallel to, but not contained in $S(x)$ either.  Right panel: paths for Eq.~\eqref{eq:Exact formula right connector}, which is the connector between $h^{(L-1)}_{n_{L-1}}$ and $y_{m_{L-1}}$, or equivalently, the connector between the $\beta_{L-1}$ and $\beta_L$ segments in Eq.~\eqref{eq:Symbolic codes y}. Similarly, note that $y_{m_{L-1}}$ is located exponentially close to, but not a member of $U(x)$. The resulting path $C^{\prime}$ is approximately parallel to $U(x)$ as well.}
\label{fig:Left_Right_Connectors}
\end{figure}     

At this point, Eq.~\eqref{eq:General trajectory action exact} gives an exact expansion of $F(y_{m_1}, y_{m_{L-1}})$ in terms of homoclinic orbit actions and phase-space areas. Although seemingly complicated, terms on the right-hand side of Eq.~\eqref{eq:General trajectory action exact} have explicit geometric interpretations. The first term, 
\begin{equation}
\label{eq:Exact formula left connector}
\int\limits_{C\left[y_{m_1},h^{(2)}_{0}\right]}\mathbf{p}\cdot\mathrm{\mathbf{d}}\mathbf{q} +  \int\limits_{U\left[h^{(2)}_0,x\right]}\mathbf{p}\cdot\mathrm{\mathbf{d}}\mathbf{q},
\end{equation}
is a phase-space integral along the path shown schematically by the left panel of Fig.~\ref{fig:Left_Right_Connectors}. It acts as an action connector between the $\beta_1$ and $\beta_2$ segments in Eq.~\eqref{eq:Symbolic codes y}. The second term,  
\begin{equation}
\label{eq:Exact formula homoclinic action}
\sum_{k=1}^{L-2} \left[ n_{k+1} {\cal F}_0 +  \Delta {\cal F}_{\lbrace h^{(k+1)}_0 \rbrace  \lbrace x \rbrace} \right], 
\end{equation}
is the sum of the contribution of all the auxiliary homoclinic orbits $\lbrace h^{(k)}_0 \rbrace$ (Eq.~\eqref{eq:Auxiliary homoclinic points}) that shadow $\beta_{k}$ for $k=2, \cdots, L-1$. The third term, 
\begin{equation}\label{eq:Exact formula homoclinic connector}
\sum_{k=1}^{L-3} {\cal A}^\circ_{CUSC^{\prime}\left[y_{m_{k+1}},h^{(k+2)}_0,x,h^{(k+1)}_{n_{k+1}}\right]},
\end{equation} 
is the sum of all action connectors between $\beta_{k+1}$ and $\beta_{k+2}$, or equivalently, $\lbrace h^{(k+1)}_0 \rbrace$ and $\lbrace h^{(k+2)}_0 \rbrace$, for $k=1, \cdots, L-3$. The fourth term, 
\begin{equation}
\label{eq:Exact formula right connector}
 \int\limits_{S\left[x, h^{(L-1)}_{n_{L-1}}\right]}\mathbf{p}\cdot\mathrm{\mathbf{d}}\mathbf{q} +  \int\limits_{C^{\prime}\left[h^{(L-1)}_{n_{L-1}},y_{m_{L-1}}\right]}\mathbf{p}\cdot\mathrm{\mathbf{d}}\mathbf{q}, 
\end{equation}
as shown schematically by the right panel of Fig.~\ref{fig:Left_Right_Connectors}, is the action connector between the $\beta_{L-1}$ and $\beta_L$ segments in Eq.~\eqref{eq:Symbolic codes y}. 

Therefore, Eq.~\eqref{eq:General trajectory action exact} provides an exact expansion of unstable trajectory actions using homoclinic orbit actions that shadow it in a piecewise fashion, and phase-space areas as connectors between successive homoclinic orbits. In applications or numerical implementations, it still requires the calculation of the trajectory points $y_{m_k}$ ($k=1, \cdots, L-1$), which may prevent resummations or be prohibitively difficult. 

\subsection{Approximate expression}
\label{Approximate general expansion}

It is possible to give an approximate expression with controlled errors for Eq.~\eqref{eq:General trajectory action exact} that does not require the numerical construction of the trajectory points, which represents a great simplification. The essential spirit of the approximation is to replace the $y_{m_k}$ ($k=1, \cdots, L-1$) trajectory points in Eq.~\eqref{eq:General trajectory action exact} with nearby homoclinic points that result in only exponentially small error corrections. In general, all $y_{m_k}$ ($k=1, \cdots, L-1$) points are replaced by auxiliary homoclinic points $g^{(k,k+1)}_0$, where the homoclinic points are identified by symbolic codes
\begin{equation}
\label{eq:Connector homoclinic point k and k+1}
g^{(k,k+1)}_0 \Rightarrow \overline{0} \beta_k \cdot \beta_{k+1} \overline{0} \ ,
\end{equation}  
which by design match the forward and backward propagated pieces of the $\lbrace y_0 \rbrace$ symbolic code closest to $y_{m_k}$ (Eq.~\eqref{eq:Symbolic codes y}).  This implies that  
\begin{equation}
\label{eq:y and g same cell general}
y_{m_k}\ ,\ g^{(k,k+1)}_0 \in H_{\beta_{k}} \cap V_{\beta_{k+1}},
\end{equation}
where $H_{\beta_k} \cap V_{\beta_{k+1}}$ is the schematically depicted phase-space cell in Fig.~\ref{fig:Shrinking_Cell_3D}.  The resulting phase-space deviation between $y_{m_k}$ and $g^{(k,k+1)}_0$ is thus bounded by the shrinking scale of $H_{\beta_k} \cap V_{\beta_{k+1}}$ (defined in Eq.~\eqref{eq:Size definition}), which is $\sim O \big( \max \lbrace e^{-n_k \mu^{(f)}_{\beta_{k}}}, e^{-n_{k+1} \mu^{(f)}_{\beta_{k+1}}} \rbrace \big) $, where $ \mu^{(f)}_{\beta_{k}}$ and $ \mu^{(f)}_{\beta_{k+1}}$ are the smallest positive stability exponents of $\overline{\beta}_k$ and $\overline{\beta}_{k+1}$, respectively.

\begin{figure}[ht]
\centering
{\includegraphics[width=8cm]{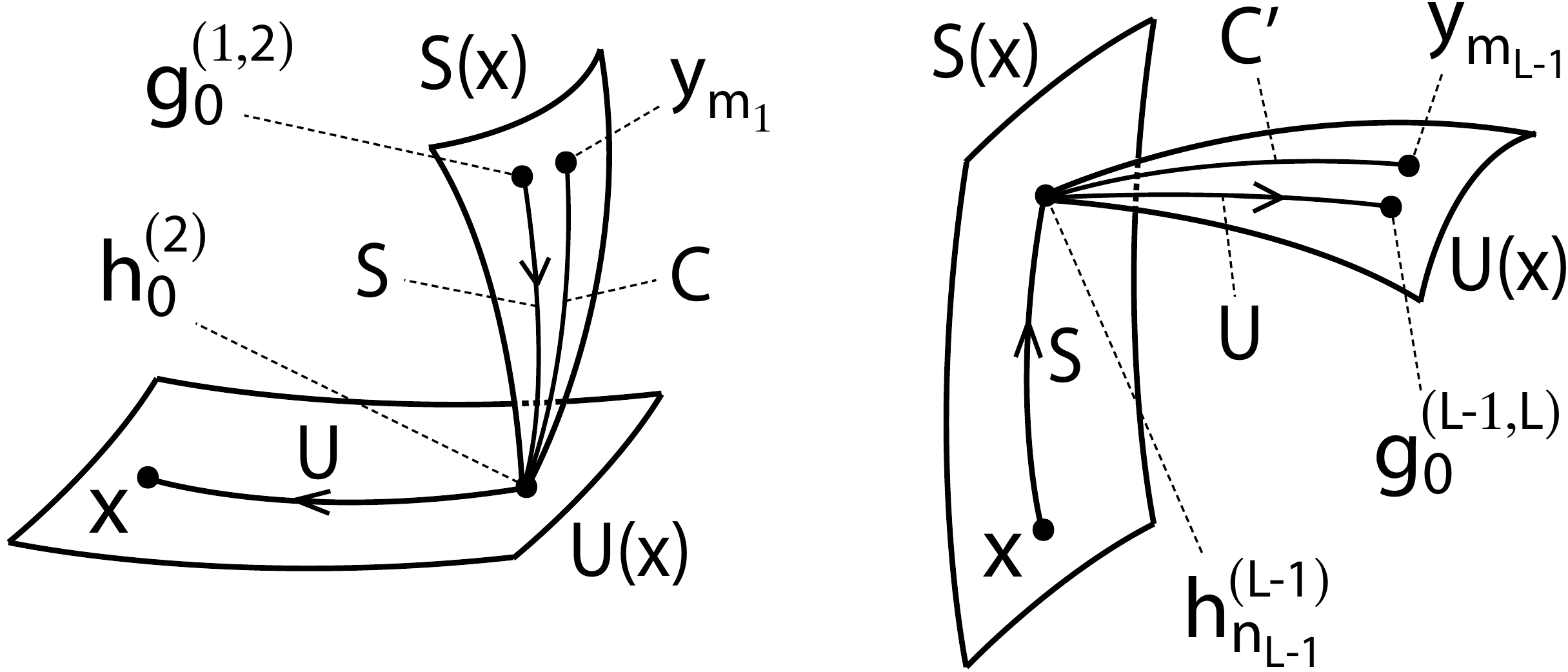}}
 \caption{The exact integration paths in Fig.~\ref{fig:Left_Right_Connectors} are replaced by the approximate integration paths in this figure resulting in exponentially small errors.  Left panel: demonstration for Eq.~\eqref{eq:Approx formula left connector}.  The path $S[g^{(1,2)}_0,h^{(2)}_0]$ is located on $S(x)$. The error is estimated to be $\sim O \big( \max \lbrace e^{-n_1 \mu^{(f)}_{\beta_{1}}}, e^{-n_{2} \mu^{(f)}_{\beta_{2}}} \rbrace \big) $.  Right panel: demonstration for Eq.~\eqref{eq:Approx formula right connector}. The path $U[h^{(L-1)}_{n_{L-1}},g^{(L-1,L)}_0]$ is located on $U(x)$. The error is estimated to be $\sim O \big( \max \lbrace e^{-n_{L-1} \mu^{(f)}_{\beta_{L-1}}}, e^{-n_{L} \mu^{(f)}_{\beta_{L}}} \rbrace \big)$ . } 
\label{fig:Left_Right_Connectors_Approximate}
\end{figure}

Starting from Eq.~\eqref{eq:Exact formula left connector}, replace $y_{m_1}$ by an auxiliary homoclinic point $g^{(1,2)}_0$ identified by symbolic code
\begin{equation}
\label{eq:Connector homoclinic point 1 and 2}
g^{(1,2)}_0 \Rightarrow \overline{0} \beta_1 \cdot \beta_2 \overline{0}\ .
\end{equation}  
The two points are necessarily exponentially close to each other.  From Eq.~\eqref{eq:y and g same cell general} 
\begin{equation}
\label{eq:y and g same cell}
y_{m_1}\ ,\ g^{(1,2)}_0 \in H_{\beta_1} \cap V_{\beta_2}\ .
\end{equation}
As shown in the left panel of Fig.~\ref{fig:Left_Right_Connectors_Approximate}, since the integration path $C\left[y_{m_1},h^{(2)}_0\right]$ is approximately parallel to the stable manifold      
that goes through $h^{(2)}_0$, it can be replaced by a simpler integration path $S\left[g^{(1,2)}_0,h^{(2)}_0\right]$ on $S(x)$, which is chosen to be approximately parallel to $C\left[y_{m_1},h^{(2)}_0\right]$.  The result is a small error comparable to the symplectic area of the gap between $C\left[y_{m_1},h^{(2)}_0\right]$ and $S\left[g^{(1,2)}_0,h^{(2)}_0\right]$, which is also exponentially small. Thus, an excellent approximation for Eq.~\eqref{eq:Exact formula left connector} is
\begin{equation}\label{eq:Approx formula left connector}
\begin{split}
&\int\limits_{C\left[y_{m_1},h^{(2)}_{0}\right]}\mathbf{p}\cdot\mathrm{\mathbf{d}}\mathbf{q} +  \int\limits_{U\left[h^{(2)}_0,x\right]}\mathbf{p}\cdot\mathrm{\mathbf{d}}\mathbf{q}\\
& = \int\limits_{S\left[g^{(1,2)}_0,h^{(2)}_0\right]}\mathbf{p}\cdot\mathrm{\mathbf{d}}\mathbf{q} +  \int\limits_{U\left[h^{(2)}_0,x\right]}\mathbf{p}\cdot\mathrm{\mathbf{d}}\mathbf{q} \\
& + O \big( \max \lbrace e^{-n_1 \mu^{(f)}_{\beta_{1}}}, e^{-n_{2} \mu^{(f)}_{\beta_{2}}} \rbrace \big).
\end{split}
\end{equation}

The great simplification is that the trajectory point $y_{m_1}$ no longer enters the calculation, and the integration paths are just curves on the stable and unstable manifolds connecting simpler homoclinic points. Moreover, since $g^{(1,2)}_0 \Rightarrow \overline{0} \beta_1 \cdot \beta_2 \overline{0}$ and $h^{(2)}_0 \Rightarrow \overline{0} \cdot \beta_2 \overline{0}$, it is easy to see that the integral is uniquely determined by the symbolic substring $\beta_1 \cdot \beta_2$. For the sake of simplicity, denote
\begin{equation}\label{eq:Approx formula left connector short notation general}
I(\beta_k \cdot \beta_{k+1}) \equiv \int\limits_{S\left[g^{(k,k+1)}_0,h^{(k+1)}_0\right]}\mathbf{p}\cdot\mathrm{\mathbf{d}}\mathbf{q} +  \int\limits_{U\left[h^{(k+1)}_0,x\right]}\mathbf{p}\cdot\mathrm{\mathbf{d}}\mathbf{q},
\end{equation}
which expresses the approximate integral over the $S$ and $U$ paths as $I(\beta_1 \cdot \beta_2)$.

This procedure applies to Eq.~\eqref{eq:Exact formula right connector} in an identical way. The trajectory point $y_{m_{L-1}}$ is replaced by the auxiliary homoclinic point $g^{(L-1,L)}_0$, where
\begin{equation}
\label{eq:Connector homoclinic point L-1 and L}
g^{(L-1,L)}_0 \Rightarrow \overline{0} \beta_{L-1} \cdot \beta_L \overline{0} \nonumber
\end{equation}  
is exponentially close to $y_{m_{L-1}}$.  The integration path $C^{\prime}\left[h^{(L-1)}_{n_{L-1}},y_{m_{L-1}}\right]$ is replaced by the simpler integration path $U\left[h^{(L-1)}_{n_{L-1}},g^{(L-1,L)}_0\right]$, similarly resulting in a exponentially small error shown schematically in the right panel of Fig.~\ref{fig:Left_Right_Connectors_Approximate}. The corresponding approximation for Eq.~\eqref{eq:Exact formula right connector} is
\begin{equation}
\label{eq:Approx formula right connector}
\begin{split}
& \int\limits_{S\left[x, h^{(L-1)}_{n_{L-1}}\right]}\mathbf{p}\cdot\mathrm{\mathbf{d}}\mathbf{q} +  \int\limits_{C^{\prime}\left[h^{(L-1)}_{n_{L-1}},y_{m_{L-1}}\right]}\mathbf{p}\cdot\mathrm{\mathbf{d}}\mathbf{q} \\
& = \int\limits_{S\left[x, h^{(L-1)}_{n_{L-1}}\right]}\mathbf{p}\cdot\mathrm{\mathbf{d}}\mathbf{q} +  \int\limits_{U\left[h^{(L-1)}_{n_{L-1}},g^{(L-1,L)}_0\right]}\mathbf{p}\cdot\mathrm{\mathbf{d}}\mathbf{q}\\
& \quad  + O \big( \max \lbrace e^{-n_{L-1} \mu^{(f)}_{\beta_{L-1}}}, e^{-n_{L} \mu^{(f)}_{\beta_{L}}} \rbrace \big).
\end{split}
\end{equation}
As in the previous case, the integral is uniquely determined by the substring $\beta_{L-1} \cdot \beta_L$.  Denoting
\begin{equation}
\label{eq:Approx formula right connector short notation general}
I^{\prime}(\beta_{k} \cdot \beta_{k+1}) \equiv \int\limits_{S[x, h^{(k)}_{n_{k}}]}\mathbf{p}\cdot\mathrm{\mathbf{d}}\mathbf{q} +  \int\limits_{U[h^{(k)}_{n_{k}},g^{(k,k+1)}_0]}\mathbf{p}\cdot\mathrm{\mathbf{d}}\mathbf{q},
\end{equation}
the approximate form of Eq.~(\ref{eq:Exact formula right connector}) is given by $I^{\prime}(\beta_{L-1} \cdot \beta_L)$.

\begin{figure}[ht]
\centering
{\includegraphics[width=8cm]{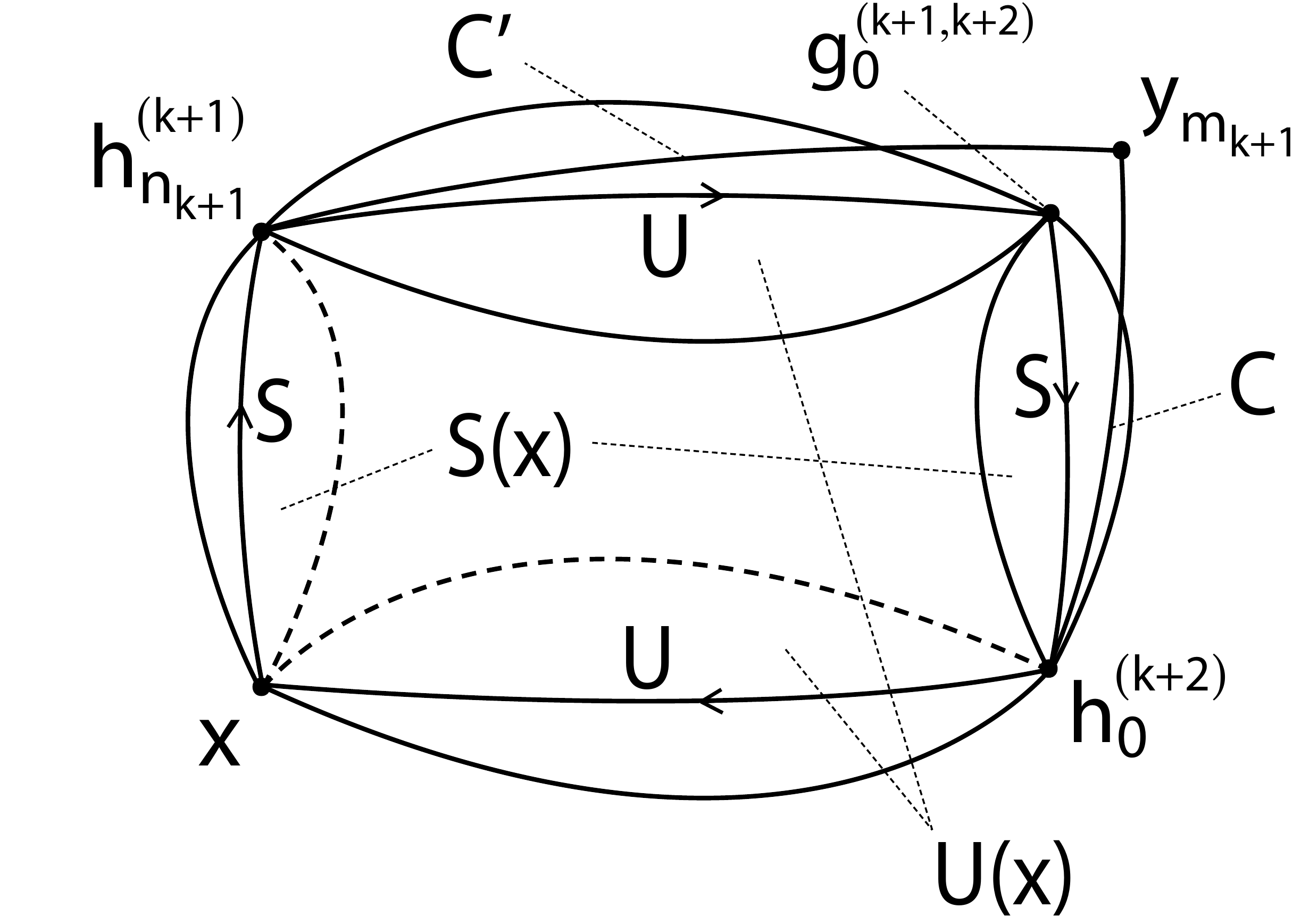}}
 \caption{Illustration of Eq.~\eqref{eq:Approx formula homoclinic connector}.  $S(x)$ is plotted as vertical surfaces, and $U(x)$ as horizontal surfaces. The integration loop ${\cal A}^\circ_{SUSU\left[g^{(k+1,k+2)}_0,h^{(k+2)}_0,x,h^{(k+1)}_{n_{k+1}}\right]}$ is marked by arrows. The error term in Eq.~\eqref{eq:Approx formula homoclinic connector} is estimated to be $\sim O \left( \max \lbrace e^{-n_{k+1} \mu^{(f)}_{\beta_{k+1}}}, e^{-n_{k+2} \mu^{(f)}_{\beta_{k+2}}} \rbrace \right)$.}
\label{fig:Action_Connector_Approx}
\end{figure}     

Substitutions follow for Eq.~\eqref{eq:Exact formula homoclinic connector} in exactly the same way:
\begin{equation}
\label{eq:Approx formula homoclinic connector replacement rule}
\begin{split}
& y_{m_{k+1}} \mapsto g^{(k+1,k+2)}_0 \\
& C\left[y_{m_{k+1}},h^{(k+2)}_0\right] \mapsto S\left[g^{(k+1,k+2)}_0,h^{(k+2)}_0\right] \\
& C^{\prime}\left[ h^{(k+1)}_{n_{k+1}},y_{m_{k+1}} \right] \mapsto U\left[ h^{(k+1)}_{n_{k+1}},g^{(k+1,k+2)}_0 \right]\ . \nonumber
\end{split}
\end{equation}
Equation~\eqref{eq:Exact formula homoclinic connector} admits the approximate form
\begin{equation}\label{eq:Approx formula homoclinic connector}
\begin{split}
& {\cal A}^\circ_{CUSC^{\prime}\left[y_{m_{k+1}},h^{(k+2)}_0,x,h^{(k+1)}_{n_{k+1}}\right]} \\
& = {\cal A}^\circ_{SUSU\left[g^{(k+1,k+2)}_0,h^{(k+2)}_0,x,h^{(k+1)}_{n_{k+1}}\right]}\\
&\quad  + O \big( \max \lbrace e^{-n_{k+1} \mu^{(f)}_{\beta_{k+1}}}, e^{-n_{k+2} \mu^{(f)}_{\beta_{k+2}}} \rbrace \big)\ ,
\end{split}
\end{equation}
where the new ${\cal A}^\circ_{SUSU[\cdots]}$ symplectic area is shown in Fig.~\ref{fig:Action_Connector_Approx}. It is important to note that the approximate symplectic area relies only on homoclinic orbits of relatively short excursions and their stable/unstable manifolds and the explicit dependence on the trajectory points $y_{m_{k+1}}$ is gone. Furthermore, the symbolic codes of the four corners of the loop have a particular simple form:
\begin{equation}\label{eq:Symbolic codes four corners}
\begin{split}
& g^{(k+1,k+2)}_0 \Rightarrow \overline{0} \beta_{k+1} \cdot \beta_{k+2} \overline{0} \\
& h^{(k+1)}_{n_{(k+1)}} \Rightarrow \overline{0} \beta_{k+1} \cdot  \overline{0} \\
& h^{(k+2)}_0 \Rightarrow \overline{0} \cdot \beta_{k+2} \overline{0} \\
& x  \Rightarrow \overline{0} \cdot \overline{0} \nonumber
\end{split}
\end{equation}
which indicates that ${\cal A}^\circ_{SUSU\left[g^{(k+1,k+2)}_0,h^{(k+2)}_0,x,h^{(k+1)}_{n_{k+1}}\right]} $ is uniquely determined by the symbolic string $\beta_{k+1} \cdot \beta_{k+2}$. Therefore, to simplify notation let
\begin{equation}
\label{eq:Approx formula homoclinic connector short notation}
 {\cal A}^\circ (\beta_{k+1} \cdot \beta_{k+2}) \equiv {\cal A}^\circ_{SUSU\left[g^{(k+1,k+2)}_0,h^{(k+2)}_0,x,h^{(k+1)}_{n_{k+1}}\right]} \ . 
\end{equation}
It turns out that
\begin{eqnarray}\label{eq:Whatever}
 {\cal A}^\circ (\beta_{k+1} \cdot \beta_{k+2}) &=& I(\beta_{k+1} \cdot \beta_{k+2}) + I^{\prime}(\beta_{k+1} \cdot \beta_{k+2})  \nonumber \\
& = & \Delta {\cal F}_{\overline{0} \beta_{k+1} \beta_{k+2} \overline{0}, \overline{0} \beta_{k+1}\overline{0}} - \Delta {\cal F}_{\overline{0} \beta_{k+2} \overline{0}, \overline{0} } \ , \nonumber \\
\end{eqnarray}
where the last equality comes from Eq.~\eqref{eq:Area-action two homoclinic pairs}. 

Substituting Eqs.~(\ref{eq:Approx formula left connector}-\ref{eq:Approx formula homoclinic connector short notation}) into Eq.~\eqref{eq:General trajectory action exact} leads to the approximate expansion for trajectory actions in general:
\begin{equation}
\label{eq:General trajectory action approx}
\begin{split}
& F\left(y_{m_1}, y_{m_{L-1}}\right) = I(\beta_1 \cdot \beta_2) + I^{\prime}(\beta_{L-1} \cdot \beta_L) \\ 
& \quad +  \sum_{k=1}^{L-2} \left[ n_{k+1}{\cal F}_0 + \Delta {\cal F}_{\overline{0} \beta_{k+1} \overline{0}, \overline{0}} \right] + \sum_{k=1}^{L-3} {\cal A}^\circ (\beta_{k+1} \cdot \beta_{k+2}) \\
& \quad + O\left( \max_{k \in [1,L]} e^{-n_k \mu^{(f)}_{\beta_k}}  \right)\ 
\end{split}
\end{equation}
where $\Delta {\cal F}_{\overline{0} \beta_{k+1} \overline{0}, \overline{0}} = \Delta {\cal F}_{\lbrace h^{(k+1)}_0 \rbrace  \lbrace x \rbrace}$ [Eq.~\eqref{eq:relative action homoclinic symbolic notation}]. 

\subsection{Loss of memory}
\label{loss of memory}

Compared to the exact expansion in Eq.~\eqref{eq:General trajectory action exact} that requires the knowledge of all the trajectory points $y_{m_k}$ ($k=1,\cdots,L-1$), Eq.~\eqref{eq:General trajectory action approx} requires only the information about the homoclinic orbits $\lbrace h^{(k)}_{0} \rbrace$ (for $k=2,\cdots,L-1$) and $\lbrace g^{(k,k+1)}_{0} \rbrace$ (for $k=1,\cdots,L-1$), constructed to have relatively short excursions (similar to the cycle expansion using short periodic orbits to represent the effects of very long periodic orbits).  If not known by some analytic means, such as in the bakers map~\cite{Ozorio91,Oconnor91,Oconnor92}, they can be calculated using very stable numerical techniques~\cite{Li17}.  The error associated with the above approximation decreases exponentially rapidly with increasing $n_k$ values, thus one can choose any desired level of accuracy for the action calculation of long trajectory segments in general.  Perhaps most importantly though, it reveals an important ``exponential memory decay" property for long trajectory segments in chaotic systems that is presumably expected, but is proven here.  Notice that the left-hand side of Eq.~\eqref{eq:General trajectory action approx} is the action function evaluated from $y_{m_1} \Rightarrow \alpha \beta_1 \cdot \beta_2 \cdots \beta_{L} \delta$ to $y_{m_{(L-1)}}  \Rightarrow \alpha \beta_1 \beta_2 \cdots \beta_{L-1} \cdot \beta_{L} \delta$.  To specify this trajectory uniquely, one must either know its entire symbolic history $\lbrace y_0 \rbrace \Rightarrow \alpha \beta \delta$ (where $y_0  \Rightarrow \alpha \cdot \beta \delta$) or know $y_0$ and its iterates with infinite precision.  However, the approximate classical action with controllable exponentially small errors depends only on the $\beta$ symbol sequence.  Not a single value of $y_0 $ or its iterates is necessary to calculate the approximation.  Thus all trajectories with the same $\beta$ sequence, independent of $\alpha$ and $\delta$ give the same exponentially accurate classical action function in shifting the present from just after $\beta_1$ to just after $\beta_{L-1}$.  The ``memory" of the past $\alpha$ and future $\delta$ fades exponentially away in the action function depending only  on the lengths of the $\beta_1$ and $\beta_L$ pieces.  Such memory loss should enable an improved matrix-product approach to semiclassical trace formulas, which is currently under investigation by the authors.

\subsection{Periodic orbits}
\label{Application to periodic orbits}

Although the exact and approximate expressions given above apply to any trajectory segment, due to their great interest in semiclassical theories~\cite{Gutzwiller90}, it is worthwhile applying the approximation procedure to long periodic orbits.  Let $\lbrace y_0 \rbrace \Rightarrow \overline{\gamma}$ with period $N$, where $\gamma$ is a symbolic string of $N$ digits.  Partition $\gamma$ into $L-2$ substrings: $\gamma = \gamma_1 \gamma_2 \cdots \gamma_{L-2}$, and denote the length of each $\gamma_k$ by $n_k$. Placing the separation dot such that 
\begin{equation}
\label{eq:Periodic orbit y_0}
y_0 \Rightarrow \overline{\gamma}  \cdot  \overline{\gamma},
\end{equation}
then
\begin{equation}
\label{eq:Periodic orbit y_T}
y_{0}=M^{N}(y_0)  \Rightarrow \overline{\gamma}  \cdot  \overline{\gamma},
\end{equation}
i.e., the mapping from $y_0$ back to itself corresponds to a shift of the dot for $N$ digits, thereby leading to identical symbolic codes. The classical action of interest is
\begin{equation}
\label{eq:Periodic orbit action y}
{\cal F}_{\gamma} = \sum_{i=0}^{N-1}F(y_{i},y_{i+1}) = F(y_0 , y_N).
\end{equation}
The connection to the notation for a general trajectory segment is
\begin{equation}
\label{eq:General to periodic substrings}
\begin{split}
& \alpha =  \overline{\gamma} \gamma_1 \gamma_2 \cdots \gamma_{L-3} \\
& \beta_1 = \gamma_{L-2} \\
& \beta_{k} = \gamma_{k-1}\ (k=2,\cdots,L-1)\\
&\beta_{L} = \gamma_1\\
&\delta = \gamma_2\gamma_3\cdots\gamma_{L-2} \overline{\gamma} \ .
\end{split}
\end{equation}
The approximation, Eq.~\eqref{eq:General trajectory action approx}, yields the periodic orbit action 
\begin{equation}
\label{eq:Periodic orbit expansion 1}
\begin{split}
& {\cal F}_{\gamma} =F(y_0 , y_N) =  I(\gamma_{L-2} \cdot \gamma_1) + I^{\prime}(\gamma_{L-2} \cdot \gamma_1)\\
&  +  \sum_{k=1}^{L-2} \left[ n_{k}{\cal F}_0 + \Delta {\cal F}_{\overline{0} \gamma_{k} \overline{0}, \overline{0}} \right]  + \sum_{k=1}^{L-3} {\cal A}^\circ (\gamma_{k} \cdot \gamma_{k+1}) \\
&  + O\left( \max_{k \in [1,L-2]} e^{-n_k \mu^{(f)}_{\gamma_k}}  \right)\ ,
\end{split}
\end{equation}
where $\mu^{(f)}_{\gamma_k}$ is the smallest positive stability exponent of the periodic orbit $\overline{\gamma}_k$. 
 
Notice that
\begin{equation}
I(\gamma_{L-2} \cdot \gamma_1) +  I^{\prime}(\gamma_{L-2} \cdot \gamma_1) = {\cal A}^\circ (\gamma_{L-2} \cdot \gamma_{1})\ . 
\end{equation}
Equation~\eqref{eq:Periodic orbit expansion 1} can be simplified:
\begin{equation}
\label{eq:Periodic orbit expansion}
\begin{split}
{\cal F}_{\gamma} = & N{\cal F}_0 + \sum_{k=1}^{L-2} \left[ \Delta {\cal F}_{\overline{0} \gamma_{k} \overline{0}, \overline{0}} +{\cal A}^\circ (\gamma_{k} \cdot \gamma_{k+1}) \right] \\
& + O\left( \max_{k \in [1,L-2]} e^{-n_k \mu^{(f)}_{\gamma_k}}  \right)
\end{split}
\end{equation}
where the $\gamma_{k}$  subscript is understood to be cyclic in $L-2$: $\gamma_{L-1}=\gamma_{1}$.  This equation provides an expansion of long periodic orbit actions in terms of homoclinic orbits $\overline{0} \gamma_{k} \overline{0}$ that shadow it in a piece-wise fashion, and symplectic areas ${\cal A}^\circ (\gamma_{k} \cdot \gamma_{k+1})$ as action connectors between successive homoclinic orbits.  Just like Eq.~\eqref{eq:General trajectory action approx}, it does not require prior numerical construction of the periodic orbits themselves.  

With the help of Eq.~\eqref{eq:Whatever}, an alternative form equivalent to Eq.~\eqref{eq:Periodic orbit expansion} can be given.  Taking into account that
\begin{equation}
\label{eq:Periodic orbit expansion area connector to action difference}
\begin{split}
 & {\cal A}^\circ (\gamma_{k} \cdot \gamma_{k+1}) =  \Delta {\cal F}_{\overline{0} \gamma_{k} \gamma_{k+1} \overline{0}, \overline{0} \gamma_{k}\overline{0}} - \Delta {\cal F}_{\overline{0} \gamma_{k+1} \overline{0}, \overline{0} } \\
& = \Delta {\cal F}_{\overline{0} \gamma_{k} \gamma_{k+1} \overline{0}, \overline{0}} - \Delta {\cal F}_{\overline{0} \gamma_{k} \overline{0}, \overline{0} } -\Delta {\cal F}_{\overline{0} \gamma_{k+1} \overline{0}, \overline{0} }
\end{split}
\end{equation}
and substituting into Eq.\eqref{eq:Periodic orbit expansion} gives
\begin{equation}
\label{eq:Periodic orbit expansion alternative}
\begin{split}
{\cal F}_{\gamma} =&   N{\cal F}_0 +\sum_{k=1}^{L-2} \Delta {\cal F}_{\overline{0} \gamma_{k} \gamma_{k+1} \overline{0}, \overline{0}\gamma_{k+1} \overline{0}}\\
& + O\left( \max_{k \in [1,L-2]} e^{-n_k \mu^{(f)}_{\gamma_k}}  \right)
\end{split}
\end{equation}
where the index $k$ is also cyclic in $L-2$: $\gamma_{L-1}=\gamma_{1}$. Eq.~\eqref{eq:Periodic orbit expansion alternative} is equivalent to Eq.~\eqref{eq:Periodic orbit expansion}, and changes the evaluation of phase-space areas into action differences between certain auxiliary homoclinic orbits. 

\subsection{Alternative view: periodic orbit expansion}
\label{Periodic orbit expansion}

To this point, homoclinic orbits have been used as the building blocks to generate the full dynamics of arbitrary trajectories.  Due to the intimate relationship between homoclinic and periodic orbits, an alternative approach can be established by using unstable periodic orbits as the building blocks. The resulting periodic orbit expansion works  equivalently well as the homoclinic orbit expansion, therefore putting periodic and homoclinic orbits on an equal footing as a scaffolding for the dynamics of chaotic systems. The original idea of a periodic-orbit expansion was pioneered by Cvitanovi{\'c} and coauthors in their studies of dynamical $\zeta$ functions in classical and quantum chaos~\cite{Cvitanovic88,Artuso90a,Artuso90b}, and has been widely know as the \textit{cycle} \textit{expansion} where the term ``cycle" stands for periodic orbits. It has since been generalized into a wide range of systems \cite{Lan10}, in particular recent applications to the state space of turbulent flows \cite{Suri17,Suri20,Yalniz20}. 

The scheme begins by specifying a sequence of periodic orbits $\lbrace z^{(k)}_0 \rbrace$ with $k= 1, \dots, L$, each identified by symbolic code 
\begin{equation}
\label{eq:Cycle expansion periodic orbits symbolic code}
z^{(k)}_0 \Rightarrow \overline{\beta}_k \cdot \overline{\beta}_k\ ,
\end{equation}
where $M^{n_k}(z^{(k)}_0) = z^{(k)}_0$. The trajectory segment $\lbrace y_{m_1}, \dots, y_{m_{L-1}} \rbrace$ is divided into short segments of transient visits to the neighborhoods of successive periodic orbits $\lbrace z^{(k)}_0 \rbrace$ for $k=2, \dots, L-1$.
\begin{figure}[ht]
\centering
{\includegraphics[width=7cm]{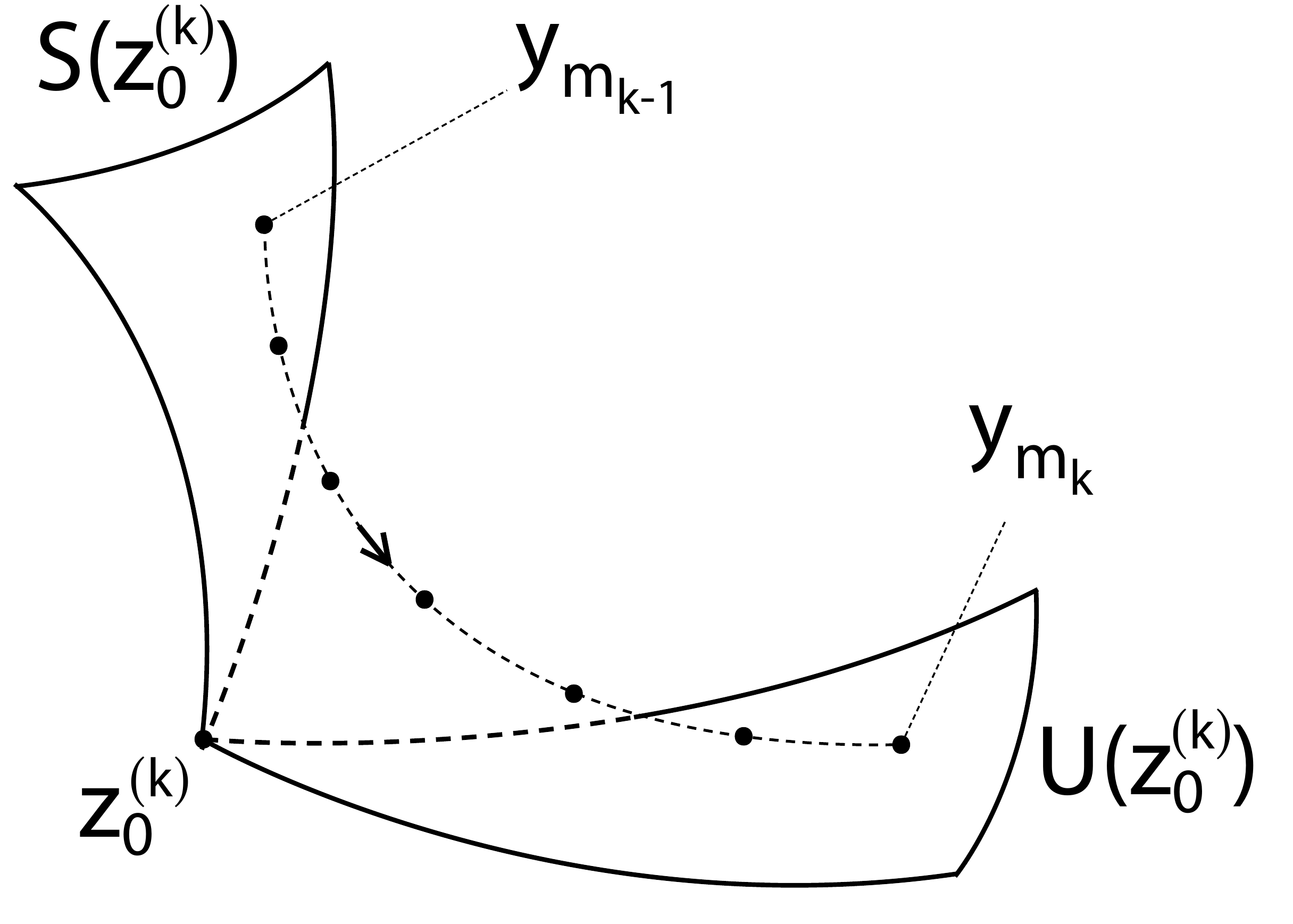}}
 \caption{The trajectory segment $\lbrace y_{m_{k-1}}, \dots, y_{m_k} \rbrace$ enters and exits the neighborhood of $\lbrace z^{(k)}_0 \rbrace$ via a region exponentially close to $S(z^{(k)}_0)$ and $U(z^{(k)}_0)$, respectively. Note that $y_{m_{k-1}}$ is $O(e^{-n_k \mu^{(f)}_{\beta_k}})$-close to $S(z^{(k)}_0)$, and $y_{m_{k}}$ is $O(e^{-n_k \mu^{(f)}_{\beta_k}})$-close to $U(z^{(k)}_0)$. }
\label{fig:Fly_by}
\end{figure}     
As illustrated by Fig.~\ref{fig:Fly_by}, under $n_k$ iterations of the map, the trajectory segment $\lbrace y_{m_{k-1}}, \dots, y_{m_k} \rbrace$ enters the neighborhood of each $\lbrace z^{(k)}_0 \rbrace$ via a region exponentially close to its stable manifold $S(z^{(k)}_0)$, makes a near fly-by with $\lbrace z^{(k)}_0 \rbrace$, and exits via a region exponentially close to its unstable manifold $U(z^{(k)}_0)$. The action $F(y_{m_1},y_{m_{L-1}})$ can be built up from the sum of the periodic-orbit actions $\sum_{k=2}^{L-1}{\cal F}_{\beta_k}$, plus correction terms $J(\beta_k \cdot \beta_{k+1})$ as action connectors between $\lbrace z^{(k)}_0 \rbrace$ and $\lbrace z^{(k+1)}_0 \rbrace$. The main purpose of this subsection is to derive an explicit expression of $J(\beta_k \cdot \beta_{k+1})$. 

Additional auxiliary homoclinic points $g^{(i,j)}_0$ are needed for the process, which are identified by generalizing Eq.~\eqref{eq:Connector homoclinic point k and k+1} into arbitrary combinations of $\beta_i$ and $\beta_j$:
\begin{equation}\label{eq:Auxilliary homoclinic point i and j}
g^{(i,j)}_0 \Rightarrow \overline{0} \beta_i \cdot \beta_j \overline{0}\ ,
\end{equation}
and in particular, $g^{(k,k)}_0 \Rightarrow \overline{0} \beta_k \cdot \beta_k \overline{0}$, which will be used extensively later.  Starting from Eq.~(27) of Ref.~\cite{Li18} and by replacing $\gamma$ with $\beta_k$, one obtains an expression for the action of each $\overline{\beta}_k$:
\begin{equation}\label{eq:Periodic orbit action Li18}
\begin{split}
{\cal F}_{\beta_k}& = n_k {\cal F}_0 + \Delta {\cal F}_{\overline{0} \beta_{k} \beta_{k} \overline{0}, \overline{0} \beta_{k}\overline{0}} +O(e^{- n_k \mu^{(f)}_{\beta_k}})\\
& = n_k {\cal F}_0 + \Delta {\cal F}_{ \lbrace g^{(k,k)}_0   \rbrace \lbrace h^{(k)}_0 \rbrace } +O(e^{- n_k \mu^{(f)}_{\beta_k}})\ .
\end{split}
\end{equation}
With the help of Eq.~\eqref{eq:Area-action two homoclinic pairs} and replacing
\begin{equation}
\begin{split}
& a_0 \mapsto g^{(k,k)}_0 \\
& c_0 \mapsto h^{(k)}_0\\
& d_0 \mapsto x \\
& b_0 \mapsto h^{(k)}_{n_k}\ ,
\end{split}
\end{equation}
gives
\begin{equation}
\label{eq:whatever}
 \Delta {\cal F}_{ \lbrace g^{(k,k)}_0   \rbrace \lbrace h^{(k)}_0 \rbrace } - \Delta {\cal F}_{ \lbrace h^{(k)}_0   \rbrace \lbrace x \rbrace } = {\cal A}^{\circ}_{SUSU[ g^{(k,k)}_0, h^{(k)}_0, x, h^{(k)}_{n_k} ]}.
\end{equation}
Substituting Eq.~\eqref{eq:whatever} into Eq.~\eqref{eq:Periodic orbit action Li18} yields
\begin{equation}\label{eq:Periodic orbit action single}
\begin{split}
{\cal F}_{\beta_k} = & n_k {\cal F}_0 +  \Delta {\cal F}_{ \lbrace h^{(k)}_0   \rbrace \lbrace x \rbrace } + {\cal A}^{\circ}_{SUSU[ g^{(k,k)}_0, h^{(k)}_0, x, h^{(k)}_{n_k} ]} \\
& + O(e^{- n_k \mu^{(f)}_{\beta_k}}) \\
= & n_k {\cal F}_0  +  \Delta {\cal F}_{ \overline{0} \beta_{k}\overline{0},\overline{0}} + {\cal A}^{\circ}_{SUSU[ g^{(k,k)}_0, h^{(k)}_0, x, h^{(k)}_{n_k} ]}\\
& + O(e^{- n_k \mu^{(f)}_{\beta_k}})\ .
\end{split}
\end{equation}
Substituting Eq.~\eqref{eq:Periodic orbit action single} into Eq.~\eqref{eq:General trajectory action approx} and accounting for cancellations between common integration paths leads to
\begin{equation}
\label{eq:General trajectory action cycle expansion}
\begin{split}
 F(y_{m_1},y_{m_{L-1}}) =& \int\limits_{S\left[g^{(1,2)}_{0},g^{(2,2)}_{0}\right]}\mathbf{p}\cdot\mathrm{\mathbf{d}}\mathbf{q} + \sum_{k=2}^{L-1}{\cal F}_{\beta_k}\\
& +   \sum_{k=2}^{L-2} \int\limits_{U\left[g^{(k,k)}_{0},g^{(k,k+1)}_{0}\right]}\mathbf{p}\cdot\mathrm{\mathbf{d}}\mathbf{q} \\
&+ \sum_{k=2}^{L-2} \int\limits_{S\left[g^{(k,k+1)}_{0},g^{(k+1,k+1)}_{0}\right]}\mathbf{p}\cdot\mathrm{\mathbf{d}}\mathbf{q}\\
&+ \int\limits_{U\left[g^{(L-1,L-1)}_{0},g^{(L-1,L)}_{0}\right]}\mathbf{p}\cdot\mathrm{\mathbf{d}}\mathbf{q}\\
&+ O\left(\max_{k\in [1,L]}e^{- n_k \mu^{(f)}_{\beta_k}}\right)\ .
\end{split}
\end{equation}
To simplify the notations, define the action connectors $J_S(\beta_k \cdot \beta_{k+1})$, $J_U(\beta_k \cdot \beta_{k+1})$, and $J(\beta_k \cdot \beta_{k+1})$ as
\begin{equation}\label{eq:General trajectory action cycle expansion J connectors}
\begin{split}
& J_S(\beta_k \cdot \beta_{k+1}) \equiv \int\limits_{S\left[g^{(k,k+1)}_{0},g^{(k+1,k+1)}_{0}\right]}\mathbf{p}\cdot\mathrm{\mathbf{d}}\mathbf{q} \\
& J_U(\beta_k \cdot \beta_{k+1}) \equiv \int\limits_{U\left[g^{(k,k)}_{0},g^{(k,k+1)}_{0}\right]}\mathbf{p}\cdot\mathrm{\mathbf{d}}\mathbf{q} \\
& J(\beta_k \cdot \beta_{k+1}) \equiv J_U(\beta_k \cdot \beta_{k+1}) + J_S(\beta_k \cdot \beta_{k+1})\ .
\end{split}
\end{equation}
Since the points $g^{(i,j)}_0$ are uniquely specified by their symbolic codes $\overline{0} \beta_i \cdot \beta_j \overline{0}$ (Eq.~\eqref{eq:Auxilliary homoclinic point i and j}), $J_S(\beta_k \cdot \beta_{k+1})$, $J_U(\beta_k \cdot \beta_{k+1})$, and $J(\beta_k \cdot \beta_{k+1})$ are also uniquely defined by substrings $\beta_k$ and $\beta_{k+1}$.  Equation~\eqref{eq:General trajectory action cycle expansion} simplifies to the form
\begin{equation}
\label{eq:General trajectory action cycle expansion final}
\begin{split}
& F(y_{m_1},y_{m_{L-1}})\\
& = J_S(\beta_1 \cdot \beta_{2}) + \sum_{k=2}^{L-1}{\cal F}_{\beta_k}\\
&\quad + \sum_{k=2}^{L-2} J(\beta_k \cdot \beta_{k+1}) + J_U(\beta_{L-1} \cdot \beta_{L})\\
&\quad  + O\left(\max_{k\in [1,L]}e^{- n_k \mu^{(f)}_{\beta_k}}\right)\ ,
\end{split}
\end{equation}
\begin{figure}[ht]
\centering
{\includegraphics[width=8cm]{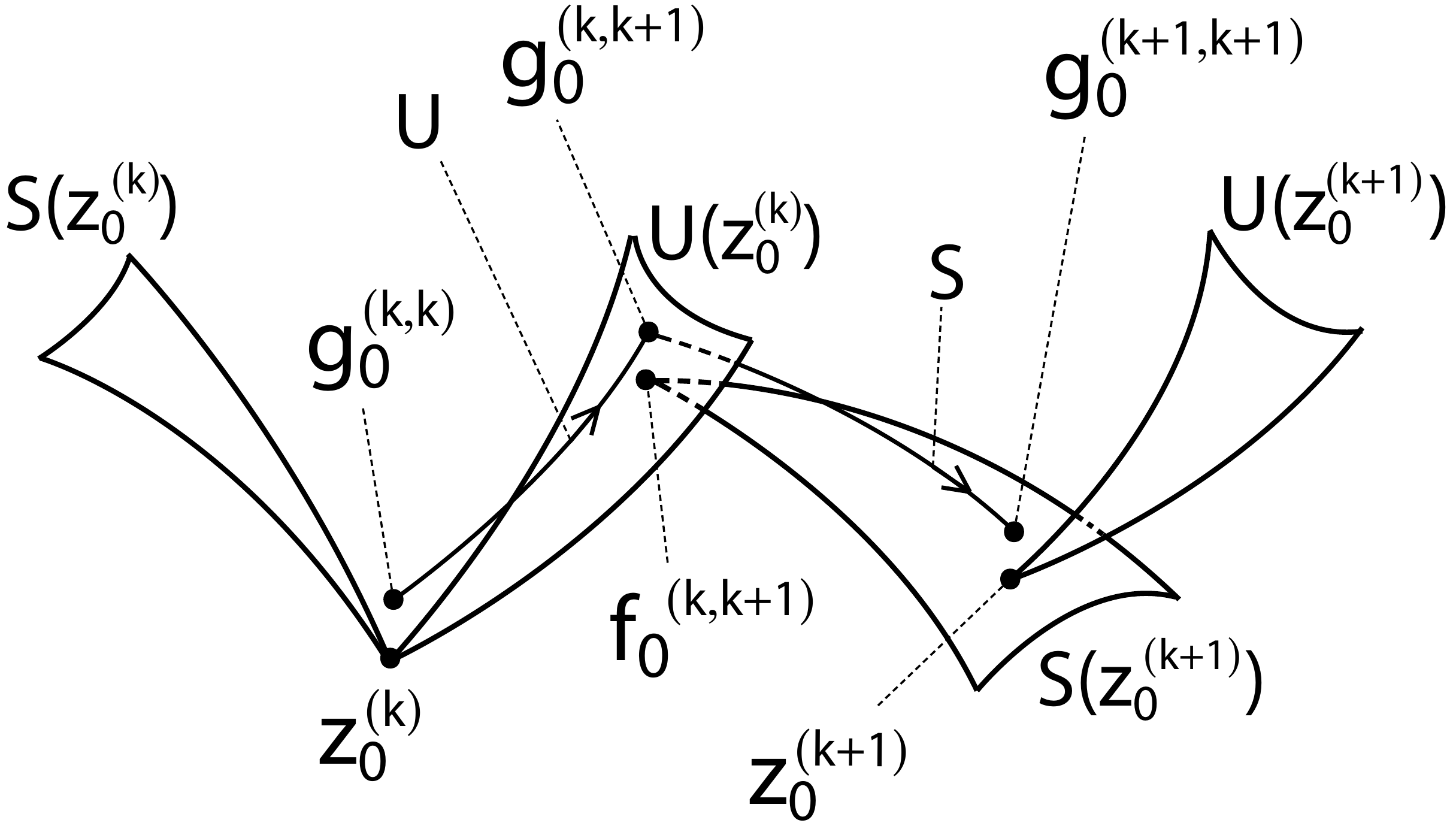}}
 \caption{The action connector $J(\beta_k \cdot \beta_{k+1})$ is indicated by arrows. $f^{(k,k+1)}_0 \in U(z^{(k)}_0) \cap S(z^{(k+1)}_0)$ is a heteroclinic point between $z^{(k)}_0$ and $z^{(k+1)}_0$, defined by code $f^{(k,k+1)}_0 \Rightarrow \overline{\beta}_k \cdot \overline{\beta}_{k+1}$. Due to the special choice of symbolic codes, $z^{(k)}_0$ and $g^{(k,k)}_0$ are $O(e^{-n_k \mu^{(f)}_{\beta_k}})$-close, while $f^{(k,k+1)}_0$ and $g^{(k,k+1)}_0$ are $O\left(\max\lbrace e^{-n_k \mu^{(f)}_{\beta_k}},e^{-n_{k+1} \mu^{(f)}_{\beta_{k+1}}}\rbrace \right)$-close. Notice that $g^{(k,k+1)}_0 \notin U(z^{(k)}_0)$. The integration paths $U\left[g^{(k,k)}_{0},g^{(k,k+1)}_{0}\right]$ and $S\left[g^{(k,k+1)}_{0},g^{(k+1,k+1)}_{0}\right]$ are arbitrary curves on $U(x)$ and $S(x)$, respectively, connecting the corresponding endpoints. Although not plotted here, the local $U(x)$ is approximately parallel to $U(z^{(k)}_0)$, and the local $S(x)$ is approximately parallel to $S(z^{(k+1)}_0)$. Therefore, the paths $U\left[g^{(k,k)}_{0},g^{(k,k+1)}_{0}\right]$ and $S\left[g^{(k,k+1)}_{0},g^{(k+1,k+1)}_{0}\right]$ are also approximately parallel to $U(z^{(k)}_0)$ and $S(z^{(k+1)}_0)$, respectively.  }
\label{fig:Cycle_expansion_connector}
\end{figure}     
\noindent which represents the periodic-orbit expansion of $F(y_{m_1},y_{m_{L-1}})$. The first term on the RHS of Eq.~\eqref{eq:General trajectory action cycle expansion final}, $J_S(\beta_1 \cdot \beta_{2}) $, is the action connector between the $\beta_1$ and $\beta_2$ segments. The second term, $ \sum_{k=2}^{L-1}{\cal F}_{\beta_k}$, is the sum of the contributions from all periodic orbits $\overline{\beta}_k$ ($k=2, \dots,L-1$). The third term, $\sum_{k=2}^{L-2} J(\beta_k \cdot \beta_{k+1})$, as illustrated in Fig.~\ref{fig:Cycle_expansion_connector}, is the sum of the action connectors between $\beta_k$ and $\beta_{k+1}$ segments. The fourth term, $J_U(\beta_{L-1} \cdot \beta_{L})$, is the final action connector between $\beta_{L-1}$ and $\beta_L$. The formula only requires the construction of simple periodic orbits $\overline{\beta}_k$ and homoclinic points $g^{(k,k)}_0$ and $g^{(k,k+1)}_0$, which can be done by stable numerical techniques.  

Similar to Sec.~\ref{Application to periodic orbits}, Eq.~\eqref{eq:General trajectory action cycle expansion final} can be used in the special case in which $\lbrace y_0 \rbrace$ is a periodic orbit.  For a long periodic orbit $\lbrace y_0\rbrace \Rightarrow \overline{\gamma}$, where $y_0 \Rightarrow \overline{\gamma}\cdot\overline{\gamma}$, we partition it in the same way as Sec.~\ref{Application to periodic orbits} into $L-2$ segments $\gamma = \gamma_1 \cdots \gamma_{L-2}$. This indicates during one full period the orbit visits the neighborhoods of simpler periodic orbits $\overline{\gamma}_k$ successively for $k=1, \dots, L-2$. Upon using the same substitutions as Eq.~\eqref{eq:General to periodic substrings}, Eq.~\eqref{eq:General trajectory action cycle expansion final} yields the periodic-orbit action expansion
\begin{equation}\label{eq:Periodic orbit action cycle expansion final}
{\cal F}_{\gamma} =  \sum_{k=1}^{L-2} \big[ {\cal F}_{\gamma_k} + J(\gamma_k \cdot \gamma_{k+1}) \big] + O\left(\max_{k\in [1,L-2]}e^{- n_k \mu^{(f)}_{\gamma_k}}\right)
\end{equation}
where the index $k$ of $\gamma_{k}$ is cyclic in $L-2$: $\gamma_{L-1} \equiv \gamma_1$. Eq.~\eqref{eq:Periodic orbit action cycle expansion final} gives the action expansion of a long periodic orbit $\overline{\gamma}$ in terms of short periodic orbits $\overline{\gamma}_k$ constructed from its substrings, and $J(\gamma_k \cdot \gamma_{k+1})$ as the action connector between $\gamma_k$ and $\gamma_{k+1}$.  An interesting fact is that the sum of the connectors, 
\begin{figure}[ht]
\centering
{\includegraphics[width=7cm]{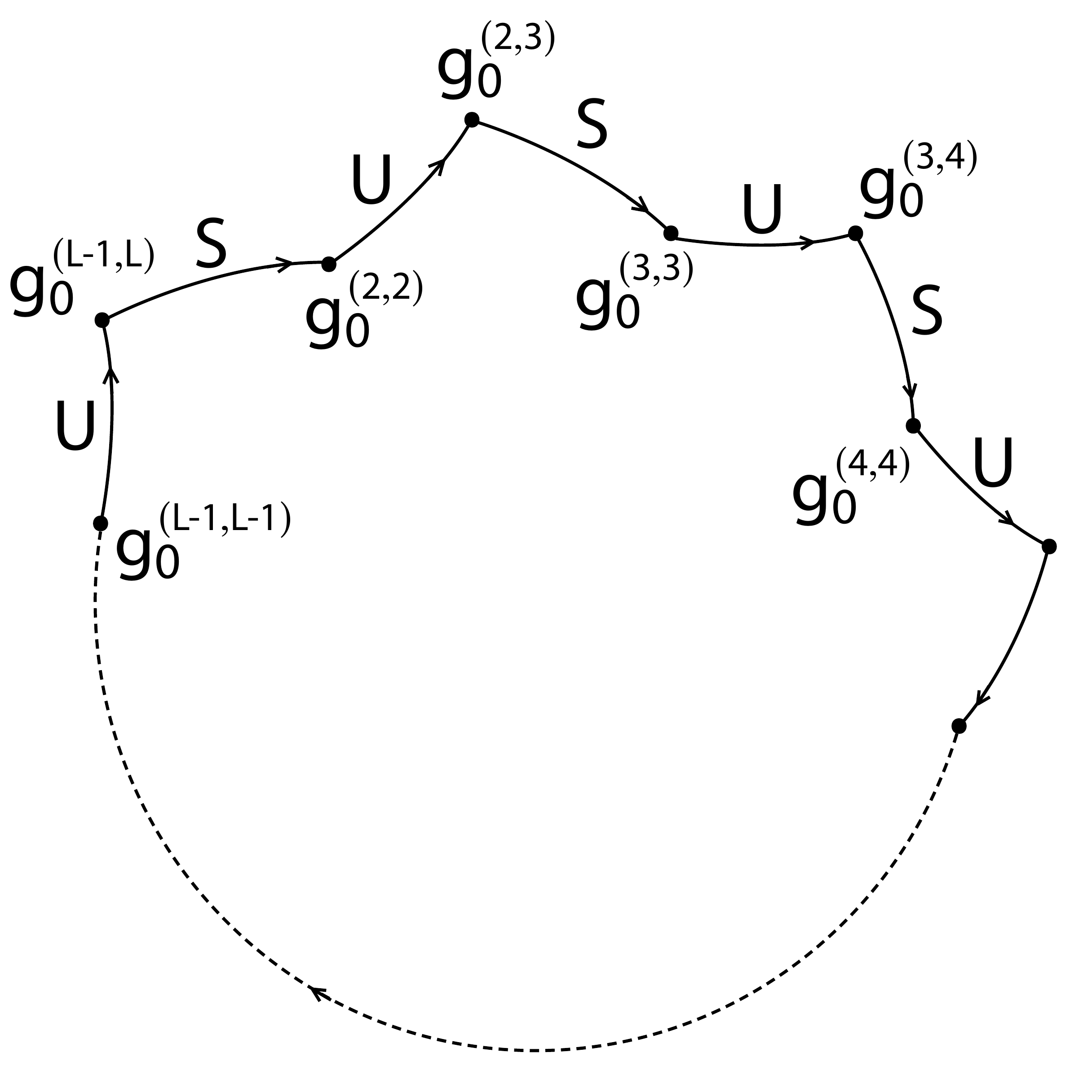}}
 \caption{(Schematic) $\sum_{k=1}^{L-2} J(\gamma_k \cdot \gamma_{k+1})$ is the symplectic area of the loop marked by arrows. Note that $g^{(i+1,j+1)}_0\Rightarrow \overline{0} \gamma_i \cdot \gamma_{j} \overline{0}$, and $g^{(L,L)}_0=g^{(2,2)}_0$. }
\label{fig:Loop_connector}
\end{figure}     
\begin{equation}\label{eq:Periodic orbit action cycle expansion connector sum}
\sum_{k=1}^{L-2} J(\gamma_k \cdot \gamma_{k+1})\ ,
\end{equation}
yields the symplectic area of the loop shown by Fig.~\ref{fig:Loop_connector}. This symplectic area is thus the main action correction between ${\cal F}_{\gamma}$ and $\sum_{k=1}^{L-2} {\cal F}_{\gamma_k}$. 

Equations~\eqref{eq:General trajectory action cycle expansion final} and \eqref{eq:Periodic orbit action cycle expansion final}, which are based on periodic-orbit expansions, are equivalent to Eqs.~\eqref{eq:General trajectory action approx} and \eqref{eq:Periodic orbit expansion}, respectively, which are based on homoclinic-orbit expansions. Therefore, homoclinic and periodic orbits are equally ideal skeletal structures for the phase space dynamics. 

\subsection{Numerical examples: approximation accuracy}
\label{Numerical verification}

The accuracy of the procedure, Eqs.~\eqref{eq:General trajectory action approx} and \eqref{eq:Periodic orbit expansion}, and its dependence on substring length is demonstrated with a numerical example from the H\'{e}non map [Eq.~\eqref{eq:Henon map}] at parameter value $a=10$. This parameter value gives rise to a complete Smale horseshoe-shaped homoclinic tangle~\cite{Smale63,Smale80} (see Appendix.~\ref{Markov partition}) with highly chaotic dynamics as it is well beyond the first tangency~\cite{Devaney79}. The generating Markov partition is a simple set of two regions $[V_0,V_1]$. The trajectory of a non-escaping initial point $z_0$ is then described by a symbolic string of binary digits,  where each digit $s_{n} \in \lbrace 0,1\rbrace$ such that $M^{n}(z_0) \in V_{s_n}$. 

\subsubsection{Accuracy expectations}
\label{Accuracy expectations}

In this two-degree-of-freedom example, all periodic orbits $\overline{\gamma}$ have one positive stability exponent, $\mu_{\gamma} > 0$.  The two simplest periodic orbits are the hyperbolic fixed points $x \Rightarrow \overline{0} \cdot \overline{0}$ and $x^{\prime} \Rightarrow \overline{1} \cdot \overline{1}$. Their stability exponents are calculated to be 
\begin{equation}
\label{eq:Henon map fixed points 0 and 1 exponents}
\begin{split}
& \mu_0 = 2.142 \\
& \mu_1 = 1.483.
\end{split}
\end{equation}
There exists a periodic orbit for any combination of zeros and ones as a symbolic string; i.e. no ``pruning front" exists in the symbolic plane~\cite{Cvitanovic88a,Cvitanovic91}.  Traversing a periodic orbit for a full period, if the current iteration is at digit $0$, it indicates that the current point belongs to the $V_0$ strip (Fig.~\ref{fig:horseshoe}).  As a rough estimate, the tangent dynamics under one iteration is approximately uniform everywhere inside $V_0$. The exponential stretching and compressing rate for the current iteration of the orbit will be close to $\mu_0$.  Following the same reasoning, tangent dynamics along the periodic orbit at digits $1$ can be characterized by $\mu_1$. Because of this, given any periodic orbit $\lbrace y \rbrace \Rightarrow \overline{\gamma}$, its stability exponent $\mu_{\gamma}$ can be estimated roughly by
 \begin{equation}\label{eq:Henon map periodic orbit exponents estimate}
 \mu_{\gamma} \sim (N_0 \mu_0 + N_1 \mu_1)/N
 \end{equation}
 where $N_0$ and $N_1$ are the numbers of $0$s and $1$s, respectively in the string $\gamma$, and $N=N_0+N_1$ is the length of $\gamma$.  We emphasize here that Eq.~\eqref{eq:Henon map periodic orbit exponents estimate} only serves as a practical estimate for the error terms in Eqs.~\eqref{eq:General trajectory action approx} and \eqref{eq:Periodic orbit expansion}, and is not intended to provide an accurate calculation of the stability exponents of the periodic orbits themselves.  As a demonstration, consider three periodic orbits, namely $\overline{1011}$, $\overline{0001}$, and $\overline{00011}$, and calculated their stability exponents:
 \begin{equation}\label{eq:Henon map periodic orbit exponents estimate numerics exact}
 \begin{split}
 & \mu_{1011} = 1.5934 \approx ( \mu_0 + 3 \mu_1)/4 = 1.6477 \\
 & \mu_{0001} = 1.9668 \approx ( 3 \mu_0 + \mu_1)/4 = 1.9772 \\
 & \mu_{00011}=1.9119   \approx ( 3 \mu_0 + 2\mu_1)/5=1.8783.
 \end{split}
 \end{equation}
The values roughly agree. The key insight of Eq.~\eqref{eq:Henon map periodic orbit exponents estimate} is for two different periodic orbits of the same lengths, the one that has more $0$s in its symbolic code tends to have the larger stability exponent. This is simply because $\mu_0 > \mu_1$.  Therefore, the magnitude of the estimated error in using the approximate form to calculate the classical action is dominated by the shortest string with the fewest $0$s in it.  The effect on the approximation accuracy of different length partitions and proportion of $0$ symbols is illustrated ahead.

\subsubsection{Partition length}
\label{Partition length}

Consider the exact trajectory $\lbrace y \rbrace = \alpha \beta \delta$ whose symbolic sequence is given by
\begin{equation}
\label{eq:General trajectory action length 15}
\begin{split}
&\alpha = \overline{0} \\
&\beta = 011110111011110 \\
&\delta = \overline{0}\ .
\end{split}
\end{equation}
The symbol length of $\beta$ has $15$ characters, which is conveniently partitioned into $5$ pieces as
\begin{equation}
\label{eq:General trajectory action approx length 3 worst codes}
\begin{split}
&\beta_1 = 011 \\
&\beta_2 = 110 \\
&\beta_3 = 111 \\
&\beta_4 = 011 \\
&\beta_5 = 110 \\
\end{split}
\end{equation}
The classical action of interest for this partition is the one given by $9$ iterations of the mapping taking the trajectory from an initial condition $y_3$ to $y_{12}$, i.e.
\begin{equation}
\label{eq:General trajectory action approx length 3 worst orbit points}
\begin{split}
& y_{m_1} = y_3 \Rightarrow \overline{0}11 \cdot 11011101111 \overline{0} \\
& y_{m_4} = y_{12} \Rightarrow \overline{0} 11110111011 \cdot 11 \overline{0}.
\end{split}
\end{equation}
The exact classical action for this trajectory segment turns out to be
\begin{equation}
\label{eq:General trajectory action approx length 3 worst F exact}
F^{(\text{exact})}(y_3, y_{12}) = -97.9401 
\end{equation}    
to the number of digits needed for comparison to the approximations.  A homoclinic orbit with a long excursion length is chosen for convenience because its action can be very accurately calculated in a fast and stable way~\cite{Li17}, but any orbit could have been selected. This orbit itself is not used in any way to calculate the approximation, only the shorter excursion homoclinic orbits defined by the partition.

Using the partition defined above, the approximation, Eq.~\eqref{eq:General trajectory action approx}, yields
\begin{equation}
\label{eq:General trajectory action approx length 3 worst F approx}
F^{(\text{approx})}(y_3, y_{12}) = -98.2363,
\end{equation}
and thus the absolute error is given by
\begin{equation}
\label{eq:General trajectory action approx length 3 worst F diff}
F^{(\text{exact})}(y_3, y_{12}) - F^{(\text{approx})}(y_3, y_{12}) = 0.2962,
\end{equation}
which is quite accurate relatively speaking with such short partition lengths.  Nevertheless, in a semiclassical theory where $\hbar$ divides the actions, small differences can lead to unwanted large phase changes.

To increase the partition length used for the approximation scheme to test how the accuracy changes, the first step is to borrow a character each from the $\alpha$ and $\delta$ codes.  This increases the length of $\beta$ to $17$ characters (an extra $0$ on the left and right, but of course the orbit is still the same).  Now consider the partition
\begin{equation}
\label{eq:General trajectory action approx length 3 to 4 worst codes}
\begin{split}
&\beta_1 = 0011 \\
&\beta_2 = 1101 \\
&\beta_3 = 11011 \\
&\beta_4 = 1100 \ , \\
\end{split}
\end{equation}
mostly of symbol length $4$ except $\beta_3$ which has length $5$.  For this new partition, the point previously denoted $y_3\Rightarrow y^\prime_4$ and $y_{12}\Rightarrow y^\prime_{13}$ (the index shifted by $1$ due to the $0$ symbol taken from $\alpha$).  The new approximate action turns out to be
\begin{equation}\label{eq:General trajectory action approx length 3 to 4 worst F approx}
F^{(\text{approx})}(y^\prime_4, y^\prime_{13}) = -97.9322.
\end{equation}
which gives an absolute error of
\begin{equation}
\label{eq:General trajectory action approx length 4 worst F approx}
F^{(\text{exact})}(y^\prime_4, y^\prime_{13}) - F^{(\text{approx})}(y^\prime_4, y^\prime_{13}) = -0.0079,
\end{equation}
which is nearly two orders of magnitude more accurate.  This illustrates the surprising rapidity of exponential convergence rates with partition length.

Finally, we mention that we have constructed other examples (not given here) that illustrate another feature of the accuracy expectations.  Note that if one keeps the trajectory segment fixed, but increases the mean partition length as in the previous example, there are necessarily fewer partitions.  If instead, one allows the trajectory segment to change, but instead one fixes the number of partitions, one can generally expect the accuracy to increase exponentially with increasing partition lengths.  This is born out although some variation is expected depending on the relative proportion of $0$ and $1$ symbols.

\subsubsection{Relative proportion of symbols}

A final meaningful comparison with the partition in Eq.~\eqref{eq:General trajectory action approx length 3 worst codes} is concerned with the relative proportion of $1$ and $0$ symbols in $\beta$.  A trajectory segment whose $\beta$ has a greater proportion of $0$s is expected to have less error than a trajectory with a smaller proportion because $\mu_0 > \mu_1$.  Swapping several of the $1$s for $0$s in the example given by Eq.~\eqref{eq:General trajectory action approx length 3 worst codes} leads to
\begin{equation}
\label{eq:General trajectory action approx length 3 optimal codes}
\begin{split}
&\beta_1 = 001 \\
&\beta_2 = 010 \\
&\beta_3 = 100 \\
&\beta_4 = 001 \\
&\beta_5 = 100 \\
\end{split}
\end{equation}
which should result in somewhat smaller approximation errors.  The exact classical action for this case is
\begin{equation}
\label{eq:General trajectory action approx length 3 optimal F exact}
F^{(\text{exact})}(y_3, y_{12}) = 59.4968
\end{equation}    
and the approximate
\begin{equation}
\label{eq:General trajectory action approx length 3 optimal F approx}
F^{(\text{approx})}(y_3, y_{12}) = 59.6026 ,
\end{equation}
giving
\begin{equation}
\label{eq:General trajectory action approx length 3 optimal F exact}
F^{(\text{exact})}(y_3, y_{12}) - F^{(\text{approx})}(y_3, y_{12}) = -0.1058
\end{equation}    
which is roughly a third of the absolute error found with the previous trajectory having the same number of partitions and partition lengths.

\subsubsection{Periodic orbits}
\label{Periodic orbits}

It is worth giving an example of the application of Eq.~\eqref{eq:Periodic orbit expansion}, which is the equivalent of Eq.~\eqref{eq:General trajectory action approx} for periodic orbits. A period-$12$ orbit $\lbrace y_0\rbrace \Rightarrow \overline{\gamma}$ with symbolic code
\begin{equation}\label{eq:Periodic orbit example symbolic code}
\begin{split}
& y_0 = y_{12} \Rightarrow \overline{\gamma} \cdot \overline{\gamma} \\ 
& \gamma = 111111011110
\end{split}
\end{equation}   
has a classical action given by
\begin{equation}\label{eq:Periodic orbit example F exact}
{\cal F}^{(\text{exact})}_{\gamma} = F(y_{0},y_{12}) = -138.6038. 
\end{equation} 
Consider the partition into $3$ length-$4$ substrings
\begin{equation}
\label{eq:Periodic orbit example symbolic code len 4 partition}
\gamma = \gamma_1 \gamma_2 \gamma_3
\end{equation}
where 
\begin{equation}\label{eq:Periodic orbit example symbolic code len 4 partition substrings}
\begin{split}
& \gamma_1 = 1111\\
& \gamma_2 = 1101\\
& \gamma_3 = 1110.
\end{split}
\end{equation}
The substrings $\gamma_i$ are chosen to be dominantly $1$s rather than $0$s so that the size of the $V_{\gamma_i}$ regions will be relatively large.  So this example is expected to be a nearly worst case scenario or a nearly upper bound for the error terms in Eq.~\eqref{eq:Periodic orbit expansion} under three length-$4$ partitions.  The approximate result is
\begin{equation}
\label{eq:Periodic orbit example len 4 partition F approx}
{\cal F}^{(\text{approx})}_{\gamma}= -138.5152
\end{equation}
which gives 
\begin{equation}
\label{eq:Periodic orbit example len 4 partition F approx diff}
{\cal F}^{(\text{exact})}_{\gamma} - {\cal F}^{(\text{approx})}_{\gamma}= -0.0886
\end{equation}
The alternative partition of $\gamma$ into two length-$6$ substrings:
\begin{equation}
\label{eq:Periodic orbit example symbolic code len 6 partition substrings}
\begin{split}
& \gamma_1 = 111111\\
& \gamma_2 = 011110,
\end{split}
\end{equation}
yields
\begin{equation}
\label{eq:Periodic orbit example len 6 partition F approx}
{\cal F}^{(\text{exact})}_{\gamma} - {\cal F}^{(\text{approx})}_{\gamma}=  0.0029 \ ,
\end{equation}
which is more than an order of magnitude more accurate. 

\section{Conclusions}
\label{Conclusions}

Special classical trajectory sets play important roles in both classical and quantum chaotic dynamics through their use in trace formulas.  It has long been known that one of these special sets, i.e.~homoclinic orbits, periodic orbits, etc., can be relevant for the calculation of dynamical averages, depending on the quantity of interest.  In fact, shown here is that all the details of the dynamics of individual trajectories can be captured by these special sets.  The results apply quite generally and are not restricted to low-dimensional chaotic dynamics.  In particular, exact formulas are given that express the classical action of any trajectory segment in terms of simpler homoclinic or periodic orbits and certain symplectic areas.  Whereas the exact formulas require construction of the trajectory segments, approximation schemes are given with controllable exponentially small errors in which the construction is not required, only a section of its symbolic sequence corresponding to the segment.  This is a great simplification.

The total number of relevant trajectories that are needed for the semiclassical trace formulas proliferates exponentially fast with increasing propagation times (or iteration numbers), rendering exponentially demanding computation times and storage spaces for standard numerical procedures. On the contrary, the relations given here make use of a much smaller set of simple homoclinic (periodic) orbits, and provides exact or extremely accurate approximate expressions of generic unstable trajectory actions.  They can be used as a starting point for understanding the action correlations in Hamiltonian chaos, corrections to cycle expansions, or the role of Richter-Sieber~\cite{Sieber01} pairs in time reversal invariant systems.

The main results in this article are expressed in terms of symbolic dynamics. Since each symbolic code corresponds to a unique phase-space trajectory, the formulas derived here will hold true for systems without a known symbolic dynamics, although more work is needed to identify the one-to-one correspondences between the trajectory segments and the auxiliary homoclinic orbits, without the help of their symbolic codes.  

Another fascinating issue is the identification of generating Markov partitions in multidimensional systems. Although the theory for their existence criteria and the mechanisms for the creation of symbolic dynamics in higher dimensions are sophisticated~\cite{Wiggins88,Katok95}, more work would be desirable on the practical identifications of the Markov partitions in such systems.  However, new methods have been developed in recent years~\cite{Rubido18,Zhang20}, which provides promising instruments for finding Markov partitions in multidimensional systems. 
  
\acknowledgments

JL acknowledges financial support from Japan Society for the Promotion of Science (JSPS) in the form of JSPS International Fellowship for Research in Japan (Standard).  

\appendix

\section{HORSESHOE, MARKOV PARTITIONS AND SYMBOLIC DYNAMICS}

\begin{figure}[ht]
\centering
{\includegraphics[width=7cm]{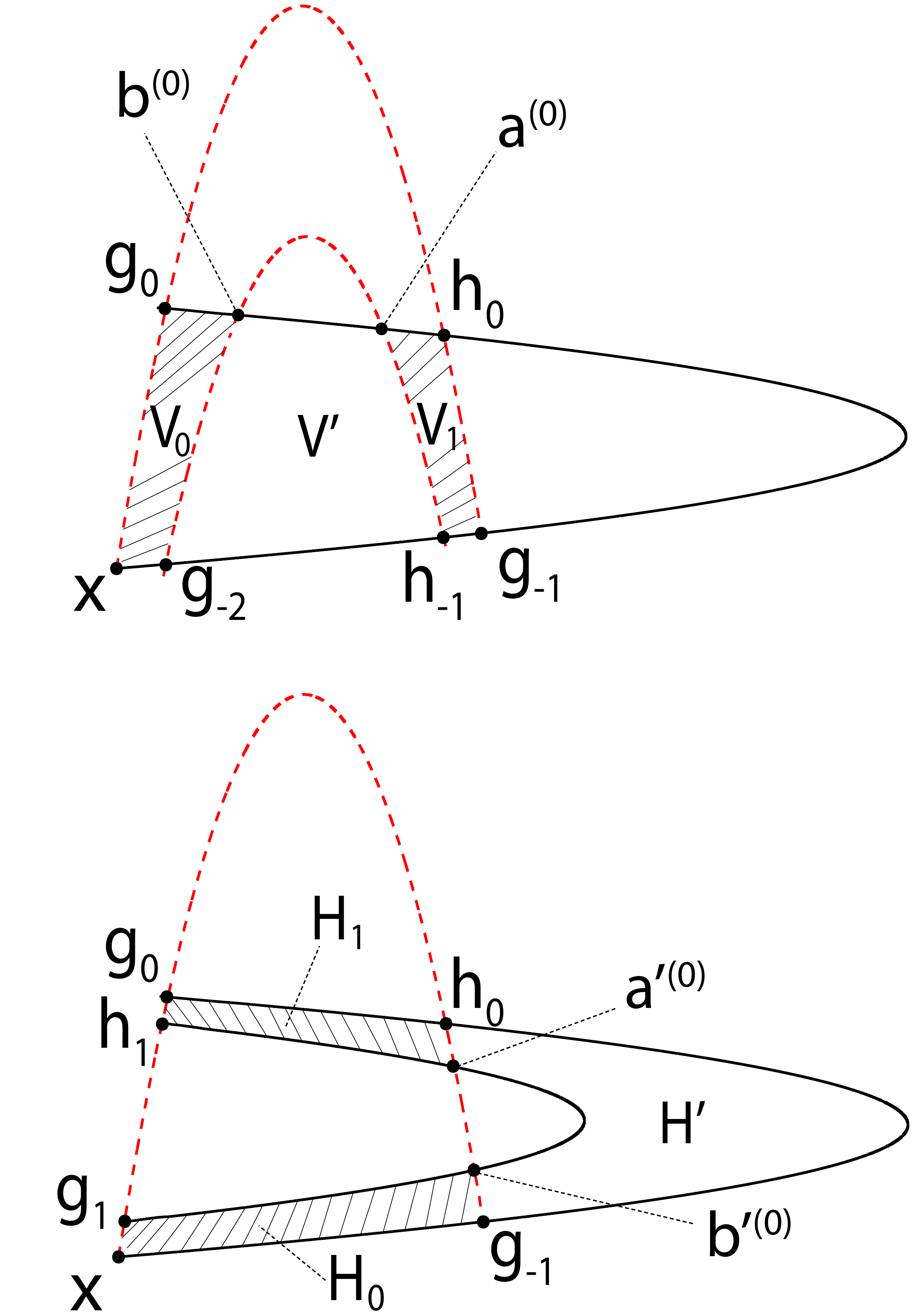}}
 \caption{Example partial homoclinic tangle from the H\'{e}non map, which forms a complete horseshoe structure. The unstable (stable) manifold of $x$ is the solid (dashed) curve. Under forward iteration, the vertical strips $V_0$ and $V_1$ (including the boundaries) from the upper panel are mapped into the horizontal strips $H_0$ and $H_1$ in the lower panel.}
\label{fig:horseshoe}
\end{figure}     

Symbolic dynamics provides a powerful technique, i.e.~the topological description of orbits in chaotic systems~\cite{Hadamard1898,Birkhoff27a,Birkhoff35,Morse38}. Perhaps the most famous model that demonstrates its elegance is the horseshoe map~\cite{Smale63,Smale80}, a two-dimensional diffeomorphism possessing an invariant Cantor set $\Omega$, which is topologically conjugate to a Bernoulli shift on symbolic strings composed by ``$0$"s and ``$1$"s. In such scenarios, the Markov partition is a simple set of two regions $[V_0,V_1]$, as shown in the upper panel of Fig.~\ref{fig:horseshoe}. Each phase-space point $z_0 \in \Omega$  can be put into an one-to-one correspondence with a bi-infinite symbolic string in Eq.~\eqref{eq:symbolic code}, where each digit $s_{n} \in {0,1}$ such that $M^{n}(z_0) \in V_{s_n}$.  A numerical realization of the horseshoe is the area-preserving H\'{e}non map \cite{Henon76} defined on the phase plane $(q,p)$, which is the simplest polynomial automorphism giving rise to chaotic dynamics~\cite{Friedland89}:
\begin{equation}\label{eq:Henon map}
\begin{split}
&p_{n+1}=q_n\\
&q_{n+1}=a-q_{n}^2-p_n.
\end{split}
\end{equation}
It follows from the work in Ref.~\cite{Devaney79} that for sufficiently large parameter values of $a$ the H\'{e}non map is topologically conjugate to a horseshoe map, therefore possessing a hyperbolic invariant set of orbits labeled by binary symbolic codes; see Chapters 23 and 24 of Ref.~\cite{Wiggins03} for a brief review of the Smale horseshoe and the corresponding symbolic dynamics. 
\label{Markov partition}

To visualize the action of the mapping $M$ (e.g. Eq.\eqref{eq:Henon map}) on the homoclinic tangle, let us consider the closed region $\cal{R}$ in Fig.~\ref{fig:horseshoe}, bounded by loop $\mathcal{L}_{USUS[x,g_{-1},h_0,g_0]}$, where $\mathcal{L}_{USUS[x,g_{-1},h_0,g_0]}=U[x,g_{-1}]+S[g_{-1},h_0]+U[h_0,g_0]+S[g_0,x]$. Under the mapping $M$, the trapezoid-shaped $\cal{R}$ is compressed along the stable direction and stretched along the unstable direction, and folded back to partially overlap with itself, with the vertical strips $V_0$ and $V_1$ mapped into the horizontal strips $H_0$ and $H_1$, respectively. Similarly, the inverse mapping $M^{-1}$ stretches $\cal{R}$ along the stable direction and fold back, with the horizontal strips $H_0$ and $H_1$ mapped into $V_0$ and $V_1$, respectively. Therefore, points in region $E_0$ bounded by $\mathcal{L}_{USUS[g_{-2},h_{-1},h^{\prime}_{-1},g^{\prime}_{-1}]}$ are mapped outside $\cal{R}$ into $E_1$ bounded by $\mathcal{L}_{USUS[g_{-1},h_{0},h^{\prime}_{0},g^{\prime}_{0}]}$ under one iteration. For open systems such as the H\'{e}non map, any point outside $\cal{R}$  never returns and escapes to infinity; there is a similar construction for inverse time.  Of great structural significance is the non-wandering set $\Omega$ of phase-space points $z$ that stay inside $\cal{R}$ for all iterations~\cite{ChaosBook,Wiggins03}:        
\begin{equation}\label{eq:Nonwandering set}
\Omega=\big\lbrace z:z\in \bigcap_{n=-\infty}^{\infty}M^{n}(\cal{R}) \big\rbrace.
\end{equation}  
In particular, we focus on the homoclinic and periodic points that belong to $\Omega$.  

Using the closed regions $V_0$ and $V_1$ in Fig.~\ref{fig:horseshoe} as Markov generating partition for the symbolic dynamics, every point $z_0$ in $\Omega$ can be labeled by an infinite symbolic string of $0$'s and $1$'s:
\begin{equation}\label{eq:symbolic code 2}
z_0 \Rightarrow \cdots s_{-2}s_{-1} \cdot s_{0}s_{1}s_{2}\cdots
\end{equation}
where each digit $s_n$ in the symbol denotes the region that $M^{n}(z_0)$ lies in: $M^{n}(z_0) = z_n \in V_{s_{n}}$, $s_n \in \lbrace 0,1\rbrace$. In that sense, the symbolic code gives an ``itinerary" of $z_0$ under successive forward and backward iterations, in terms of the regions $V_0$ and $V_1$ that each iteration lies in. The semi-infinite segment ``$s_{0}s_{1}s_{2}\cdots$" (resp. ``$\cdots s_{-2}s_{-1}$") from the symbolic code is referred to as the $\mathit{head}$ (resp. $\mathit{tail}$) of the orbit with initial condition $z_0$~\cite{Hagiwara04}, and the dot separating the head and the tail denotes the region ($V_{s_0}$) that the current iteration $z_0$ belongs to. Let $\Sigma$ denote the symbolic space of all such bi-infinite symbolic strings. Strings in $\Sigma$ are then in 1-to-1 correspondence with points in $\Omega$, and the mapping $M$ in phase space is topological conjugate to a Bernoulli shift in the symbolic space.  Therefore, forward iterations of $z_0$ move its dot towards the right side of the symbolic string, and backward iterations move it towards the left side. 

\begin{figure}
 \subfigure{
   \label{fig:Markov_1}
   \includegraphics[width=5.5cm]{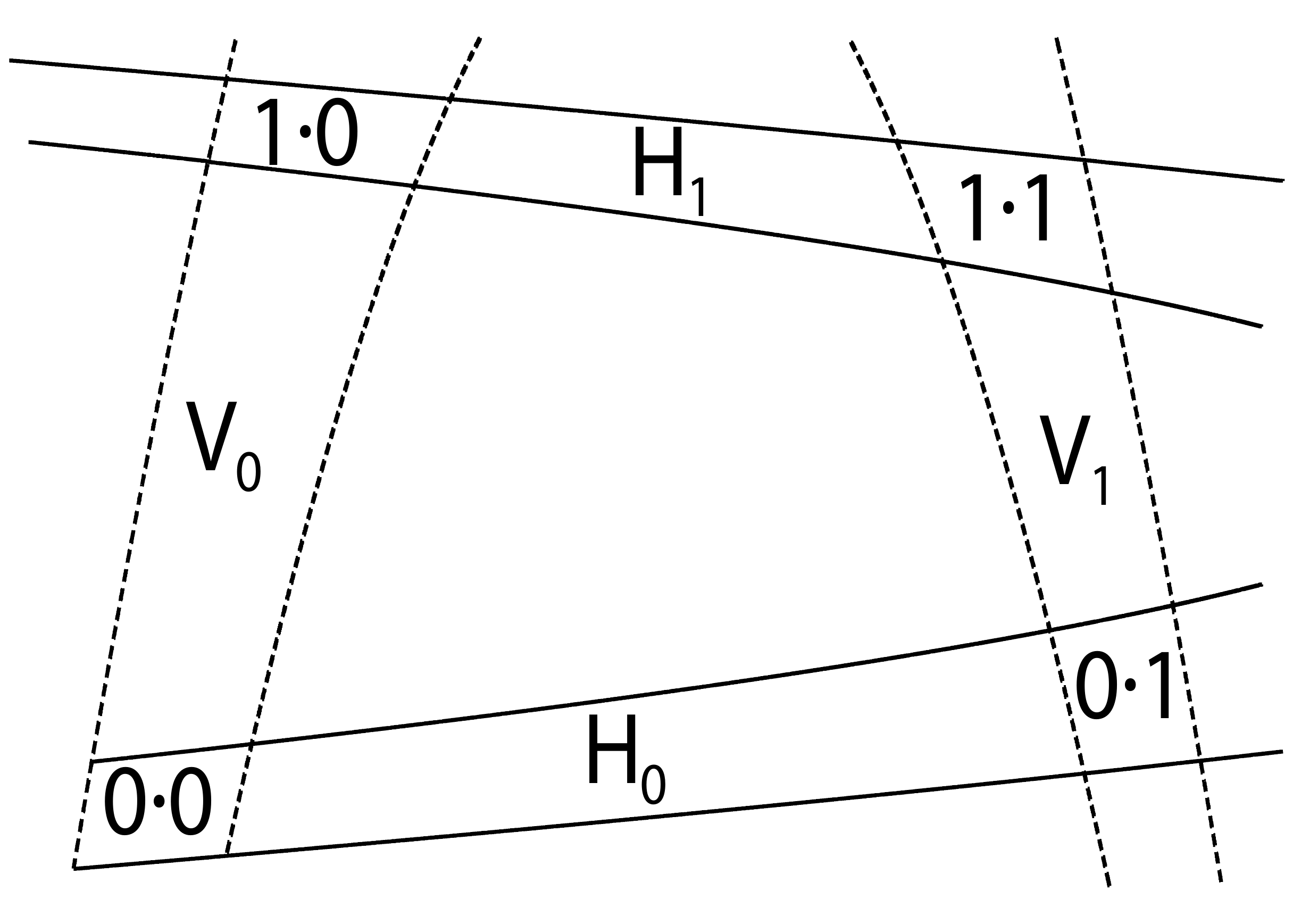}}
 \subfigure{
   \label{fig:Markov_2}
   \includegraphics[width=5.5cm]{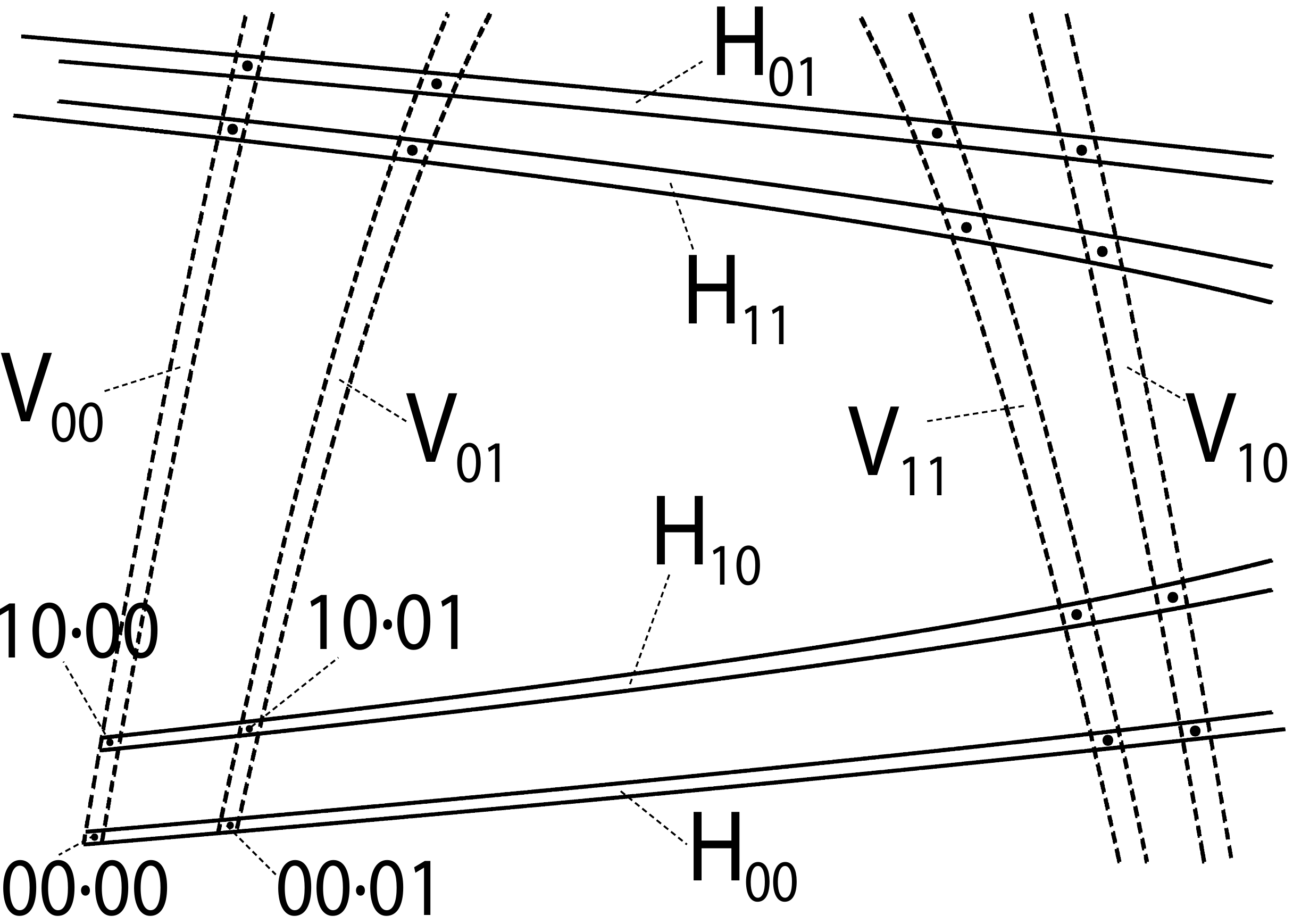}}
\caption{Markov partitions constructed in the H\'{e}non map. Upper panel: The $V_{s_{0}}$ and $H_{s_{-1}}$ regions corresponds to the same regions in Fig.~\ref{fig:horseshoe}. The four cells $H_{s_{-1}}\cap V_{s_0} \Rightarrow s_{-1} \cdot s_{0}$ are the Markov partitions of lengths $2$. Lower panel: Markov partitions of length $4$. The horizontal and vertical strips are created as $H_{s_{-2}s_{-1}}=M(H_{s_{-2}})\cap H_{s_{-1}}$ and $V_{s_0 s_1}=V_{s_0} \cap M^{-1}(V_{s_1})$. The $H$ and $V$ strips intersect at sixteen cells $H_{s_{-2}s_{-1}} \cap V_{s_0 s_1} \Rightarrow s_{-2} s_{-1} \cdot s_0 s_1$, as indicated by a black dot inside each of them. For the sake of clarity, we only explicitly labeled four cells in the lower left corner. Any point from $\Omega$ with symbolic string of fixed central block $\cdots s_{-2} s_{-1} \cdot s_0 s_1 \cdots$ must either locate inside or on the boundary of the $s_{-2}s_{-1}\cdot s_0 s_1$ cell. The sizes of the cells shrink exponentially with increasing string lengths.  } 
\label{fig:Markov}
\end{figure}         

Besides elegant topological conjugacy, the symbolic strings also contain information about the location of points in phase space. Following a standard procedure \cite{Wiggins88}, subsequent Markov partitions \cite{Bowen75,Gaspard98} can be constructed from the generating partitions $[V_0,V_1]$, which specifies the phase-space regions that points with certain central blocks of fixed lengths must locate within. Starting from $V_0$ and $V_1$, define recursively an ever-shrinking family of vertical strips $V_{s_0\cdots s_{n-1}}$ in phase space, such that:
\begin{equation}\label{eq:Markov partition vertical strips}
V_{s_0\cdots s_{n-1}}\equiv V_{s_0} \bigcap M^{-1}(V_{s_1\cdots s_{n-1}})
\end{equation}  
where $s_{i}\in \lbrace 0,1\rbrace$ for $i=0,\cdots,n-1$. Similarly, starting from $H_0$ and $H_1$, an ever-shrink family of horizontal strips $H_{s_{-n}\cdots s_{-1}}$ can be defined:
\begin{equation}\label{eq:Markov partition horizontal strips}
H_{s_{-n}\cdots s_{-1}} \equiv M(H_{s_{-n}\cdots s_{-2}}) \bigcap H_{s_{-1}} 
\end{equation}  
where $s_{-j}\in \lbrace 0,1\rbrace$ for $j=1,\cdots,n$. The horizontal strips are just forward images of the corresponding vertical strips: $H_{s_0\cdots s_{n-1}} =  M^{n}(V_{s_0\cdots s_{n-1}})$. Under $n$ iterations of the map, $V_{s_0\cdots s_{n-1}}$ is compressed along the stable manifold, at the meantime stretched along the unstable manifold while keeping its total area unchanged, and eventually deformed into $H_{s_0\cdots s_{n-1}}$. Denoting the symbolic string $s_0\cdots s_{n-1}$ by Greek letter $\gamma$: $\gamma = s_0\cdots s_{n-1}$, the exponential stretching rate for the entire process can be estimated using the stability exponent of the periodic orbit $\overline{\gamma}$, namely $\mu_{\gamma}$, which leads to an estimate for the size of the areas of $V_{\gamma}$ and $H_{\gamma}$: 
\begin{equation}\label{eq:Markov partition size estimates}
V_{\gamma},\ H_{\gamma} \sim O(e^{-n_{\gamma}\mu_{\gamma}})
\end{equation} 
where $n_{\gamma}$ is the length of $\gamma$. Typical periodic orbits $\overline{\gamma}$ in chaotic systems will have positive $\mu_{\gamma}$, and thus the sizes of $V_{\gamma}$ and $H_{\gamma}$ shrink exponentially rapidly with the length of $\gamma$. 

The horizontal and vertical strips intersect at curvy ``rectangular" cells, which can be labeled by a finite string of symbols: 
\begin{equation}\label{eq:Markov partition cells}
H_{\gamma_1} \bigcap V_{\gamma_2} \Rightarrow \gamma_1 \cdot \gamma_2
\end{equation}
where $\gamma_1 = s_{-n}\cdots s_{-1}$ and $\gamma_2 = s_0\cdots s_{n-1}$ denote the length-$n$ symbolic strings. These cells are Markov partitions of central block lengths $2n$, in the sense that any point from $\Omega$ with coinciding central blocks $\gamma_1 \cdot \gamma_2$ must locate inside (or on the boundary of) the corresponding cell. Shown in the upper and lower panels of Fig.~\ref{fig:Markov} are two examples of Markov partitions of lengths $2$ and $4$, respectively, numerically generated from the H\'{e}non map. Take the cell $10 \cdot 01$ from the lower panel as example, any point with symbolic string of the form: $\cdots s_{-4}s_{-3} 10 \cdot 01 s_2 s_3 \cdots$ must either locate inside or on the boundary of $10 \cdot 01$. 

\begin{figure}[ht]
\centering
{\includegraphics[width=6.5cm]{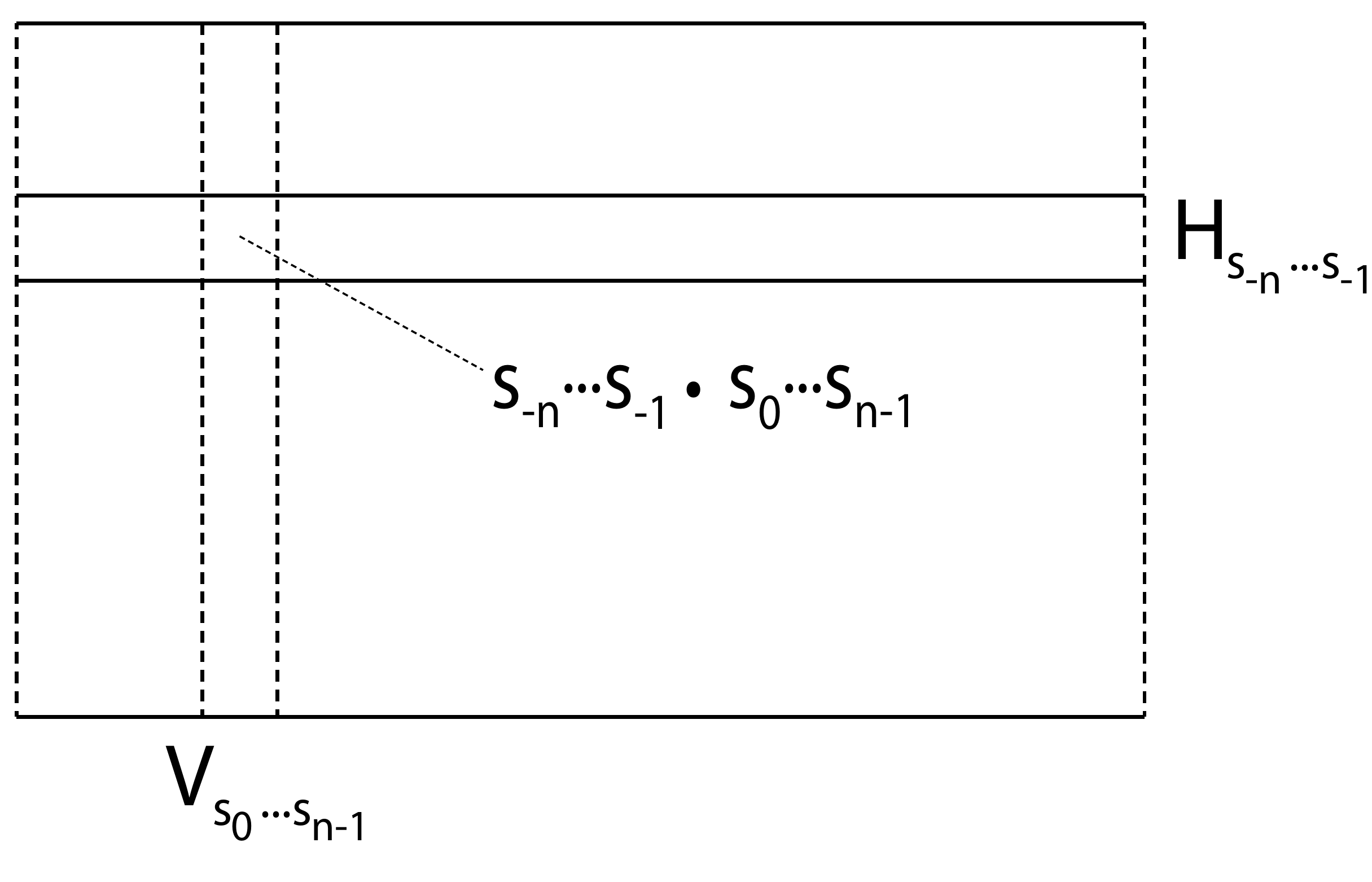}}
 \caption{(Schematic) $\gamma_1 = s_{-n}\cdots s_{-1}$ and $\gamma_2 = s_0\cdots s_{n-1}$. The width of $H_{\gamma_1}$ is $\sim O(e^{-n\mu_{\gamma_1}})$, and the width of $V_{\gamma_2}$ is $\sim O(e^{-n\mu_{\gamma_2}})$, so the cell area $\gamma_1 \cdot \gamma_2$ is $\sim O(e^{-(n\mu_{\gamma_1} +n\mu_{\gamma_2})})$. }
\label{fig:Shrinking_Cells}
\end{figure}  

Closeness between two symbolic strings imply closeness between the corresponding points in phase space. Because of the compressing and stretching nature of the horseshoe map, the widths of the horizontal and vertical strips becomes exponentially small with increasing block lengths, and so do the cell areas they intersect. Without loss of generality, we assume, in Fig.~\ref{fig:horseshoe}, that the area ${\cal A}^{\circ}_{SUSU[x,g_0,h_0,g_{-1}]}$ is of order $\sim O(1)$. Then the resulting area of the cell $\gamma_1 \cdot \gamma_2$ is of order $\sim O(e^{-(n\mu_{\gamma_1}+n\mu_{\gamma_2})})$, where $\mu_{\gamma_1}$ is the stability exponent of the periodic orbit $\overline{\gamma_1}$, and $\mu_{\gamma_2}$ is the stability exponent of the periodic orbit $\overline{\gamma_2}$. Averaging over all possible combinations of $\gamma_1$ and $\gamma_2$ , the area of the cell $\gamma_1 \cdot \gamma_2$ can be estimated as $\sim O(e^{-2n \mu})$, where $\mu$ is the Lyapunov exponent of the system, an exponentially small area for large $n$ values. This geometry \cite{Wiggins88} is shown by Fig.~\ref{fig:Shrinking_Cells}. Therefore, any two points from $\Omega$ with identical central blocks of length $2n$ must locate in the same exponentially small cell. Consider two points $h \Rightarrow \cdots s_{-n} \cdots s_{-1}\cdot s_0 \cdots s_n \cdots$ and $h^{\prime} \Rightarrow \cdots s^{\prime}_{-n} \cdots s^{\prime}_{-1}\cdot s^{\prime}_0 \cdots s^{\prime}_n \cdots$, if $h$ and $h^{\prime}$ agree on a central block of length $2n$, i.e., $s^{\prime}_{-n} \cdots s^{\prime}_{-1}\cdot s^{\prime}_0 \cdots s^{\prime}_{n-1}=s_{-n}\cdots s_{-1}\cdot s_0\cdots s_{n-1}$, they must both located in same cell labeled by $s_{-n} \cdots s_{-1}\cdot s_0 \cdots s_{n-1}$
\begin{equation}\label{eq:Matching central block lengths area estimate}
h,h^{\prime} \in H_{s_{-n}\cdots s_{-1}} \bigcap V_{s_0\cdots s_{n-1}} \Rightarrow s_{-n}\cdots s_{-1}\cdot s_0\cdots s_{n-1}
\end{equation}
the area of which is $\sim O(e^{-2n\mu})$. Therefore, by specifying longer and longer central block lengths of a point's symbolic string, we can narrow down its possible location in phase space with smaller and smaller cells from the Markov partition.

\section{MACKAY-MEISS-PERCIVAL ACTION PRINCIPLE}
\label{MacKay-Meiss-Percival}

\begin{figure}[ht]
\centering
{\includegraphics[width=8cm]{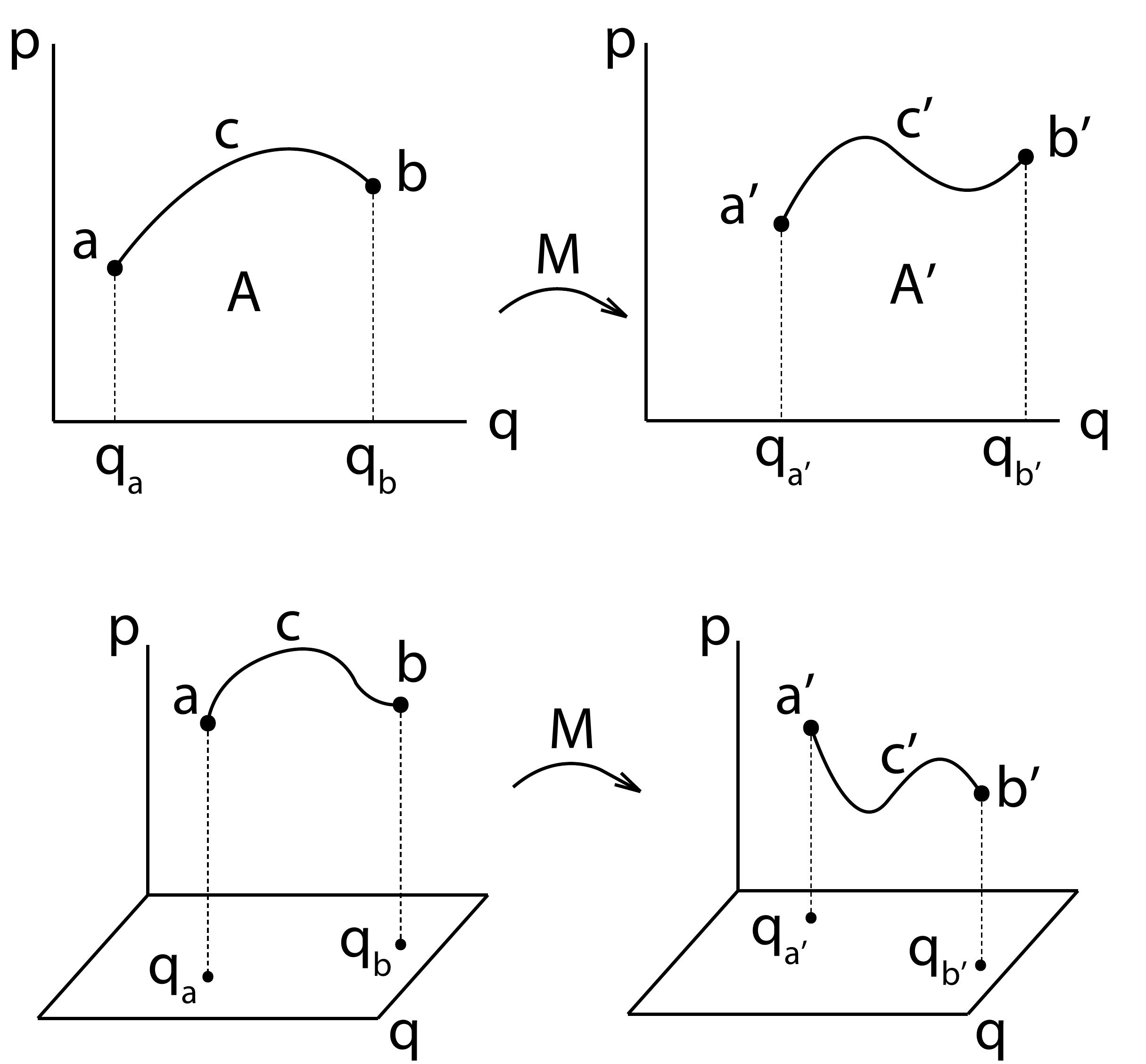}}
 \caption{(Schematic) $a$ and $b$ are arbitrary points and $c$ is a curve connecting them.  $a'=M(a)$, $b'=M(b)$ and $c'=M(c)$.  Upper panel: two-dimensional version. $A'-A=F(q_{b},q_{b'})-F(q_{a},q_{a'})$. Lower panel: multidimensional version. }
\label{fig:Area_under_curve}
\end{figure}  

The MacKay-Meiss-Percival action principle discussed in this section was first developed in \cite{MacKay84a} for transport theory.  A comprehensive review can be found in \cite{Meiss92}. Generalization of the original principle beyond the ``twist" and area-preserving conditions is discussed in \cite{Easton91}, and we only give a brief outline of the theory in this appendix. Shown in Fig.~\ref{fig:Area_under_curve} are two arbitrary points $a=(\mathbf{q_{a}},\mathbf{p_{a}})$, $b=(\mathbf{q_{b}},\mathbf{p_{b}})$ and their images $a'=M(a)$, $b'=M(b)$.  Let $c$ be an arbitrary curve connecting $a$ and $b$, which is mapped to a curve $c'=M(c)$ connecting $a'$ and $b'$. Shown in Fig.~\ref{fig:Area_under_curve} are the two-dimensional (upper panel) and multidimensional (lower panel) scenarios of the action principle. For two-dimensional cases, let $A$ and $A'$ denote the algebraic area under $c$ and $c'$ respectively.  Then the difference between these areas is
\begin{equation}\label{eq:Meiss92}
\begin{split}
A'-A&=\int_{c'}p\mathrm{d}q-\int_{c}p\mathrm{d}q\\
&=F(q_{b},q_{b'})-F(q_{a},q_{a'})
\end{split}
\end{equation}
i.e., the difference between the two algebraic areas gives the difference between the action functions for one iteration of the map. 

For $2f$-dimensional phase space we have similarly
\begin{equation}\label{eq:Meiss92 multidimensional}
\begin{split}
& F(\mathbf{q_{b}},\mathbf{q_{b'}})-F(\mathbf{q_{a}},\mathbf{q_{a'}})\\
& =\int\limits_{c^{\prime}[a^{\prime},b^{\prime}]} \sum_{j=1}^{f} p_j \mathrm{d}q_j - \int\limits_{c[a,b]} \sum_{j=1}^{f} p_j\mathrm{d}q_j  \\
&= \int\limits_{c^{\prime}[a^{\prime},b^{\prime}]} \mathbf{p}\cdot\mathrm{\mathbf{d}}\mathbf{q} - \int\limits_{c[a,b]}\mathbf{p}\cdot\mathrm{\mathbf{d}}\mathbf{q}\ . 
\end{split}
\end{equation}

Starting from this, MacKay $\mathit{et}$ $\mathit{al.}$ \cite{MacKay84a} derived a formula on the action difference between a pair of homoclinic orbits, namely $\lbrace a_0 \rbrace$ and $\lbrace b_0 \rbrace$, for which
\begin{figure}[ht]
\centering
{\includegraphics[width=8cm]{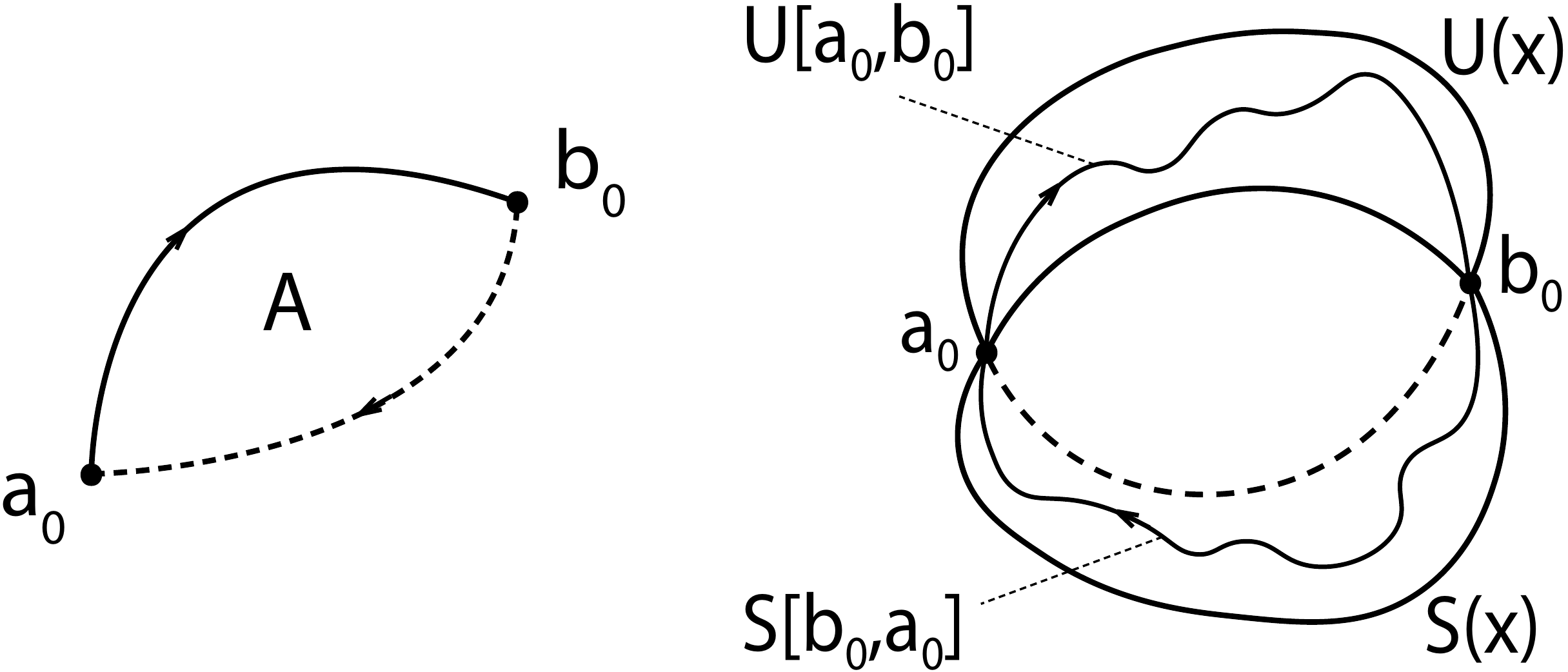}}
 \caption{Homoclinic orbit pair $a_0$ and $b_0$. Left panel: two-dimensional phase space. They are connected by an unstable segment $U[a_{0},b_{0}]$ (solid) and a stable segment $S[b_{0},a_{0}]$ (dashed).  Then the action difference between the homoclinic orbit pair is $\Delta {\cal F}_{\lbrace b_0\rbrace \lbrace a_0\rbrace}=A$. Right panel: $2f$-dimensional phase space. $U(x)$ and $S(x)$ are $f$-dimensional surfaces. $U[a_0,b_0] \subset U(x)$ and $S[b_0,a_0] \subset S(x)$ are arbitrary paths between $a_0$ and $b_0$. Together they form a loop $US[a_0,b_0]$ which gives rise to the symplectic area in Eq.~\eqref{eq:Homoclinic action difference appendix}.     }
\label{fig:Homoclinic_pair_action}
\end{figure}  
\begin{equation}\label{eq:asymptotic pair}
a_{\pm\infty} = b_{\pm\infty} = x,
\end{equation}
where $x$ is a hyperbolic fixed point. Then as shown by Fig.~\ref{fig:Homoclinic_pair_action}, $a_0$ and $b_0$ are connected by $U(x)$ and $S(x)$. Let $U[a_{0},b_{0}] \subset U(x)$ and $S[b_{0},a_{0}] \subset S(x)$ be arbitrary paths between the two points, we first apply Eq.~\eqref{eq:Meiss92 multidimensional} repeatedly to the semi-infinite pair of homoclinic orbit segments $\lbrace a_{-\infty},\cdots,a_{0}\rbrace$ and $\lbrace b_{-\infty},\cdots,b_0\rbrace$, and get:
\begin{equation}\label{eq:Homoclinic action difference infinite past appendix}
\begin{split}
&\lim_{N\to\infty} \sum_{i=-N}^{-1} [F(b_{i},b_{i+1})-F(a_{i},a_{i+1})]\\
&=\int\limits_{U[a_{0},b_{0}]}\mathbf{p}\cdot\mathrm{\mathbf{d}}\mathbf{q}-\int\limits_{U[a_{-\infty},b_{-\infty}]}\mathbf{p}\cdot\mathrm{\mathbf{d}}\mathbf{q}=\int\limits_{U[a_{0},b_{0}]}\mathbf{p}\cdot\mathrm{\mathbf{d}}\mathbf{q}
\end{split}
\end{equation}
where $\int\limits_{U[a_{-\infty},b_{-\infty}]}\mathbf{p}\cdot\mathrm{\mathbf{d}}\mathbf{q}=0$ since $a_{-\infty} \to b_{-\infty}$. Similarly for the semi-infinite pairs $\lbrace a_{0},\cdots,a_{\infty}\rbrace$ and $\lbrace b_{0},\cdots,b_{\infty}\rbrace$ we have:
\\
\begin{equation}\label{eq:Homoclinic action difference infinite future appendix}
\begin{split}
&\lim_{N\to\infty} \sum_{i=0}^{N-1} [F(b_{i},b_{i+1})-F(a_{i},a_{i+1})]\\
&=\int\limits_{S[a_{\infty},b_{\infty}]}\mathbf{p}\cdot\mathrm{\mathbf{d}}\mathbf{q}-\int\limits_{S[a_{0},b_{0}]}\mathbf{p}\cdot\mathrm{\mathbf{d}}\mathbf{q}=\int\limits_{S[b_{0},a_{0}]}\mathbf{p}\cdot\mathrm{\mathbf{d}}\mathbf{q}\ .
\end{split}
\end{equation}
Adding up Eqs.~\eqref{eq:Homoclinic action difference infinite past appendix} and \eqref{eq:Homoclinic action difference infinite future appendix} we have: 
\\
\begin{equation}\label{eq:Homoclinic action difference appendix}
\begin{split}
\Delta {\cal F}_{\lbrace b_0\rbrace \lbrace a_0\rbrace}&=\lim_{N\to\infty} \sum_{i=-N}^{N-1}[F(b_i,b_{i+1}) - F(a_i, a_{i+1})]\\
&=\int\limits_{U[a_{0},b_{0}]}\mathbf{p}\cdot\mathrm{\mathbf{d}}\mathbf{q}+\int\limits_{S[b_{0},a_{0}]}\mathbf{p}\cdot\mathrm{\mathbf{d}}\mathbf{q}\\
&={\cal A}^\circ_{US[a_0,b_0]}\ .
\end{split}
\end{equation}
\\
For two-dimensional systems, ${\cal A}^\circ_{US[a_0,b_0]}$ reduces to the area $A$ shown in the left panel of Fig.~\ref{fig:Homoclinic_pair_action}. For systems with $2f$-dimensional phase space ($f \geq 2$), ${\cal A}^\circ_{US[a_0,b_0]}$ is the symplectic area of the loop shown in the right panel of Fig.~\ref{fig:Homoclinic_pair_action}.

\bibliography{classicalchaos,quantumchaos}

\begin{thebibliography}{53}%
\makeatletter
\providecommand \@ifxundefined [1]{%
 \@ifx{#1\undefined}
}%
\providecommand \@ifnum [1]{%
 \ifnum #1\expandafter \@firstoftwo
 \else \expandafter \@secondoftwo
 \fi
}%
\providecommand \@ifx [1]{%
 \ifx #1\expandafter \@firstoftwo
 \else \expandafter \@secondoftwo
 \fi
}%
\providecommand \natexlab [1]{#1}%
\providecommand \enquote  [1]{``#1''}%
\providecommand \bibnamefont  [1]{#1}%
\providecommand \bibfnamefont [1]{#1}%
\providecommand \citenamefont [1]{#1}%
\providecommand \href@noop [0]{\@secondoftwo}%
\providecommand \href [0]{\begingroup \@sanitize@url \@href}%
\providecommand \@href[1]{\@@startlink{#1}\@@href}%
\providecommand \@@href[1]{\endgroup#1\@@endlink}%
\providecommand \@sanitize@url [0]{\catcode `\\12\catcode `\$12\catcode
  `\&12\catcode `\#12\catcode `\^12\catcode `\_12\catcode `\%12\relax}%
\providecommand \@@startlink[1]{}%
\providecommand \@@endlink[0]{}%
\providecommand \url  [0]{\begingroup\@sanitize@url \@url }%
\providecommand \@url [1]{\endgroup\@href {#1}{\urlprefix }}%
\providecommand \urlprefix  [0]{URL }%
\providecommand \Eprint [0]{\href }%
\providecommand \doibase [0]{http://dx.doi.org/}%
\providecommand \selectlanguage [0]{\@gobble}%
\providecommand \bibinfo  [0]{\@secondoftwo}%
\providecommand \bibfield  [0]{\@secondoftwo}%
\providecommand \translation [1]{[#1]}%
\providecommand \BibitemOpen [0]{}%
\providecommand \bibitemStop [0]{}%
\providecommand \bibitemNoStop [0]{.\EOS\space}%
\providecommand \EOS [0]{\spacefactor3000\relax}%
\providecommand \BibitemShut  [1]{\csname bibitem#1\endcsname}%
\let\auto@bib@innerbib\@empty
\bibitem [{\citenamefont {Poincar\'e}(1899)}]{Poincare99}%
  \BibitemOpen
  \bibfield  {author} {\bibinfo {author} {\bibfnamefont {H.}~\bibnamefont
  {Poincar\'e}},\ }\href@noop {} {\emph {\bibinfo {title} {Les m\'ethodes
  nouvelles de la m\'ecanique c\'eleste}}},\ Vol.~\bibinfo {volume} {3}\
  (\bibinfo  {publisher} {Gauthier-Villars et fils},\ \bibinfo {address}
  {Paris},\ \bibinfo {year} {1899})\BibitemShut {NoStop}%
\bibitem [{\citenamefont {Artuso}\ \emph
  {et~al.}(1990{\natexlab{a}})\citenamefont {Artuso}, \citenamefont {Aurell},\
  and\ \citenamefont {Cvitanovi\'{c}}}]{Artuso90a}%
  \BibitemOpen
  \bibfield  {author} {\bibinfo {author} {\bibfnamefont {R.}~\bibnamefont
  {Artuso}}, \bibinfo {author} {\bibfnamefont {E.}~\bibnamefont {Aurell}}, \
  and\ \bibinfo {author} {\bibfnamefont {P.}~\bibnamefont {Cvitanovi\'{c}}},\
  }\href@noop {} {\bibfield  {journal} {\bibinfo  {journal} {Nonlinearity}\
  }\textbf {\bibinfo {volume} {3}},\ \bibinfo {pages} {325} (\bibinfo {year}
  {1990}{\natexlab{a}})}\BibitemShut {NoStop}%
\bibitem [{\citenamefont {Artuso}\ \emph
  {et~al.}(1990{\natexlab{b}})\citenamefont {Artuso}, \citenamefont {Aurell},\
  and\ \citenamefont {Cvitanovi\'{c}}}]{Artuso90b}%
  \BibitemOpen
  \bibfield  {author} {\bibinfo {author} {\bibfnamefont {R.}~\bibnamefont
  {Artuso}}, \bibinfo {author} {\bibfnamefont {E.}~\bibnamefont {Aurell}}, \
  and\ \bibinfo {author} {\bibfnamefont {P.}~\bibnamefont {Cvitanovi\'{c}}},\
  }\href@noop {} {\bibfield  {journal} {\bibinfo  {journal} {Nonlinearity}\
  }\textbf {\bibinfo {volume} {3}},\ \bibinfo {pages} {361} (\bibinfo {year}
  {1990}{\natexlab{b}})}\BibitemShut {NoStop}%
\bibitem [{\citenamefont {Cvitanovi\'{c}}(1991)}]{Cvitanovic91}%
  \BibitemOpen
  \bibfield  {author} {\bibinfo {author} {\bibfnamefont {P.}~\bibnamefont
  {Cvitanovi\'{c}}},\ }\href@noop {} {\bibfield  {journal} {\bibinfo  {journal}
  {Physica~D}\ }\textbf {\bibinfo {volume} {51}},\ \bibinfo {pages} {138}
  (\bibinfo {year} {1991})}\BibitemShut {NoStop}%
\bibitem [{\citenamefont {Du}\ and\ \citenamefont
  {Delos}(1988{\natexlab{a}})}]{Du88a}%
  \BibitemOpen
  \bibfield  {author} {\bibinfo {author} {\bibfnamefont {M.~L.}\ \bibnamefont
  {Du}}\ and\ \bibinfo {author} {\bibfnamefont {J.~B.}\ \bibnamefont {Delos}},\
  }\href@noop {} {\bibfield  {journal} {\bibinfo  {journal} {Phys.~Rev.~A}\
  }\textbf {\bibinfo {volume} {38}},\ \bibinfo {pages} {1896} (\bibinfo {year}
  {1988}{\natexlab{a}})}\BibitemShut {NoStop}%
\bibitem [{\citenamefont {Du}\ and\ \citenamefont
  {Delos}(1988{\natexlab{b}})}]{Du88b}%
  \BibitemOpen
  \bibfield  {author} {\bibinfo {author} {\bibfnamefont {M.~L.}\ \bibnamefont
  {Du}}\ and\ \bibinfo {author} {\bibfnamefont {J.~B.}\ \bibnamefont {Delos}},\
  }\href@noop {} {\bibfield  {journal} {\bibinfo  {journal} {Phys.~Rev.~A}\
  }\textbf {\bibinfo {volume} {38}},\ \bibinfo {pages} {1913} (\bibinfo {year}
  {1988}{\natexlab{b}})}\BibitemShut {NoStop}%
\bibitem [{\citenamefont {Friedrich}\ and\ \citenamefont
  {Wintgen}(1989)}]{Friedrich89}%
  \BibitemOpen
  \bibfield  {author} {\bibinfo {author} {\bibfnamefont {H.}~\bibnamefont
  {Friedrich}}\ and\ \bibinfo {author} {\bibfnamefont {D.}~\bibnamefont
  {Wintgen}},\ }\href@noop {} {\bibfield  {journal} {\bibinfo  {journal}
  {Phys.~Rep.}\ }\textbf {\bibinfo {volume} {183}},\ \bibinfo {pages} {37}
  (\bibinfo {year} {1989})}\BibitemShut {NoStop}%
\bibitem [{\citenamefont {Tomsovic}\ and\ \citenamefont
  {Heller}(1991)}]{Tomsovic91b}%
  \BibitemOpen
  \bibfield  {author} {\bibinfo {author} {\bibfnamefont {S.}~\bibnamefont
  {Tomsovic}}\ and\ \bibinfo {author} {\bibfnamefont {E.~J.}\ \bibnamefont
  {Heller}},\ }\href@noop {} {\bibfield  {journal} {\bibinfo  {journal}
  {Phys.~Rev.~Lett.}\ }\textbf {\bibinfo {volume} {67}},\ \bibinfo {pages}
  {664} (\bibinfo {year} {1991})}\BibitemShut {NoStop}%
\bibitem [{\citenamefont {Tomsovic}\ and\ \citenamefont
  {Heller}(1993)}]{Tomsovic93}%
  \BibitemOpen
  \bibfield  {author} {\bibinfo {author} {\bibfnamefont {S.}~\bibnamefont
  {Tomsovic}}\ and\ \bibinfo {author} {\bibfnamefont {E.~J.}\ \bibnamefont
  {Heller}},\ }\href@noop {} {\bibfield  {journal} {\bibinfo  {journal}
  {Phys.~Rev.~E}\ }\textbf {\bibinfo {volume} {47}},\ \bibinfo {pages} {282}
  (\bibinfo {year} {1993})}\BibitemShut {NoStop}%
\bibitem [{\citenamefont {Li}\ and\ \citenamefont
  {Tomsovic}(2017{\natexlab{a}})}]{Li17a}%
  \BibitemOpen
  \bibfield  {author} {\bibinfo {author} {\bibfnamefont {J.}~\bibnamefont
  {Li}}\ and\ \bibinfo {author} {\bibfnamefont {S.}~\bibnamefont {Tomsovic}},\
  }\href@noop {} {\bibfield  {journal} {\bibinfo  {journal} {Phys.~Rev.~E}\
  }\textbf {\bibinfo {volume} {95}},\ \bibinfo {pages} {062224} (\bibinfo
  {year} {2017}{\natexlab{a}})},\ \bibinfo {note} {arXiv:1703.07045
  [nlin.CD]}\BibitemShut {NoStop}%
\bibitem [{\citenamefont {Li}\ and\ \citenamefont {Tomsovic}(2018)}]{Li18}%
  \BibitemOpen
  \bibfield  {author} {\bibinfo {author} {\bibfnamefont {J.}~\bibnamefont
  {Li}}\ and\ \bibinfo {author} {\bibfnamefont {S.}~\bibnamefont {Tomsovic}},\
  }\href@noop {} {\bibfield  {journal} {\bibinfo  {journal} {Phys.~Rev.~E}\
  }\textbf {\bibinfo {volume} {97}},\ \bibinfo {pages} {022216} (\bibinfo
  {year} {2018})},\ \bibinfo {note} {arXiv:1712.05568 [nlin.CD]}\BibitemShut
  {NoStop}%
\bibitem [{\citenamefont {Li}\ and\ \citenamefont {Tomsovic}(2019)}]{Li19a}%
  \BibitemOpen
  \bibfield  {author} {\bibinfo {author} {\bibfnamefont {J.}~\bibnamefont
  {Li}}\ and\ \bibinfo {author} {\bibfnamefont {S.}~\bibnamefont {Tomsovic}},\
  }\href@noop {} {\bibfield  {journal} {\bibinfo  {journal} {Phys.~Rev.~E}\
  }\textbf {\bibinfo {volume} {100}},\ \bibinfo {pages} {052202} (\bibinfo
  {year} {2019})},\ \bibinfo {note} {arXiv:1909.00544 [nlin.CD]}\BibitemShut
  {NoStop}%
\bibitem [{\citenamefont {Rom-Kedar}(1990)}]{Rom-Kedar90}%
  \BibitemOpen
  \bibfield  {author} {\bibinfo {author} {\bibfnamefont {V.}~\bibnamefont
  {Rom-Kedar}},\ }\href@noop {} {\bibfield  {journal} {\bibinfo  {journal}
  {Physica~D}\ }\textbf {\bibinfo {volume} {43}},\ \bibinfo {pages} {229}
  (\bibinfo {year} {1990})}\BibitemShut {NoStop}%
\bibitem [{\citenamefont {Li}\ and\ \citenamefont
  {Tomsovic}(2017{\natexlab{b}})}]{Li17}%
  \BibitemOpen
  \bibfield  {author} {\bibinfo {author} {\bibfnamefont {J.}~\bibnamefont
  {Li}}\ and\ \bibinfo {author} {\bibfnamefont {S.}~\bibnamefont {Tomsovic}},\
  }\href@noop {} {\bibfield  {journal} {\bibinfo  {journal} {J.~Phys.~A:
  Math.~Theor.}\ }\textbf {\bibinfo {volume} {50}},\ \bibinfo {pages} {135101}
  (\bibinfo {year} {2017}{\natexlab{b}})},\ \bibinfo {note} {arXiv:1507.06455
  [nlin.CD]}\BibitemShut {NoStop}%
\bibitem [{\citenamefont {Gutzwiller}(1971)}]{Gutzwiller71}%
  \BibitemOpen
  \bibfield  {author} {\bibinfo {author} {\bibfnamefont {M.~C.}\ \bibnamefont
  {Gutzwiller}},\ }\href@noop {} {\bibfield  {journal} {\bibinfo  {journal}
  {J.~Math.~Phys.}\ }\textbf {\bibinfo {volume} {12}},\ \bibinfo {pages} {343}
  (\bibinfo {year} {1971})},\ \bibinfo {note} {and references
  therein}\BibitemShut {NoStop}%
\bibitem [{\citenamefont {Cvitanovi\'c}\ \emph {et~al.}()\citenamefont
  {Cvitanovi\'c}, \citenamefont {Artuso}, \citenamefont {Dahlqvist},
  \citenamefont {Mainieri}, \citenamefont {Tanner}, \citenamefont {Vattay},
  \citenamefont {Whelan},\ and\ \citenamefont {Wirzba}}]{Chaosbook1}%
  \BibitemOpen
  \bibfield  {author} {\bibinfo {author} {\bibfnamefont {P.}~\bibnamefont
  {Cvitanovi\'c}}, \bibinfo {author} {\bibfnamefont {R.}~\bibnamefont
  {Artuso}}, \bibinfo {author} {\bibfnamefont {P.}~\bibnamefont {Dahlqvist}},
  \bibinfo {author} {\bibfnamefont {R.}~\bibnamefont {Mainieri}}, \bibinfo
  {author} {\bibfnamefont {G.}~\bibnamefont {Tanner}}, \bibinfo {author}
  {\bibfnamefont {G.}~\bibnamefont {Vattay}}, \bibinfo {author} {\bibfnamefont
  {N.}~\bibnamefont {Whelan}}, \ and\ \bibinfo {author} {\bibfnamefont
  {A.}~\bibnamefont {Wirzba}},\ }\href@noop {} {\bibinfo  {journal}
  {chaosbook.org}\ ,\ \bibinfo {pages} {1}}\BibitemShut {NoStop}%
\bibitem [{\citenamefont {Oseledec}(1968)}]{Oseledec68}%
  \BibitemOpen
\bibfield  {journal} {  }\bibfield  {author} {\bibinfo {author} {\bibfnamefont
  {V.~I.}\ \bibnamefont {Oseledec}},\ }\href@noop {} {\bibfield  {journal}
  {\bibinfo  {journal} {Trudy~Moskov.~Mat.~Ob\v{s}\v{c}.}\ }\textbf {\bibinfo
  {volume} {19}},\ \bibinfo {pages} {197} (\bibinfo {year} {1968})}\BibitemShut
  {NoStop}%
\bibitem [{\citenamefont {Ginelli}\ \emph {et~al.}(2007)\citenamefont
  {Ginelli}, \citenamefont {Poggi}, \citenamefont {Turchi}, \citenamefont
  {Chat\'{e}}, \citenamefont {Livi},\ and\ \citenamefont {Politi}}]{Ginelli07}%
  \BibitemOpen
  \bibfield  {author} {\bibinfo {author} {\bibfnamefont {F.}~\bibnamefont
  {Ginelli}}, \bibinfo {author} {\bibfnamefont {P.}~\bibnamefont {Poggi}},
  \bibinfo {author} {\bibfnamefont {A.}~\bibnamefont {Turchi}}, \bibinfo
  {author} {\bibfnamefont {H.}~\bibnamefont {Chat\'{e}}}, \bibinfo {author}
  {\bibfnamefont {R.}~\bibnamefont {Livi}}, \ and\ \bibinfo {author}
  {\bibfnamefont {A.}~\bibnamefont {Politi}},\ }\href@noop {} {\bibfield
  {journal} {\bibinfo  {journal} {Phys.~Rev.~Lett.}\ }\textbf {\bibinfo
  {volume} {99}},\ \bibinfo {pages} {130601} (\bibinfo {year}
  {2007})}\BibitemShut {NoStop}%
\bibitem [{\citenamefont {Ginelli}\ \emph {et~al.}(2013)\citenamefont
  {Ginelli}, \citenamefont {Chat\'{e}}, \citenamefont {Livi},\ and\
  \citenamefont {Politi}}]{Ginelli13}%
  \BibitemOpen
  \bibfield  {author} {\bibinfo {author} {\bibfnamefont {F.}~\bibnamefont
  {Ginelli}}, \bibinfo {author} {\bibfnamefont {H.}~\bibnamefont {Chat\'{e}}},
  \bibinfo {author} {\bibfnamefont {R.}~\bibnamefont {Livi}}, \ and\ \bibinfo
  {author} {\bibfnamefont {A.}~\bibnamefont {Politi}},\ }\href@noop {}
  {\bibfield  {journal} {\bibinfo  {journal} {J.~Phys.~A: Math.~Theor.}\
  }\textbf {\bibinfo {volume} {46}},\ \bibinfo {pages} {254005} (\bibinfo
  {year} {2013})}\BibitemShut {NoStop}%
\bibitem [{\citenamefont {Kuptsov}\ and\ \citenamefont
  {Parlitz}(2012)}]{Kuptsov12}%
  \BibitemOpen
  \bibfield  {author} {\bibinfo {author} {\bibfnamefont {P.~V.}\ \bibnamefont
  {Kuptsov}}\ and\ \bibinfo {author} {\bibfnamefont {U.}~\bibnamefont
  {Parlitz}},\ }\href@noop {} {\bibfield  {journal} {\bibinfo  {journal}
  {J.~Nonlinear.~Sci.}\ }\textbf {\bibinfo {volume} {22}},\ \bibinfo {pages}
  {727} (\bibinfo {year} {2012})}\BibitemShut {NoStop}%
\bibitem [{\citenamefont {Easton}(1986)}]{Easton86}%
  \BibitemOpen
  \bibfield  {author} {\bibinfo {author} {\bibfnamefont {R.~W.}\ \bibnamefont
  {Easton}},\ }\href@noop {} {\bibfield  {journal} {\bibinfo  {journal}
  {Trans.~Am.~Math.~Soc.}\ }\textbf {\bibinfo {volume} {294}},\ \bibinfo
  {pages} {719} (\bibinfo {year} {1986})}\BibitemShut {NoStop}%
\bibitem [{\citenamefont {Bowen}(1975)}]{Bowen75}%
  \BibitemOpen
  \bibfield  {author} {\bibinfo {author} {\bibfnamefont {R.}~\bibnamefont
  {Bowen}},\ }\href@noop {} {\emph {\bibinfo {title} {Lect. Notes in Math. Vol.
  470.}}}\ (\bibinfo  {publisher} {Springer-Verlag},\ \bibinfo {address}
  {Berlin},\ \bibinfo {year} {1975})\BibitemShut {NoStop}%
\bibitem [{\citenamefont {Gaspard}(1998)}]{Gaspard98}%
  \BibitemOpen
  \bibfield  {author} {\bibinfo {author} {\bibfnamefont {P.}~\bibnamefont
  {Gaspard}},\ }\href@noop {} {\emph {\bibinfo {title} {Chaos, Scattering and
  Statistical Mechanics}}}\ (\bibinfo  {publisher} {Cambridge University
  Press},\ \bibinfo {address} {Cambridge, UK},\ \bibinfo {year}
  {1998})\BibitemShut {NoStop}%
\bibitem [{\citenamefont {Hadamard}(1898)}]{Hadamard1898}%
  \BibitemOpen
  \bibfield  {author} {\bibinfo {author} {\bibfnamefont {J.}~\bibnamefont
  {Hadamard}},\ }\href@noop {} {\bibfield  {journal} {\bibinfo  {journal}
  {J.~Math.~Pures~Appl.~series 5}\ }\textbf {\bibinfo {volume} {4}},\ \bibinfo
  {pages} {27} (\bibinfo {year} {1898})}\BibitemShut {NoStop}%
\bibitem [{\citenamefont {Birkhoff}(1927)}]{Birkhoff27a}%
  \BibitemOpen
  \bibfield  {author} {\bibinfo {author} {\bibfnamefont {G.~D.}\ \bibnamefont
  {Birkhoff}},\ }\href@noop {} {\emph {\bibinfo {title} {A.M.S. Coll.
  Publications, vol. 9}}}\ (\bibinfo  {publisher} {American Mathematical
  Society},\ \bibinfo {address} {Providence},\ \bibinfo {year}
  {1927})\BibitemShut {NoStop}%
\bibitem [{\citenamefont {Birkhoff}(1935)}]{Birkhoff35}%
  \BibitemOpen
  \bibfield  {author} {\bibinfo {author} {\bibfnamefont {G.~D.}\ \bibnamefont
  {Birkhoff}},\ }\href@noop {} {\bibfield  {journal} {\bibinfo  {journal}
  {Mem.~Pont.~Acad.~Sci.~Novi~Lyncaei}\ }\textbf {\bibinfo {volume} {1}},\
  \bibinfo {pages} {85} (\bibinfo {year} {1935})}\BibitemShut {NoStop}%
\bibitem [{\citenamefont {Morse}\ and\ \citenamefont
  {Hedlund}(1938)}]{Morse38}%
  \BibitemOpen
  \bibfield  {author} {\bibinfo {author} {\bibfnamefont {M.}~\bibnamefont
  {Morse}}\ and\ \bibinfo {author} {\bibfnamefont {G.~A.}\ \bibnamefont
  {Hedlund}},\ }\href@noop {} {\bibfield  {journal} {\bibinfo  {journal}
  {Amer.~J.~Math.}\ }\textbf {\bibinfo {volume} {60}},\ \bibinfo {pages} {815}
  (\bibinfo {year} {1938})}\BibitemShut {NoStop}%
\bibitem [{\citenamefont {Wiggins}(1988)}]{Wiggins88}%
  \BibitemOpen
  \bibfield  {author} {\bibinfo {author} {\bibfnamefont {S.}~\bibnamefont
  {Wiggins}},\ }\href@noop {} {\emph {\bibinfo {title} {Global Bifurcations and
  Chaos}}}\ (\bibinfo  {publisher} {Springer-Verlag},\ \bibinfo {address} {New
  York, Berlin, Heidelberg},\ \bibinfo {year} {1988})\BibitemShut {NoStop}%
\bibitem [{\citenamefont {Cvitanovi\'{c}}\ \emph {et~al.}(1988)\citenamefont
  {Cvitanovi\'{c}}, \citenamefont {Gunaratne},\ and\ \citenamefont
  {Procaccia}}]{Cvitanovic88a}%
  \BibitemOpen
  \bibfield  {author} {\bibinfo {author} {\bibfnamefont {P.}~\bibnamefont
  {Cvitanovi\'{c}}}, \bibinfo {author} {\bibfnamefont {G.}~\bibnamefont
  {Gunaratne}}, \ and\ \bibinfo {author} {\bibfnamefont {I.}~\bibnamefont
  {Procaccia}},\ }\href@noop {} {\bibfield  {journal} {\bibinfo  {journal}
  {Phys.~Rev.~A}\ }\textbf {\bibinfo {volume} {38}},\ \bibinfo {pages} {1503}
  (\bibinfo {year} {1988})}\BibitemShut {NoStop}%
\bibitem [{\citenamefont {Hagiwara}\ and\ \citenamefont
  {Shudo}(2004)}]{Hagiwara04}%
  \BibitemOpen
  \bibfield  {author} {\bibinfo {author} {\bibfnamefont {R.}~\bibnamefont
  {Hagiwara}}\ and\ \bibinfo {author} {\bibfnamefont {A.}~\bibnamefont
  {Shudo}},\ }\href@noop {} {\bibfield  {journal} {\bibinfo  {journal}
  {J.~Phys.~A: Math.~Gen.}\ }\textbf {\bibinfo {volume} {37}},\ \bibinfo
  {pages} {10521–10543} (\bibinfo {year} {2004})}\BibitemShut {NoStop}%
\bibitem [{\citenamefont {MacKay}\ \emph {et~al.}(1984)\citenamefont {MacKay},
  \citenamefont {Meiss},\ and\ \citenamefont {Percival}}]{MacKay84a}%
  \BibitemOpen
  \bibfield  {author} {\bibinfo {author} {\bibfnamefont {R.~S.}\ \bibnamefont
  {MacKay}}, \bibinfo {author} {\bibfnamefont {J.~D.}\ \bibnamefont {Meiss}}, \
  and\ \bibinfo {author} {\bibfnamefont {I.~C.}\ \bibnamefont {Percival}},\
  }\href@noop {} {\bibfield  {journal} {\bibinfo  {journal} {Physica~D}\
  }\textbf {\bibinfo {volume} {13}},\ \bibinfo {pages} {55} (\bibinfo {year}
  {1984})}\BibitemShut {NoStop}%
\bibitem [{\citenamefont {Meiss}(1992)}]{Meiss92}%
  \BibitemOpen
  \bibfield  {author} {\bibinfo {author} {\bibfnamefont {J.~D.}\ \bibnamefont
  {Meiss}},\ }\href@noop {} {\bibfield  {journal} {\bibinfo  {journal}
  {Rev.~Mod.~Phys.}\ }\textbf {\bibinfo {volume} {64}},\ \bibinfo {pages} {795}
  (\bibinfo {year} {1992})}\BibitemShut {NoStop}%
\bibitem [{\citenamefont {de~Almeida}\ and\ \citenamefont
  {Saraceno}(1991)}]{Ozorio91}%
  \BibitemOpen
  \bibfield  {author} {\bibinfo {author} {\bibfnamefont {A.~M.~O.}\
  \bibnamefont {de~Almeida}}\ and\ \bibinfo {author} {\bibfnamefont
  {M.}~\bibnamefont {Saraceno}},\ }\href@noop {} {\bibfield  {journal}
  {\bibinfo  {journal} {Ann.~Phys.}\ }\textbf {\bibinfo {volume} {210}},\
  \bibinfo {pages} {1} (\bibinfo {year} {1991})}\BibitemShut {NoStop}%
\bibitem [{\citenamefont {O'Connor}\ and\ \citenamefont
  {Tomsovic}(1991)}]{Oconnor91}%
  \BibitemOpen
  \bibfield  {author} {\bibinfo {author} {\bibfnamefont {P.~W.}\ \bibnamefont
  {O'Connor}}\ and\ \bibinfo {author} {\bibfnamefont {S.}~\bibnamefont
  {Tomsovic}},\ }\href@noop {} {\bibfield  {journal} {\bibinfo  {journal}
  {Ann.~Phys. (N.Y.)}\ }\textbf {\bibinfo {volume} {207}},\ \bibinfo {pages}
  {218 } (\bibinfo {year} {1991})}\BibitemShut {NoStop}%
\bibitem [{\citenamefont {O'Connor}\ \emph {et~al.}(1992)\citenamefont
  {O'Connor}, \citenamefont {Tomsovic},\ and\ \citenamefont
  {Heller}}]{Oconnor92}%
  \BibitemOpen
  \bibfield  {author} {\bibinfo {author} {\bibfnamefont {P.~W.}\ \bibnamefont
  {O'Connor}}, \bibinfo {author} {\bibfnamefont {S.}~\bibnamefont {Tomsovic}},
  \ and\ \bibinfo {author} {\bibfnamefont {E.~J.}\ \bibnamefont {Heller}},\
  }\href@noop {} {\bibfield  {journal} {\bibinfo  {journal} {Physica D}\
  }\textbf {\bibinfo {volume} {55}},\ \bibinfo {pages} {340} (\bibinfo {year}
  {1992})}\BibitemShut {NoStop}%
\bibitem [{\citenamefont {Gutzwiller}(1990)}]{Gutzwiller90}%
  \BibitemOpen
  \bibfield  {author} {\bibinfo {author} {\bibfnamefont {M.~C.}\ \bibnamefont
  {Gutzwiller}},\ }\href@noop {} {\emph {\bibinfo {title} {Chaos in Classical
  and Quantum Mechanics}}}\ (\bibinfo  {publisher} {Springer-Verlag},\ \bibinfo
  {address} {New York},\ \bibinfo {year} {1990})\BibitemShut {NoStop}%
\bibitem [{\citenamefont {Cvitanovi\'{c}}(1988)}]{Cvitanovic88}%
  \BibitemOpen
  \bibfield  {author} {\bibinfo {author} {\bibfnamefont {P.}~\bibnamefont
  {Cvitanovi\'{c}}},\ }\href@noop {} {\bibfield  {journal} {\bibinfo  {journal}
  {Phys.~Rev.~Lett.}\ }\textbf {\bibinfo {volume} {61}},\ \bibinfo {pages}
  {2729} (\bibinfo {year} {1988})}\BibitemShut {NoStop}%
\bibitem [{\citenamefont {Lan}(2010)}]{Lan10}%
  \BibitemOpen
  \bibfield  {author} {\bibinfo {author} {\bibfnamefont {Y.}~\bibnamefont
  {Lan}},\ }\href@noop {} {\bibfield  {journal} {\bibinfo  {journal}
  {Commun.~Nonlinear.~Sci.~Numer.~Simulat}\ }\textbf {\bibinfo {volume} {15}},\
  \bibinfo {pages} {502} (\bibinfo {year} {2010})}\BibitemShut {NoStop}%
\bibitem [{\citenamefont {Suri}\ \emph {et~al.}(2017)\citenamefont {Suri},
  \citenamefont {Tithof}, \citenamefont {Grigoriev},\ and\ \citenamefont
  {Schatz}}]{Suri17}%
  \BibitemOpen
  \bibfield  {author} {\bibinfo {author} {\bibfnamefont {B.}~\bibnamefont
  {Suri}}, \bibinfo {author} {\bibfnamefont {J.}~\bibnamefont {Tithof}},
  \bibinfo {author} {\bibfnamefont {R.~O.}\ \bibnamefont {Grigoriev}}, \ and\
  \bibinfo {author} {\bibfnamefont {M.~F.}\ \bibnamefont {Schatz}},\
  }\href@noop {} {\bibfield  {journal} {\bibinfo  {journal} {Phys.~Rev.~Lett.}\
  }\textbf {\bibinfo {volume} {118}},\ \bibinfo {pages} {114501} (\bibinfo
  {year} {2017})}\BibitemShut {NoStop}%
\bibitem [{\citenamefont {Suri}\ \emph {et~al.}(2020)\citenamefont {Suri},
  \citenamefont {Kageorge}, \citenamefont {Grigoriev},\ and\ \citenamefont
  {Schatz}}]{Suri20}%
  \BibitemOpen
  \bibfield  {author} {\bibinfo {author} {\bibfnamefont {B.}~\bibnamefont
  {Suri}}, \bibinfo {author} {\bibfnamefont {L.}~\bibnamefont {Kageorge}},
  \bibinfo {author} {\bibfnamefont {R.~O.}\ \bibnamefont {Grigoriev}}, \ and\
  \bibinfo {author} {\bibfnamefont {M.~F.}\ \bibnamefont {Schatz}},\
  }\href@noop {} {\bibfield  {journal} {\bibinfo  {journal} {Phys.~Rev.~Lett.}\
  }\textbf {\bibinfo {volume} {125}},\ \bibinfo {pages} {064501} (\bibinfo
  {year} {2020})}\BibitemShut {NoStop}%
\bibitem [{\citenamefont {Yaln{\i}z}\ \emph {et~al.}(2020)\citenamefont
  {Yaln{\i}z}, \citenamefont {Hof},\ and\ \citenamefont {Budanur}}]{Yalniz20}%
  \BibitemOpen
  \bibfield  {author} {\bibinfo {author} {\bibfnamefont {G.}~\bibnamefont
  {Yaln{\i}z}}, \bibinfo {author} {\bibfnamefont {B.}~\bibnamefont {Hof}}, \
  and\ \bibinfo {author} {\bibfnamefont {N.~B.}\ \bibnamefont {Budanur}},\
  }\href@noop {} {\bibfield  {journal} {\bibinfo  {journal} {arXiv:2007.02584
  [physics.flu-dyn]}\ } (\bibinfo {year} {2020})}\BibitemShut {NoStop}%
\bibitem [{\citenamefont {Smale}(1963)}]{Smale63}%
  \BibitemOpen
  \bibfield  {author} {\bibinfo {author} {\bibfnamefont {S.}~\bibnamefont
  {Smale}},\ }\href@noop {} {\emph {\bibinfo {title} {Differential and
  Combinatorial Topology}}},\ edited by\ \bibinfo {editor} {\bibfnamefont
  {S.~S.}\ \bibnamefont {Cairns}}\ (\bibinfo  {publisher} {Princeton University
  Press},\ \bibinfo {address} {Princeton},\ \bibinfo {year} {1963})\BibitemShut
  {NoStop}%
\bibitem [{\citenamefont {Smale}(1980)}]{Smale80}%
  \BibitemOpen
  \bibfield  {author} {\bibinfo {author} {\bibfnamefont {S.}~\bibnamefont
  {Smale}},\ }\href@noop {} {\emph {\bibinfo {title} {The Mathematics of Time:
  Essays on Dynamical Systems, Economic Processes and Related Topics}}}\
  (\bibinfo  {publisher} {Springer-Verlag},\ \bibinfo {address} {New York,
  Heidelberg, Berlin},\ \bibinfo {year} {1980})\BibitemShut {NoStop}%
\bibitem [{\citenamefont {Devaney}\ and\ \citenamefont
  {Nitecki}(1979)}]{Devaney79}%
  \BibitemOpen
  \bibfield  {author} {\bibinfo {author} {\bibfnamefont {R.}~\bibnamefont
  {Devaney}}\ and\ \bibinfo {author} {\bibfnamefont {Z.}~\bibnamefont
  {Nitecki}},\ }\href@noop {} {\bibfield  {journal} {\bibinfo  {journal}
  {Comm.~Math.~Phys.}\ }\textbf {\bibinfo {volume} {67}},\ \bibinfo {pages}
  {137} (\bibinfo {year} {1979})}\BibitemShut {NoStop}%
\bibitem [{\citenamefont {Sieber}\ and\ \citenamefont
  {Richter}(2001)}]{Sieber01}%
  \BibitemOpen
  \bibfield  {author} {\bibinfo {author} {\bibfnamefont {M.}~\bibnamefont
  {Sieber}}\ and\ \bibinfo {author} {\bibfnamefont {K.}~\bibnamefont
  {Richter}},\ }\href@noop {} {\bibfield  {journal} {\bibinfo  {journal}
  {Physica Scripta}\ }\textbf {\bibinfo {volume} {T90}},\ \bibinfo {pages}
  {128} (\bibinfo {year} {2001})}\BibitemShut {NoStop}%
\bibitem [{\citenamefont {Katok}\ and\ \citenamefont
  {Hasselblatt}(1995)}]{Katok95}%
  \BibitemOpen
  \bibfield  {author} {\bibinfo {author} {\bibfnamefont {A.}~\bibnamefont
  {Katok}}\ and\ \bibinfo {author} {\bibfnamefont {B.}~\bibnamefont
  {Hasselblatt}},\ }\href@noop {} {\emph {\bibinfo {title} {Introduction to the
  Modern Theory of Dynamical Systems}}}\ (\bibinfo  {publisher} {Cambridge
  University Press},\ \bibinfo {address} {Cambridge},\ \bibinfo {year}
  {1995})\BibitemShut {NoStop}%
\bibitem [{\citenamefont {Rubido}\ \emph {et~al.}(2018)\citenamefont {Rubido},
  \citenamefont {Grebogi},\ and\ \citenamefont {Baptista}}]{Rubido18}%
  \BibitemOpen
  \bibfield  {author} {\bibinfo {author} {\bibfnamefont {N.}~\bibnamefont
  {Rubido}}, \bibinfo {author} {\bibfnamefont {C.}~\bibnamefont {Grebogi}}, \
  and\ \bibinfo {author} {\bibfnamefont {M.~S.}\ \bibnamefont {Baptista}},\
  }\href@noop {} {\bibfield  {journal} {\bibinfo  {journal} {Chaos}\ }\textbf
  {\bibinfo {volume} {28}},\ \bibinfo {pages} {033611} (\bibinfo {year}
  {2018})}\BibitemShut {NoStop}%
\bibitem [{\citenamefont {Zhang}\ and\ \citenamefont {Lan}(2020)}]{Zhang20}%
  \BibitemOpen
  \bibfield  {author} {\bibinfo {author} {\bibfnamefont {C.}~\bibnamefont
  {Zhang}}\ and\ \bibinfo {author} {\bibfnamefont {Y.}~\bibnamefont {Lan}},\
  }\href@noop {} {\bibfield  {journal} {\bibinfo  {journal} {arXiv:2007.11236
  [nlin.CD]}\ } (\bibinfo {year} {2020})}\BibitemShut {NoStop}%
\bibitem [{\citenamefont {H\'enon}(1976)}]{Henon76}%
  \BibitemOpen
  \bibfield  {author} {\bibinfo {author} {\bibfnamefont {M.}~\bibnamefont
  {H\'enon}},\ }\href@noop {} {\bibfield  {journal} {\bibinfo  {journal}
  {Comm.~Math.~Phys.}\ }\textbf {\bibinfo {volume} {50}},\ \bibinfo {pages}
  {69} (\bibinfo {year} {1976})}\BibitemShut {NoStop}%
\bibitem [{\citenamefont {Friedland}\ and\ \citenamefont
  {Milnor}(1989)}]{Friedland89}%
  \BibitemOpen
  \bibfield  {author} {\bibinfo {author} {\bibfnamefont {S.}~\bibnamefont
  {Friedland}}\ and\ \bibinfo {author} {\bibfnamefont {J.}~\bibnamefont
  {Milnor}},\ }\href@noop {} {\bibfield  {journal} {\bibinfo  {journal}
  {Ergod.~Th.~\&~Dynam.~Sys.}\ }\textbf {\bibinfo {volume} {9}},\ \bibinfo
  {pages} {67} (\bibinfo {year} {1989})}\BibitemShut {NoStop}%
\bibitem [{\citenamefont {Wiggins}(2003)}]{Wiggins03}%
  \BibitemOpen
  \bibfield  {author} {\bibinfo {author} {\bibfnamefont {S.}~\bibnamefont
  {Wiggins}},\ }\href@noop {} {\emph {\bibinfo {title} {Introduction to Applied
  Nonlinear Dynamical Systems and Chaos, Second Edition}}}\ (\bibinfo
  {publisher} {Springer-Verlag},\ \bibinfo {address} {New York, Berlin,
  Heidelberg},\ \bibinfo {year} {2003})\BibitemShut {NoStop}%
\bibitem [{\citenamefont {Cvitanovi{\'c}}\ \emph {et~al.}(2016)\citenamefont
  {Cvitanovi{\'c}}, \citenamefont {Artuso}, \citenamefont {Mainieri},
  \citenamefont {Tanner},\ and\ \citenamefont {Vattay}}]{ChaosBook}%
  \BibitemOpen
  \bibfield  {author} {\bibinfo {author} {\bibfnamefont {P.}~\bibnamefont
  {Cvitanovi{\'c}}}, \bibinfo {author} {\bibfnamefont {R.}~\bibnamefont
  {Artuso}}, \bibinfo {author} {\bibfnamefont {R.}~\bibnamefont {Mainieri}},
  \bibinfo {author} {\bibfnamefont {G.}~\bibnamefont {Tanner}}, \ and\ \bibinfo
  {author} {\bibfnamefont {G.}~\bibnamefont {Vattay}},\ }\href
  {http://ChaosBook.org/} {\emph {\bibinfo {title} {Chaos: Classical and
  Quantum}}}\ (\bibinfo  {publisher} {Niels Bohr Inst.},\ \bibinfo {address}
  {Copenhagen},\ \bibinfo {year} {2016})\BibitemShut {NoStop}%
\bibitem [{\citenamefont {Easton}(1991)}]{Easton91}%
  \BibitemOpen
  \bibfield  {author} {\bibinfo {author} {\bibfnamefont {R.}~\bibnamefont
  {Easton}},\ }\href@noop {} {\bibfield  {journal} {\bibinfo  {journal}
  {Nonlinearity}\ }\textbf {\bibinfo {volume} {4}},\ \bibinfo {pages} {583}
  (\bibinfo {year} {1991})}\BibitemShut {NoStop}%
\end{thebibliography}%

\end{document}